%% file: ExpSummary-hoecker-MoriondEW2016.tex
\documentclass{moriond}

\usepackage{xspace}
\usepackage{graphicx}
\usepackage{wrapfig}
\usepackage{parskip}

\input Definitions
\newcommand{\Photo}{}

\begin{document}

\vspace*{4cm}
\title{MORIOND ELECTROWEAK \& UNIFIED THEORIES 2016 \\[0.1cm] --- EXPERIMENTAL SUMMARY ---}

\author{Andreas Hoecker}

\address{CERN, CH-1211 Geneva 23, Switzerland \vspace{0.5cm} }

\maketitle\abstracts{
Summary of the experimental results presented at the 51$^{\rm st}$ edition 
of the Moriond Electroweak and Unified Theories conference held in March 2016 
at La Thuile, Italy. 
}

\section{Introduction}
\label{sec:introduction}

The 51$^{\rm st}$ Moriond Electroweak and Unified Theories conference 
featured, as is tradition,
a vibrant snapshot of newest results and trends in the fields of neutrino physics, 
astrophysics and cosmology, gravitational waves (!),  dark matter and collider 
physics (it became the promised LHC feast). There were 53 beautifully prepared talks 
in addition to young scientist presentations reporting a wealth of new experimental 
results that demonstrated once again that our field lives in data-driven times. The 
following is an attempt for a (necessarily incomplete) summary of the results presented. 

\section{Neutrino Physics}

The year 2015 has seen yet another Nobel Price for particle physics, and another 
one for neutrino oscillation. It was awarded jointly to Takaaki Kajita and Arthur B. 
McDonald from the Super-Kamiokande and Sudbury Neutrino Observatory
experiments, respectively, ``for the discovery of neutrino oscillations, 
which shows that neutrinos have mass''.$\,$\cite{nobel}

Since these dramatic developments at the turn of the millennium neutrino physics 
has come a long way. Beyond the established facts that neutrinos are massive 
fermions with three active flavours and mass eigenstates that are mixed flavour 
states, there are, however, yet critical questions. 
\begin{itemize}
\item[--]  What is the nature of the neutrinos, are they Majorana fermions?
\item[--]  While the absolute mass splitting, $\Delta m_{ij}^2$, and mixing angles, 
$\theta_{12}$, $\theta_{13}$, $\theta_{23}$, are known to about 3\% and 
3--7\%, respectively, the mass hierachy is not. By convention {\em normal hierarchy}
is dubbed the case where $m_{3}^2\gg m_{2}^2>m_{1}^2$ and 
{\em inverted hierarchy} stands for $m_{2}^2>m_{1}^2\gg m_{3}^2$.
\item[--] {\em CP} violation in the neutrino sector, described by the phase $\delta_{CP}$ for 
flavour-changing transitions in the Pontecorvo-Maki-Nakagawa-Sakata (PMNS) neutrino 
mixing matrix, is unknown so far.
\item[--]  Are there sterile neutrinos, i.e., neutrinos that interact only with gravity but 
  are singlets with respect to the Standard Model interactions? 
  Are there heavy additional right-handed neutrinos? 
  If so, are they in reach of current experiments?
\end{itemize}

And also: neutrino cross section and flux measurements and their theoretical 
predictions need to be improved. 

The experimental tools to get handles on these questions are neutrino oscillation 
measurements (short and long baseline),
single beta decay measurements, searches for neutrinoless double-beta decay, and
cosmology.$\:$\footnote{The combination of Lyman-$\alpha$, CMB and BAO data allows
to set the upper limit$\,$\cite{GrazianoRossi} $\sum m_\mu < 0.12\;{\rm eV}$.}
Neutrinos also serve as messengers in astronomy, Sun and Geo science, as 
well as for phenomena such as 
grand unification, lepto/baryogenesis and physics beyond the Standard Model. 
Given the amount and importance of the open questions, and the variety of the 
available tools, neutrino physics benefits from an exciting experimental programme. 

\subsection{Results from short-baseline neutrino experiments}

Low-energy scattering interactions of electron neutrinos or antineutrinos with matter 
has been a longstanding source of uncertainty. Apart from the controversial LSND 
result,$\,$\cite{LNSD} there was the 2013 electron-neutrino appearance 
measurement by the MiniBooNE experiment at Fermilab that revealed in both 
neutrino and antineutrino beam modes$\,$\cite{MiniBoone} an excess of events 
in the 0.2--0.4$\;$GeV electron-neutrino energy range over the  expectation, 
which is composed of (in order of importance) $\pi^0$ misidentification, 
$\Delta\to N\gamma$, muon and kaon decays, and other background sources. 
The excess appears electron-like in MiniBooNE's Cherenkov detector, which cannot
separate the signal from photon backgrounds. It is therefore important 
to have precise alternative low-energy cross-section
measurements. This is the task of the new MicroBooNE experiment at 
Fermilab that is installed $\sim$500$\;$m from the Booster Neutrino Beamline (BNB)
(anti)muon-neutrino beam, and is dedicated to low-energy neutrino cross sections 
measurements of (anti)electron appearance.$\,$\cite{SarahLockwitz} Because
the LAr-TPC tracking-calorimeter technique is similar to that of the future 
large-scale DUNE (LBNF) neutrino experiment, featuring a kiloton of such a 
detector, MicroBooNE also represents a pilot project of 
that experiment. In a LAr-TPC a charged particle interacts with the liquid argon,
wire planes detect drifting ionisation electrons ($\to$ tracks), photomultipliers 
detect scintillation light, and $dE/dx$ is used to separate between electrons and 
photons. Very first and promising commissioning results from October 2015 
with muon-neutrino beam scattering reactions in MicroBooNE's 170 ton LAr-TPC 
were presented at this conference. 

The MINERvA experiment at Fermilab performs detailed studies of neutrino 
interactions in varying nuclear targets (C, Pb, Fe, H$_2$O) with the aim to help
improve the modelling of these processes.$\,$\cite{DanielRutherbories}
For example, electron-neutrino
quasi-elastic charged-current (CCQE) scattering is an oscillation signal, but
only little low-energy cross-section data are available. Can the $\numu\to\nue$ 
cross-section measurements be universally trusted?
MINERvA sits on-axis at a short baseline along the NuMI (Neutrinos at the Main Injector) 
muon-neutrino beam, approximately 1$\;$km after the NuMI target. 
During the low-energy NuMI running the beam peaks at 3.1$\;$GeV muon-neutrino
energy. The MINERvA detector features charged particle as well as electromagnetic
and hadronic energy reconstruction, particle identification, and it uses the MINOS 
near detector as muon spectrometer. The exclusive measurement of 
flux-integrated differential cross sections for $\nue$ and $\nueb$ CCQE-like 
interactions ($\nue n \to e^-p$ and $\nueb p \to e^+n$) on nucleons in a 
hydrocarbon target by MINERvA and comparison with modelling expectations 
(from the neutrino event generator GENIE) exhibits sufficiently good modelling 
for the current needs of  the neutrino oscillation 
experiments.$\,$\cite{MinervaCCQE} A nearly three times larger dataset has been
already collected. The next step in the experimental programme consists of
measurements at higher neutrino beam energy.

\subsection{Results from long-baseline neutrino experiments}

There are three present programmes for long-baseline neutrino 
experiments at Fermilab (MINOS, NOvA), in Japan (Tokai-to-Kamioka --- T2K) 
and at CERN (OPERA). Long-baseline experiments measure  muon-neutrino 
disappearance and $\numu\to\nue$ 
appearance, as well as their anti-processes. Their probabilities depend on 
${\rm sin}^2(2\theta_{13})$, which is well measured and large, on 
${\rm sin}^2(2\theta_{23})$, $\Delta m^2_{32}$, and $\delta_{CP}$, and on 
the sign of $\Delta m^2_{31}$ that sets the mass hierarchy. All these 
properties can be experimentally addressed. 

\begin{figure}[t]
\centerline{\includegraphics[width=0.6\linewidth]{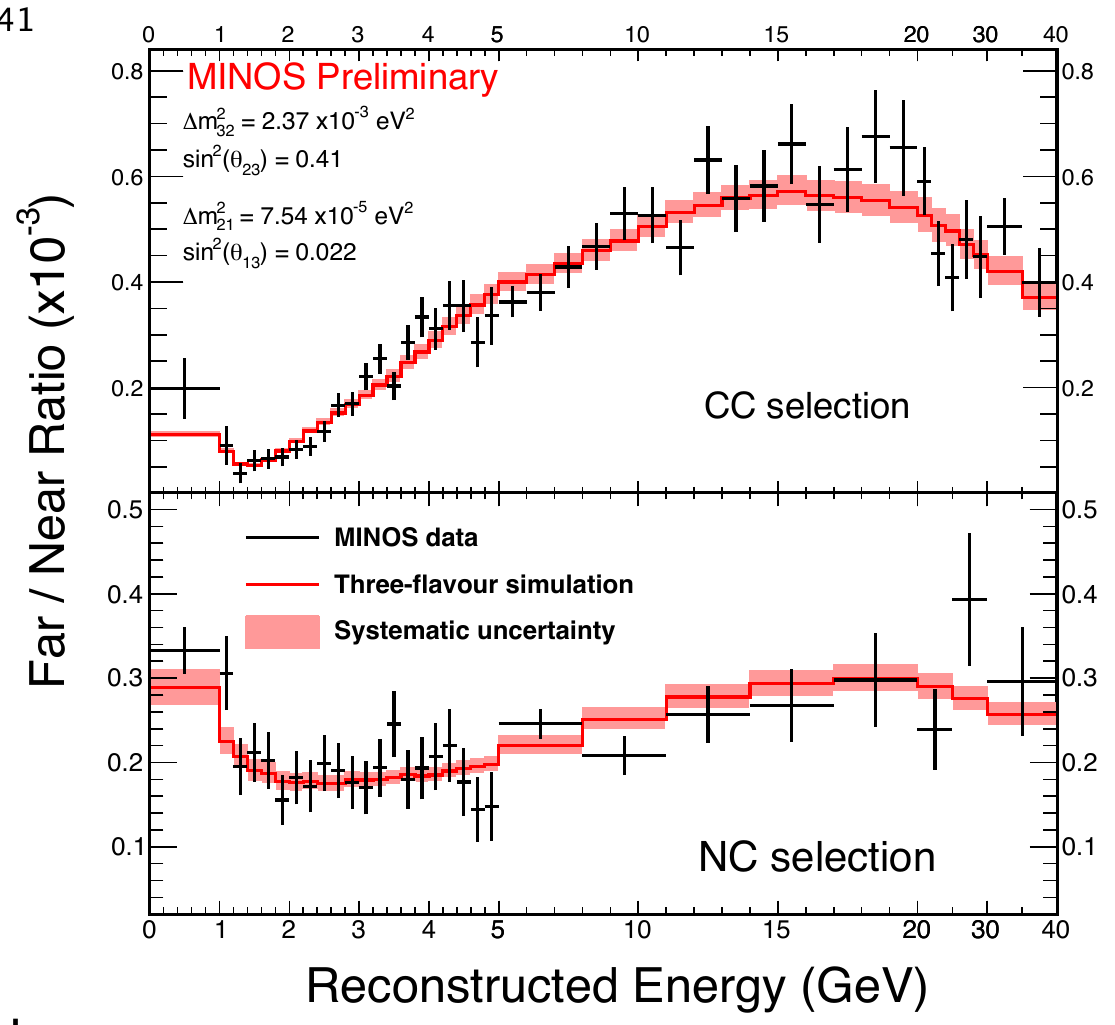}}
\vspace{-0.0cm}
\caption[]{Ratios of the far-to-near detector counts versus the reconstructed 
  neutrino energy for the charged-current (top panel) and 
  neutral-current (bottom panel) selected events. The red band shows the prediction
  of the three-neutrino-flavour model with  systematic uncertainty.
  \label{fig:MinosSterile}}
\end{figure}
The MINOS experiment consists of a 24 ton near detector (ND), placed about 1$\;$km 
from the NuMI beam target, and a 4.2 kiloton far detector (FD) installed 
735$\;$km away from the target and 705$\;$m underground in the Soudan mine.
Both near and far detectors are magnetised tracking/sampling calorimeters, 
segmented into planes of steel and scintillator strips. The detectors are 
designed to have equivalent functionality so that systematic uncertainties 
in the neutrino flux modelling and interaction cross sections cancel in the ratio.
MINOS released in May 2014 a combined analysis of its muon-neutrino 
disappearance and $\numu\to\nue$ appearance data with results for 
$\Delta m^2_{32}$ and ${\rm sin}^2\theta_{23}$. At this conference
MINOS reported on a search for sterile neutrino using the muon-neutrino 
beam.$\,$\cite{AshleyTimmons,MinosSterile} Presence of a fourth (sterile) 
neutrino ($\nu_{\rm sterile}$)  requires to introduce six new parameters to 
the PMNS matrix (three plus one flavour model). 
For simplicity the additional $CP$ phases and $\theta_{14}$
are set to zero, and the fit to data determines simultaneously the parameters
$\Delta m^2_{32}$, $\Delta m^2_{41}$, $\theta_{23}$, $\theta_{24}$, $\theta_{34}$.
Because $\nu_{\rm active}$--$\nu_{\rm sterile}$ mixing may affect the ND 
reference measurement, which conventionally is assumed not to be affected 
by neutrino oscillation, a combined fit of the FD/ND ratio is performed. 
That fit shows agreement with the three-flavour model (c.f. Fig.~\ref{fig:MinosSterile}) 
allowing to derive limits
on the additional four-flavour sterile neutrino parameters that improve over 
constraints from other experiments.

\begin{figure}[t]
\centerline{\includegraphics[width=\linewidth]{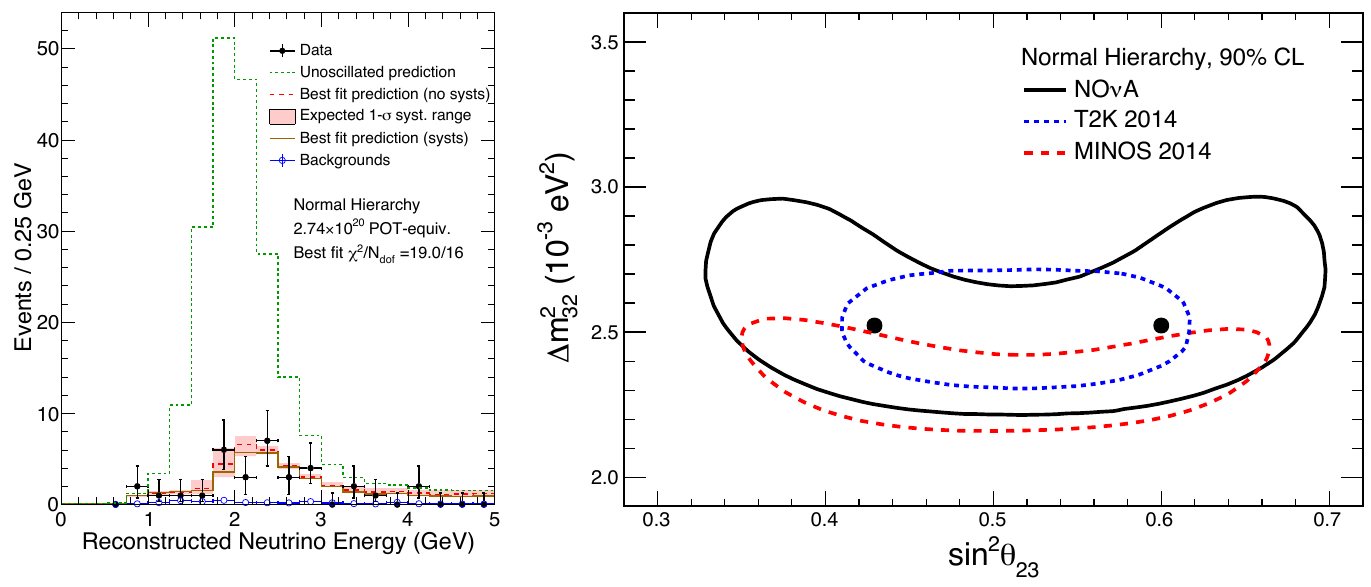}}
\vspace{-0.0cm}
\caption[]{The left panel shows the reconstructed neutrino energy distribution in the 
  NOvA far detector. The green dotted line indicates the expected distribution
  without $\numu$ disappearance. The data are significantly lower and well fitted
  with an oscillating signal. The oscillation parameter constraints obtained from these
  data are shown in the right panel, compared to other experiments. 
  \label{fig:NOvA}}
\end{figure}
The new NOvA long-baseline neutrino experiment at Fermilab consists of a
14 kiloton FD, 810$\;$km away from target, installed on surface, and a 0.3 kiloton 
ND, both using fine-grained tracking-calorimeter 
technology.$\,$\cite{JeffHartnell} NOvA is placed
$0.8^\circ$ off-axis from the NuMI beam so that the muon-neutrino 
beam$\,$\footnote{With magnetic horns focusing on positive mesons the NuMI 
beam is composed of 97.6\% $\numu$, 1.7\% $\numub$, 0.7\% $\nue$ and 
$\nueb$ for neutrino energies between 1 and 3$\;$GeV.} energy spread is 
reduced with peak at about 2$\;$GeV close to the maximum muon-neutrino 
disappearance and electron-neutrino appearance probabilities. NOvA allows
to identify electron-neutrino reactions. First NOvA results are based on data taken
between November 2014 and June 2015 with a low-intensity ($<500\;$kW) beam. 
Electron-neutrino cross-section measurements found somewhat larger values than 
T2K and Gargamelle, which is input to the GENIE modelling. 
An initial  measurement of muon-neutrino disappearance$\,$\cite{NOvADisap} 
provided a first constraint 
on $\Delta m^2_{32}$ and ${\rm sin}^2\theta_{23}$, both in agreement with 
earlier results from MINOS and T2K, but not yet reaching their 
precision (see Fig.~\ref{fig:NOvA}). A first $\numu\to\nue$ appearance 
measurement$\,$\cite{NOvAAp}
resulted in 6/11 events observed with the use of two different analysis methods 
(LID/LEM) in the FD for about one expected background event (estimate based on ND 
measurements). This corresponds to an excess of 3.3/5.3$\sigma$, respectively, 
with the LEM result being less compatible with the inverted hierarchy.
NOvA results with a twice larger dataset are forthcoming. Data with increased 
beam power (700$\;$kW) are expected to be taken in 2016. 

\begin{figure}[t]
\centerline{\includegraphics[width=\linewidth]{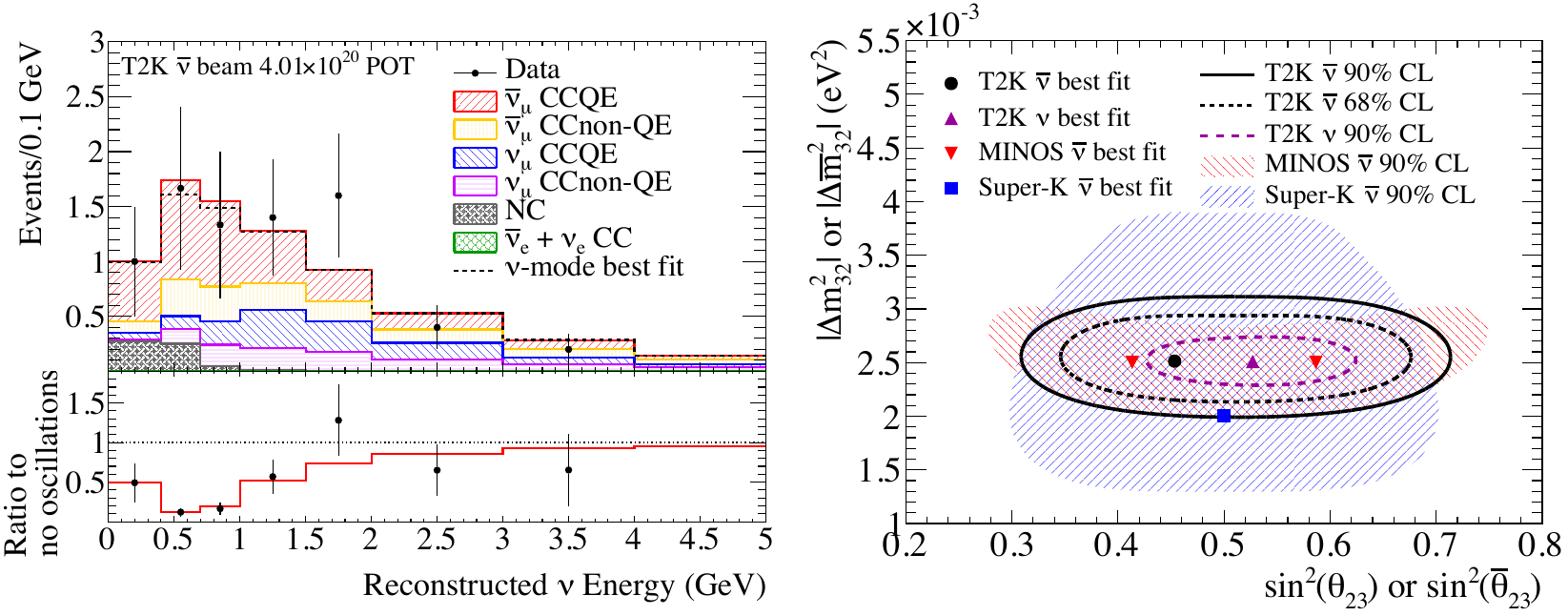}}
\vspace{-0.0cm}
\caption[]{Anti-muon-neutrino disappearance signal measured by T2K in 
  a (dominantly) anti-muon-neutrino beam. The left panel shows the 34 data events
  seen in the far detector compared to the best fit prediction. 
  The right panel shows the extracted (anti-)oscillation parameters. 
  \label{fig:T2K}}
\end{figure}
The Japan-based experiment T2K$\,$\cite{ChristineNielsen}
 has a 295$\;$km long baseline, using as FD 
Super-Kamiokande a Cherenkov detector with pure water as active material, 
and the NDs INGRID (on axis) and ND280 (off-axis), featuring different target 
materials, though currently only carbon was deployed. T2K is placed 2.5$^\circ$ 
off beam axis providing a narrow neutrino energy at a peak value of about 0.6 GeV. 
A combined $\numu$ disappearance and $\nue$ appearance analysis using T2K's
2010--2013 data provided the world's best measurements of 
$\Delta m^2_{32}$ and ${\rm sin}^2\theta_{23}$. During the 2014/2015 runs
T2K operated in $\numub$ beam mode with 390$\;$kW beam power collecting a
total of $11\cdot10^{20}$ protons-on-target (POT). The antineutrino beam being
less pure, the larger wrong-sign background must be measured in the ND giving
about 10\% flux and cross section systematic uncertainty. This is improved
with the use of a combined fit of the neutrino flux model together with external 
and ND280 data as input to the oscillation fit. Such a complex extraction is required 
because FD and ND use different target and measurement technologies. 
The measurement of $\numub$ disappearance yielded a significant deficit with 
only 34 muon events (c.f. Fig.~\ref{fig:T2K}), 
hence a clear sign of oscillation, while that of $\nueb$ 
appearance with 3 electron events seen is not yet significant.$\,$\cite{T2K}

The European long-baseline programme concentrated on the search for  
tau-neutrino appearance from the conventional muon-neutrino beam sent from
CERN to the 732$\;$km away OPERA detector in the Italian Gran-Sasso Laboratory 
(CNGS). A breakthrough for this experiment was achieved with the July 2015 
observation of a fifth tau-neutrino candidate exceeding the threshold of $5\sigma$
for the $\numu\to\nutau$ appearance observation.$\,$\cite{OPERA}
In OPERA charged-current 
neutrino interactions ($(\numu\to)\nutau+N\to\tau^-(\to e,\mu,h)+X$) 
are recorded in detectors (bricks) of lead and emulsion 
film with sub-micron resolution. The total target size consists of of 150 thousand 
bricks. OPERA features additional target trackers and muon spectrometers.
Tau-neutrino candidates are identified by tracks with a large impact parameter
from the tau decay and no muon from the primary interaction vertex.
Data between 2008 and 2012 were used, corresponding to $18\cdot10^{19}$ 
POT giving 20 thousand neutrino interactions in the detector of which 6.7 thousand 
were fully analysed.$\,$\cite{AlessandraPastore} The five identified tau candidates
consist of three one-prong and one three-prong hadronic decays, and 
one muon decay. The pure muon decay candidate has a very small background 
expectation of $0.004 \pm 0.001$ events. The overall background expectation 
is estimated to be $0.25 \pm 0.05$ events, the expected signal $2.64\pm0.53$
events, which is compatible with the observed five events. The signal 
significance is 5.1$\sigma$ hence establishing the observation of tau-neutrino 
appearance. OPERA  also set  limits on sterile neutrinos. The OPERA physics 
programme has now ended.

\subsection{Results from (short-baseline) reactor experiments}

New neutrino measurements from experiments placed close to nuclear reactors 
in China (Daya Bay) and France (Double Chooz) were reported. Nuclear reactors 
represent powerful $\nueb$ sources from beta-decay of the nuclear fission 
products. Detectors installed in their ${\cal O}({\rm km})$ vicinity can measure 
the mixing angle $\theta_{13}$ from the $\nueb$ survival probability 
that is dominated by the $\Delta m^2_{32}$ term. 

The Daya Bay detector has completed its full assembly. It consists of two near 
experimental areas (effective baselines 512$\;$m and 561$\;$m from the 
17.4$\;$GW thermal power reactor near Hong Kong) and one far area (1.6$\;$km).
The detection of $\nueb$ occurs through the inverse beta decay (IBD) reaction
$\nueb+p\to e^++n$ in gadolinium (Gd) doped liquid scintillators. The prompt 
$e^+$ annihilation photon and delayed 8$\;$MeV photons from the neutron 
capture are measured. The $\nueb$ flux uncertainty is largely eliminated by 
simultaneous measurements at the three different detector sites. Daya Bay 
already provided
the world's most precise measurement ${\rm sin}^2(2\theta_{13}) =0.084 \pm 0.005$ 
using data taken between October 2012 and November 2013 and using two third of 
the total of eight antineutrino detectors.$\,$\cite{DayaBay1} 
The new analysis presented at this conference used neutrons captured 
by hydrogen (instead of Gd) providing an additional $\theta_{13}$ measurement
as the data sample is largely independent and the systematic uncertainties different.
It found$\,$\cite{YimingZhang,DayaBay2} 
${\rm sin}^2(2\theta_{13}) =0.071 \pm 0.011$ (nH) and, when averaged 
with the Gd result, ${\rm sin}^2(2\theta_{13}) =0.082 \pm 0.004$ (nGd \& nH).

\begin{figure}[t]
\centerline{\includegraphics[width=\linewidth]{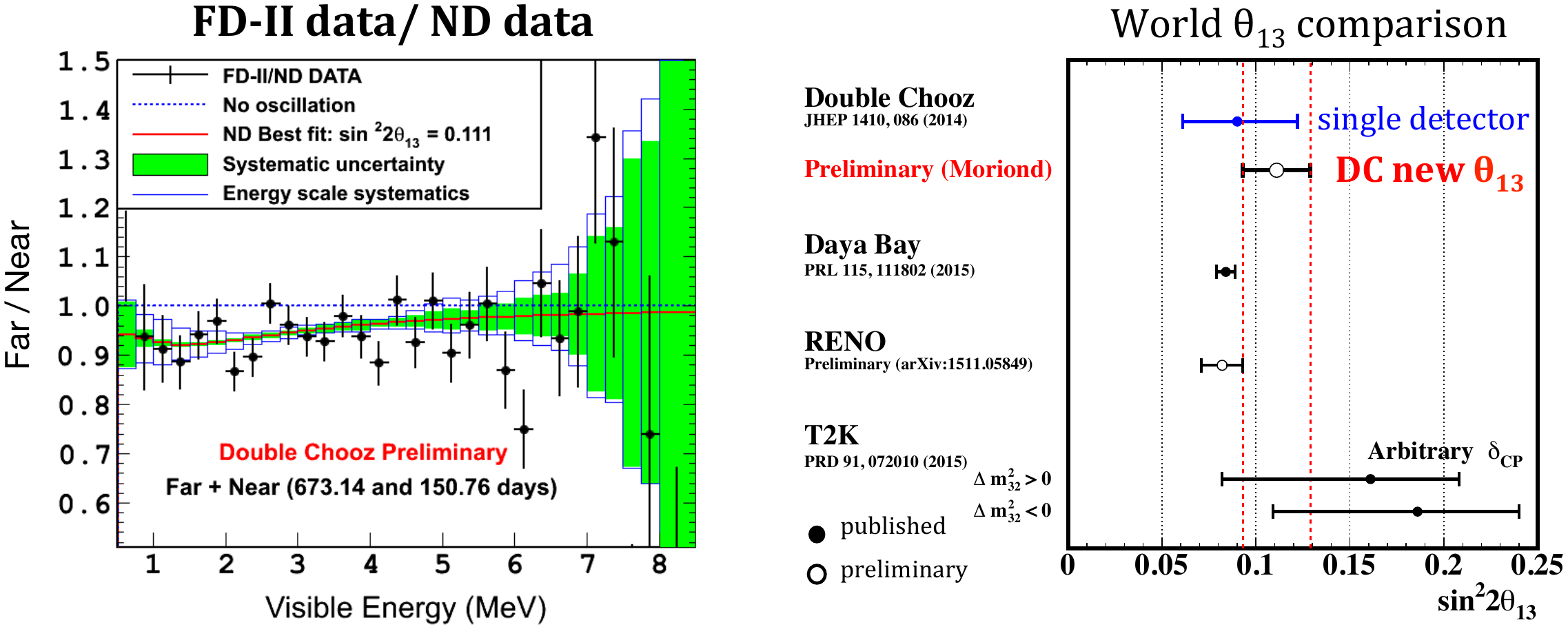}}
\vspace{-0.0cm}
\caption[]{{\bf Left:} ratio of far (FD-II) to near detector yields versus visible
  energy measured by Double Chooz. 
  Overlaid is the best fit result. The no-oscillation hypothesis is clearly
  excluded. {\bf Right:} comparison of ${\rm sin}^2(2\theta_{13})$ 
  measurements.$\,$\cite{MasakiIshitsuka}
  \label{fig:DoubleChooz}}
\end{figure}
The Double Chooz collaboration presented their brand new oscillation measurement 
at this conference.$\,$\cite{MasakiIshitsuka} 
Double Chooz installed at the Chooz nuclear power plant in France (close to the 
Belgium border) with two operating units (B1 nd B2) has terminated its 
multi-detector setup with a near detector 
(0.4$\;$km from the nuclear cores, available since 2015) and a far detector 
(1.1$\;$km, available since 2011). The nearly iso-flux setup of the detectors 
reduces the flux uncertainty to less than 0.1\%. The uncorrelated detection 
systematic uncertainty is lower than 0.3\%. Double Chooz performs a combined 
parameter fit to the FD-I, FD-II and ND data (c.f. left panel of 
Fig.~\ref{fig:DoubleChooz} for the ratio FD-II/ND), including also reactor-off data 
to constrain backgrounds. The preliminary result  
${\rm sin}^2(2\theta_{13}) =0.118 \pm 0.018$ has a significance of 5.8$\sigma$
and is in agreement with previous measurements. 
The right panel in Fig.~\ref{fig:DoubleChooz} shows a comparison of 
${\rm sin}^2(2\theta_{13})$ measurements (not including the latest 
Daya Bay combination).

\subsubsection*{Reactor flux anomalies}

Daya Bay reports a recent $\nueb$ flux measurement$\,$\cite{DayaBayFlux} 
using 340 thousand near-detector IBD candidates, with better 
than 1\% energy calibration, and comparison with 
model predictions: an overall deficit in data of about 2$\sigma$ is found and
a significant local deviation at around 5$\;$MeV antineutrino energy. 
While an overall deficit may seem like disappearance to a sterile neutrino, the 
local deviation does not. These findings are consistent with the reactor neutrino anomaly 
picture emerging from earlier short baseline measurements (Daya Bay, Reno, 
Double Chooz) that found a ratio of measured to expected $\nueb$  flux 
of about 0.94. It was reported at this conference$\,$\cite{AnnaHayes} 
that much caution is needed             
when interpreting these results as  systematic uncertainties in the flux
modelling may in total cover the observed deficit. The 5$\;$MeV 
bump may be due to several fission daughter isotopes (e.g.,  uranium 238
or plutonium). Therefore one cannot currently consider seriously new physics 
claims based on absolute reactor flux comparisons. There is ample literature
about the reactor flux anomaly.$\,$\cite{ReactorAnomaly} 

Approximately ten very-short-baseline experiments are currently in their construction 
or planning phases with the aim 
to provide additional absolute flux measurements. Among these is the 
SoLid experiment, a 3 ton highly segmented plastic scintillation detector coated with 
Lithium-6, designed to measure flux and energy of $\nueb$ at distances 
between 6--10$\;$m from the compact BR2 test reactor with a highly-enriched 
uranium core in Mol (Belgium). The main experimental challenges are the suppression 
of background in the proximity of the reactor (requiring a good separation of 
captured-neutrons versus $e/\gamma$) and the precise location of the IBD products. 
To achieve this, not only the time difference
but also spatial information is used to reconstruct IBD events. The  goal
of SoLid is to run the experiment for three years to resolve the reactor neutrino 
anomaly without relying on theoretical modelling.$\,$\cite{NickvanRemortel}

\subsection{Neutrinos from the Sun}

The Borexino collaboration reported new measurements$\,$\cite{SandraZavatarelli}  
after their 2014 breakthrough evidence for detection of the Sun's primary 
proton--proton fusion neutrinos, found within 10\% precision to have a yield 
consistent with the Sun's photon luminosity.$\,$\cite{BorexinoSun}
Borexino was initially designed for studying the 0.86$\;$MeV Be-7 solar
electron-neutrinos via $\nue$--$e$ scattering and electron recoil measurements 
(also IBD). The experiment consists of a 270$\;$ton liquid scintillator, surrounded 
by 890$\;$ton buffer fluid. It is installed in a 9.5$\;$m diameter nylon vessel, 
1.3$\;$km underground at the Gran Sasso Laboratory (LGNS). 
The extremely high radiopurity of Borexino 
allows for a 250$\;$keV neutrino energy threshold. Since that seminal 2014 result
Borexino focused on the highly challenging detection (proof) of the catalytic CNO 
cycle in the Sun, a complex chain of CNCNONC transitions involving different 
C, N, O isotopes and believed to be the dominant energy source in stars more 
massive than the Sun. Borexino also performed tests of electron charge 
conservation through the search 
for $e\to\gamma\nu, \nu\nu\nu$ decays achieving the world's best electron lifetime
sensitivity $\tau_e > 6.6\cdot10^{28}\;$years; and the 5.9$\sigma$
observation  of geological $\nueb$ for which the largest background stems 
from nuclear reactors.\cite{Borexino-charge,Borexino-geo}

\subsection{Neutrino astronomy}

Cosmic rays have been measured over eleven orders of magnitude in energy, but 
their highest-energy sources are not well known yet. Several favourable conditions 
make neutrinos from outer space to excellent astronomical probes for the study of 
cosmic rays. Neutrinos are not deflected by astrophysical foreground and therefore
point back to their sources. Moreover, owing to their characteristic scattering signatures, 
the flavour of neutrinos can be reconstructed in a large detector providing information
about their origin. 

IceCube$\,$\cite{JanAuffenberg} is a spectacular experiment buried between 
1.5--2.5$\;$km deep in South Pole ice. It has and active volume of about 
1$\;$km$^3$ distributed among 86 
strings. IceCube measures Cherenkov light ``track'' and ``cascade'' (shower) 
signatures that are characteristic charged-current interactions in ice of muon-neutrinos
and electron-neutrinos, respectively. A so-called ``double-bang'' event would be
signature for a tau-neutrino in which a produced tau lepton
of PeV energy penetrates  50$\;$m ice in average before it decays leaving a hadronic
or electromagnetic shower. IceCube detected interactions from about 100 thousand 
neutrinos with larger than 200$\;$GeV energy per year, among which a few dozens 
are of astrophysical origin, and the majority stems from atmospheric muon and 
muon-neutrino background. 

\begin{figure}[t]
\centerline{\includegraphics[width=0.5\linewidth]{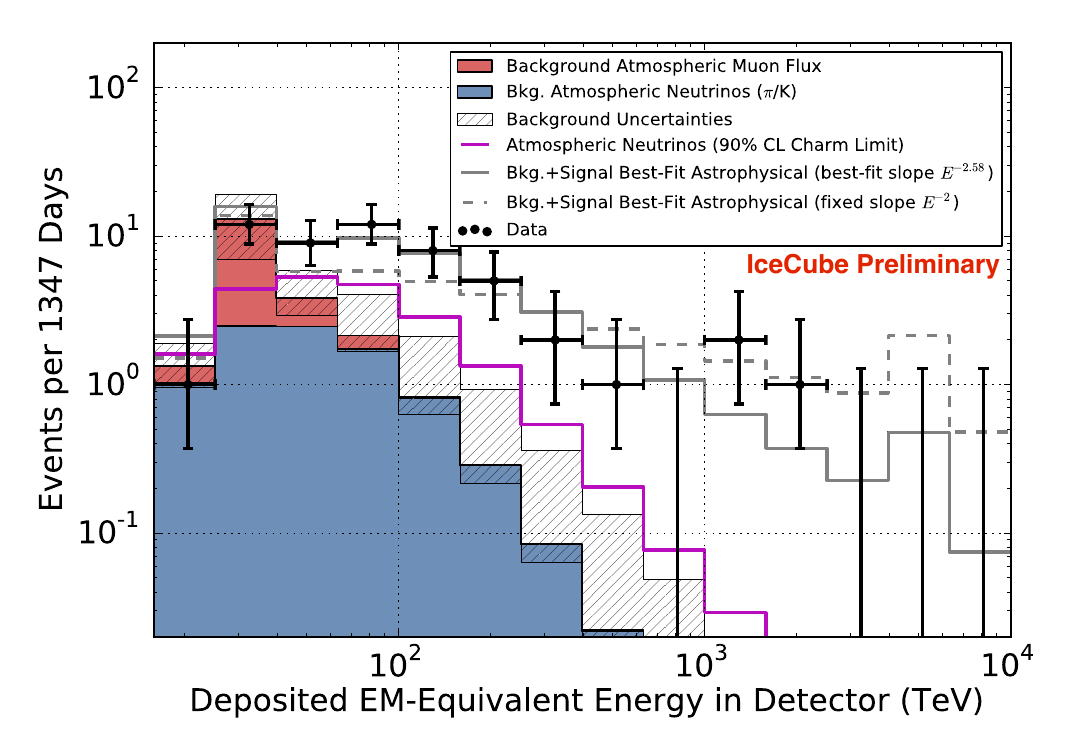}
                 \includegraphics[width=0.5\linewidth]{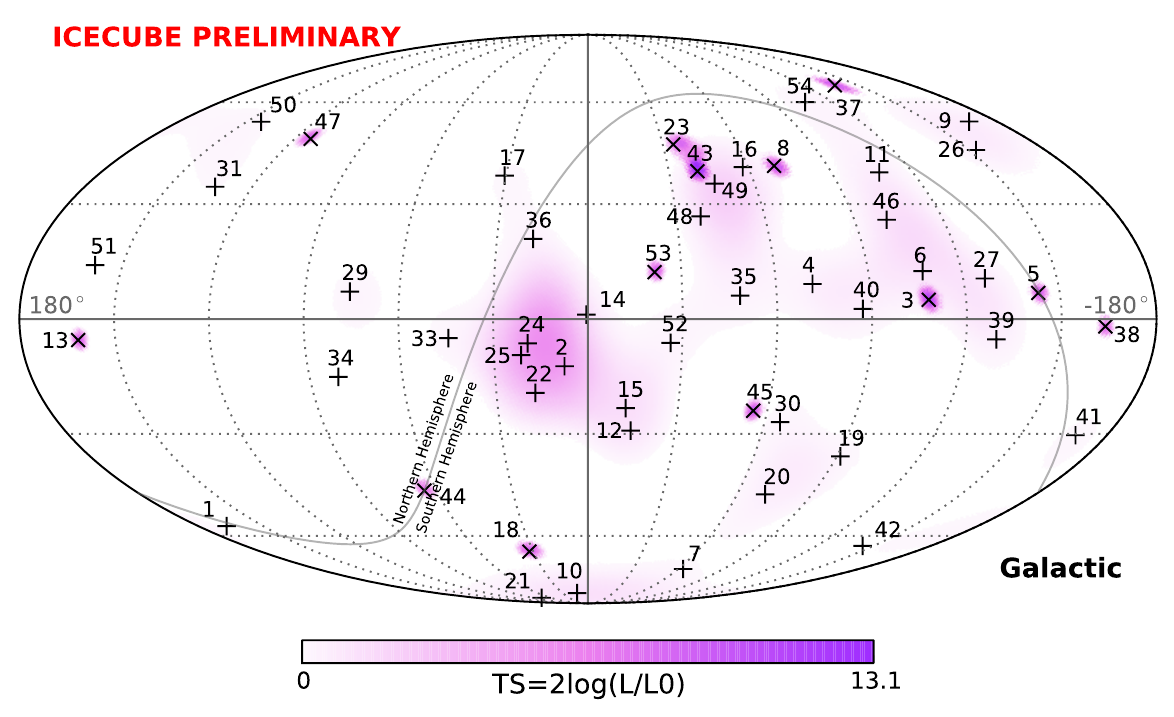}}
\vspace{-0.0cm}
\caption[]{{\bf Left:} inclusive neutrino energy spectrum measured by IceCube after 
  four years of data taking. Also shown are the estimated atmospheric 
  backgrounds. {\bf Right:} arrival directions of the neutrino events in 
  galactic coordinates.  Shower-like events are marked with '+' and those containing 
  tracks with '$\times$'. Colours show the test statistics value for point-source 
  clustering at each location. No significant clustering was found. 
  Both figures are taken from Ref.$\,$\cite{IceCube}.
  \label{fig:IceCube}}
\end{figure}
IceCube measured the inclusive neutrino energy spectrum above 60$\;$TeV 
during four years of data taking (see left panel of Fig.~\ref{fig:IceCube}). 
A total of 53 good events were found up to around 2$\;$PeV energy with 
a significance of $6.5\sigma$ for a signal of astrophysical neutrinos. 
The energy spectrum was found to be somewhat harder than expected indicating that the 
canonical $E^{-2}$ model may be insufficient to describe the data.$\,$\cite{IceCube} 
IceCube also sees 
a 5.9$\sigma$ excess of up-going muon-neutrinos (charged current only) in the 
0.2--8.3$\;$PeV energy regime over atmospheric background normalised to data at 
100$\;$TeV neutrino energy. A possible pattern in the spectral index versus the 
neutrino energy cannot be excluded. The measured arrival directions of the observed 
astrophysical neutrinos do not exhibit clustering that would hint to a point source 
(see the example in the right panel of Fig.~\ref{fig:IceCube}).

The reconstruction of the neutrino flavour can provide information about the 
source of the astrophysical neutrinos. Pion decay should produce relative neutrino
abundances of $\nue:\numu:\nutau=1:1:1$ on earth, if muons are suppressed
due to, e.g., large magnetic fields in space the relative abundances would be
$1:1.8:1.8$, and if the neutrinos originate from neutron decay one would expect
to see a pattern of $2.5:1:1$. The current IceCube data are consistent 
with the first two but exclude the third pattern. No hint for tau neutrinos was 
found yet but is expected to occur in the accumulated data sample. 

IceCube also belongs to the elite of experiments who have observed neutrino 
oscillation through muon-neutrino disappearance. The measurement of  
$\Delta m^2_{32}$ and $\theta_{23}$ is consistent with that from other experiments. 
IceCube also searched for sterile neutrinos, heavy dark matter annihilation, and solar 
flares.$\,$\cite{GwenhaeldeWasseige} 
A next generation experiment, IceCube-Gen2, covering an active area of about 
10$\;$km$^3$, is currently in its R\&D phase. 

\subsection{Of which quantum nature are neutrinos?}

\begin{wrapfigure}{R}{0.5\textwidth}
\centering
\vspace{-0.4cm}
\includegraphics[width=0.24\textwidth]{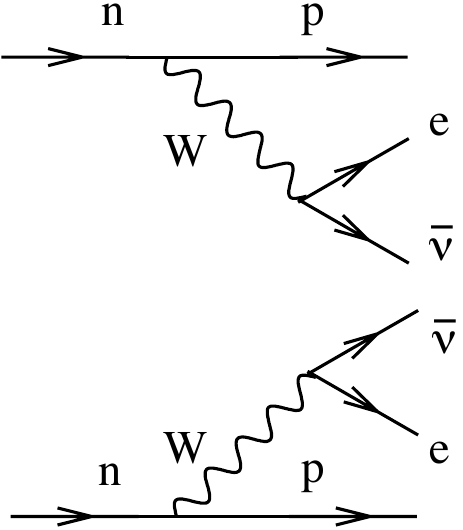}
\includegraphics[width=0.24\textwidth]{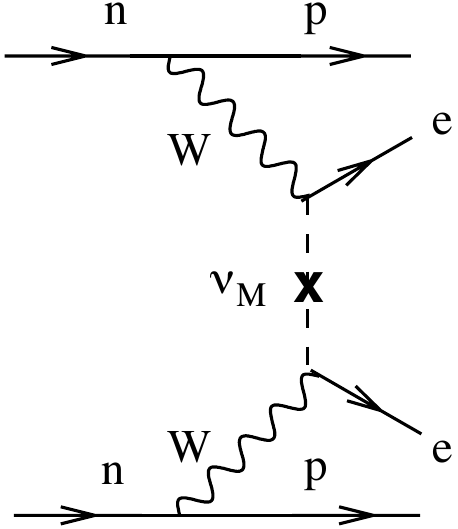}
\caption[.]{Representative Feynman diagrams for two-neutrino double beta 
  decay (left) and neutrinoless double beta decay (right). 
  Figures taken from Ref.$\,$\cite{DoubleBetaFigures}. 
  \label{fig:2betadecay}}
\vspace{-0.2cm}
\end{wrapfigure}
The yet unrevealed Majorana or Dirac nature of neutrinos can be 
addressed experimentally by detecting neutrinoless double beta ($\zn$) decay
(c.f. diagrams in Fig.~\ref{fig:2betadecay}), 
which would indicate that there is a non-zero Majorana mass term as Dirac
neutrino masses do not mix neutrinos and antineutrinos.$\,$\footnote{Majorana 
masses cannot originate from a Yukawa coupling to the Standard Model 
Brout-Englert-Higgs (BEH) field and thus would make neutrinos very different 
from the other known fermions.} Through the relation 
$\Gamma_{\zn}\propto|M_{\zn}|^2 \langle m_{\beta\beta}\rangle^2$ and theory 
input for the nuclear matrix element one can via the measurement of or bound 
on $\Gamma_{\zn}$ infer information on the neutrino 
mass and hierarchy. Experiments searching for $\zn$ decay require large 
mass, high isotopic abundance, good energy resolution, high efficiency and 
low background. Results from the EXO-200 and CUORE-0 experiments where 
reported at this conference.$\,$\cite{YungRueyYen,PaoloGorla}

EXO-200 is a detector that uses an enriched (81\%) liquid-xenon TPC 
($^{136}{\rm Xe}\to^{136}{\rm Ba} + 2e^-$) and is installed in a nuclear waste 
isolation plant in New Mexico, US. EXO-200 presented in 2014 a result using data 
corresponding to 100$\;{\rm kg}\cdot{\rm years}$ of $^{136}{\rm Xe}$ 
exposure$\,$\cite{EXO-200Xe} with no evidence for $\zn$ decay giving a half-life 
lower limit of $1.1\cdot10^{25}\;$years at 90\% CL. This corresponds to 
$\langle m_{\beta\beta}\rangle < (190$--$470)\;{\rm meV}$, where the range
is due to different theoretical assumptions on the nuclear matrix element. 
A recent analysis$\,$\cite{EXO200,YungRueyYen}
reported at this conference searched for the $\tn$ decay
of $^{136}{\rm Xe}$ to the 0$_1^+$ excited state of $^{136}$Ba, which 
de-excites via two photons.
No significant signal was found in that search.

The LGNS based experiment CUORE-0, a prototype of the full CUORE experiment,
employs a bolometric technique using an array of tellurium dioxide crystals 
($^{130}{\rm Te} \to ^{130}{\rm Xe} + 2e^-$) cooled down to remarkable 10$\;$mK. 
The bolometer benefits from excellent energy resolution but no particle identification
capability. A first CUORE-0 measurement$\,$\cite{CUORE-0,PaoloGorla} 
using a $^{130}{\rm Te}$ exposure of $9.8\;{\rm kg}\cdot {\rm years}$ 
revealed no signal at the expected $Q_{\beta\beta}$ value of 2528$\;$keV, giving, 
when combined with a previous Cuoricino result, the 90\% CL limit 
$\langle m_{\beta\beta}\rangle < (270$--$760)\;{\rm meV}$, where again the 
range reflects the matrix element uncertainty. 

\section{Proton decay --- GUT messengers}

It is not possible to  reach energies in the laboratory that would allow to directly 
study the physics at the expected grand unification scale. 
Even Enrico Fermi's ``Globatron'' (that was to be built in 1994) would with current
LHC magnet technology ``only'' reach insufficient 20 PeV 
proton--proton centre-of-mass energy. 
Proton decay is among the greatest mysteries in elementary particle physics. 
It is required for baryogenesis and predicted by grand unified theories (GUT).
Its discovery could therefore provide a probe of GUT scale physics. 

All current limits are dominated by searches at the Super-Kamiokande (SK) 
experiment.$\,$\footnote{We recall that ``KamiokaNDE'' stands for 
``Kamioka Nucleon Decay Experiment''.}
New results from SK combining all SK I--IV data (1996--now) were presented at 
this conference.$\,$\cite{VolodymyrTakhistov} 
No significant excess was found leading to the following strong limits: 
$\tau(p\to e^+\pi^0) > 1.7\cdot10^{34}\;{\rm years}$ (no events seen 
in the signal regions R1/2, for 0.07/0.54 background events  expected), 
$\tau(p\to \mu^+\pi^0) > 7.8\cdot10^{33}\;{\rm years}$ 
(less sensitive because the $\mu^+$ is detected through its decay to $e^+$,  
0/2 events seen in R1/2, for 0.05/0.82 background events expected), 
$\tau(p\to K^+\nu) > 6.6\cdot10^{33}\;{\rm years}$ 
(the $K^+$ being below Cherenkov threshold is detected through its decay,
no events seen in signal regions SB/C, for 0.39/0.56 background events expected), 
The SK collaboration also looked for more exotic phenomena. 

An order of magnitude gain in sensitivity on $\tau(p\to e^+\pi^0)$ is expected 
from the Hyper-Kamiokande project which has 25 times the size of SK (SK holds 
50 kiloton of pure water) and has an expected begin of construction in 2018. 

\section{Direct dark matter searches}

Direct dark matter experiments search for elastic collisions of a weakly interacting 
massive particle (WIMP) from the galactic halo with a target nucleus at rest in 
the laboratory. With an assumed WIMP average speed of about 220$\;$km per second 
the collision is expected to lead to a measurable nuclear recoil of about 20$\;$keV.
The effective scattering Lagrangian may have scalar (spin-independent, 
${\rm SI}\propto A^2$, where $A$ is the atomic number of the target nucleus) 
or axial-vector (spin-dependent, ${\rm SD}\propto {\rm nuclear}$ spin, no 
coherent amplification) terms. The dominant background stems from electrons 
recoiling after X-ray or $\gamma$-ray interactions.
Direct dark matter experiments have similar challenges to overcome as neutrino
experiments. They must be deep underground, have excellent radiopurity, 
must be shielded around the active detector volume and they require redundant 
signal detection technologies.

The CDMSlite experiment (CDMSlite stands for CDMS low ionisation threshold 
experiment) is located at the US Soudan Underground 
Laboratory. CDMS looks for keV-scale recoils from elastic scattering of WIMPs 
off target nuclei. It uses up to 19 Ge and 11 Si detectors. Ionisation charges 
and phonons (heat) are measured and used to discriminate electron from 
nuclear recoils. CDMSlite operates one Ge detector at increased bias voltage 
to amplify the phonon signal and gain sensitivity to lower WIMP masses.
Two runs were taken with the second benefiting from reduced acoustic noise 
(hence a lower threshold) and longer exposure. The 
newly$\,$\cite{CDMSlite,EliasLopezAsamar}  excluded parameter 
space for SI WIMP--nucleon interaction extends to WIMP masses down to 
1.6--5.5$\;$GeV and cross sections between $10^{-37}$--$10^{-41}\;$cm$^2$.
With this measurement the SuperCDMS programme has ended. It will be followed
up by SuperCDMS at SNOLAB (2.1 km deep underground).

\begin{wrapfigure}{R}{0.63\textwidth}
\centering
\vspace{-0.4cm}
\includegraphics[width=0.60\textwidth]{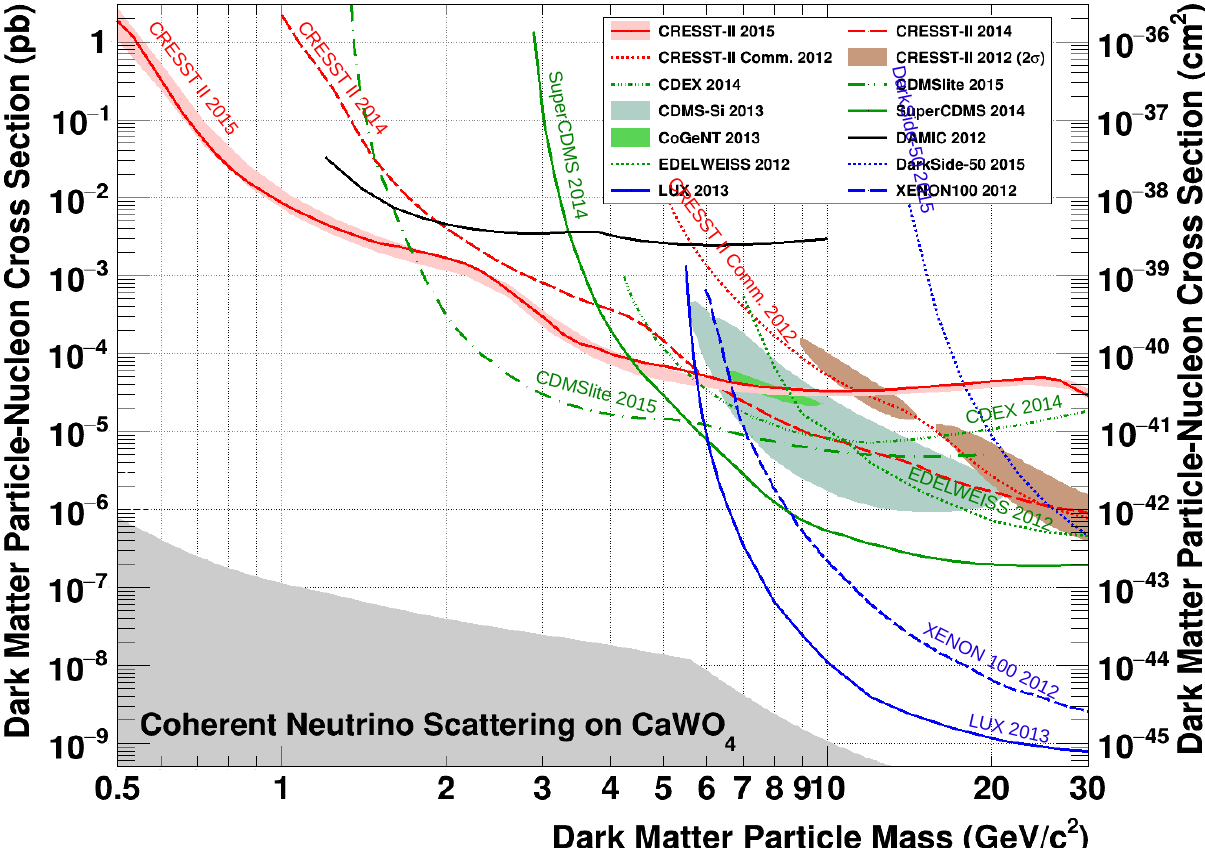}
\vspace{-0.0cm}
\caption[.]{Summary of low-mass WIMP--nucleon cross-section limits (in pb)
  versus WIMP mass.$\,$\cite{FranzPröbst} 
  \label{fig:CRESST}}
\vspace{-0.2cm}
\end{wrapfigure}
The CRESST-II experiment at the LNGS has further improved its sensitivity to 
even lower mass WIMPs. Energy threshold is key for this search. CRESST-II 
uses cryogenic calcium tungstate (CaWO$_4$) crystals to measure scintillation 
light and phonons to separate electron from neutron recoils. Transition edge sensors 
(TES, about 15 mK) and a squid system measure, amplify and read out the signal,
allowing sub-keV energy thresholds and a high-precision energy reconstruction. 
Combined with the light target nuclei, CRESST-II has the potential to probe 
$< 1\;$GeV dark matter particles. At 0.5$\;$GeV a limit of about
$10^{-36}\;$cm$^2$ was obtained.$\,$\cite{CRESST,FranzPröbst} Systematic 
studies on the current data sample are still ongoing. The follow-up 
programme with 50--100$\;$eV threshold starts in April 2016.

XENON100 is the second phase of the XENON dark matter experiment at LNGS 
running since 2009.$\,$\cite{ConstanzeHasterok} 
It is the predecessor of the ambitious programme XENON1T.
XENON100 uses a 61 (100) kg target (active veto) liquid-gas xenon (LXe) filled TPC. 
Liquid xenon as target material features a high density, high atomic number, sensitivity 
to spin-dependent interactions through approximately 50\% odd isotopes, low 
threshold due to high ionisation and scintillation yield, low backgrounds, 
and a self-shielding target. The liquid xenon scintillation light is measured by 
photomultiplier tubes (PMT). Light from prompt scintillation (S1) and delayed 
ionisation (S2) allows to discriminate electron from nuclear recoil. The primary 
results were published$\,$\cite{XENON100} by XENON100 in 2012 providing 
powerful limits on the WIMP--nucleon interaction cross section down to about 
$2\cdot10^{-45}\;$cm$^2$ for a WIMP mass of  50$\;$GeV at 90\% CL for 
SI interactions.$\,$\cite{XE100-1} SD results were released in 2013 with best 
limits of about $10^{-38}\;$cm$^2$ and $4\cdot10^{-40}\;$cm$^2$ for proton 
and neutron cross sections, respectively.$\,$\cite{XE100-2}  A recent XENON100 
analysis$\,$\cite{XENON100Mod} addresses 
the periodic signal reported by the DAMA collaboration. XENON100 does not find 
significant periodicity, excluding DAMA's phase and amplitude at 4.8$\sigma$. 
DAMA-like dark matter models are excluded to at least 3.6$\sigma$. 
The experimental follow-up programme XENON1T has its commissioning almost 
completed. First results are expected in the course of 2016. 

Dark matter searches with the LUX experiment were also 
presented.$\,$\cite{PaoloBeltrame} LUX is 
a liquid-Xe experiment located at the Sanford Underground Research Facility in
South Dakota, US, about 1.5$\;$km deep. LUX is very similar to XENON100 
based on a dual-phase liquid Xe target. It has a larger active target and 
lower threshold than XENON100 (3$\;$keV vs. 6.6$\;$keV) and hence 
sensitivity to lower WIMP masses.  A reanalysis of the 2013 data (95 live 
days, 145$\;$kg fiducial mass) with improved calibration, event 
reconstruction and background modelling increases the sensitivity 
especially at low WIMP masses.$\,$\cite{LUXRe} The best SI limit at WIMP masses of 
around 40$\;$GeV reaches down to approximately $7\cdot10^{-46}\;$cm$^2$ 
WIMP--nucleon cross section. LUX has also recently published SD 
limits using the same dataset.$\,$\cite{LUXSD} Their follow-up programme 
LZ = LUX + ZEPLIN is entering CD-2 review and has a planned start 
for 2025.

\begin{figure}[t]
\centerline{\includegraphics[width=0.72\linewidth]{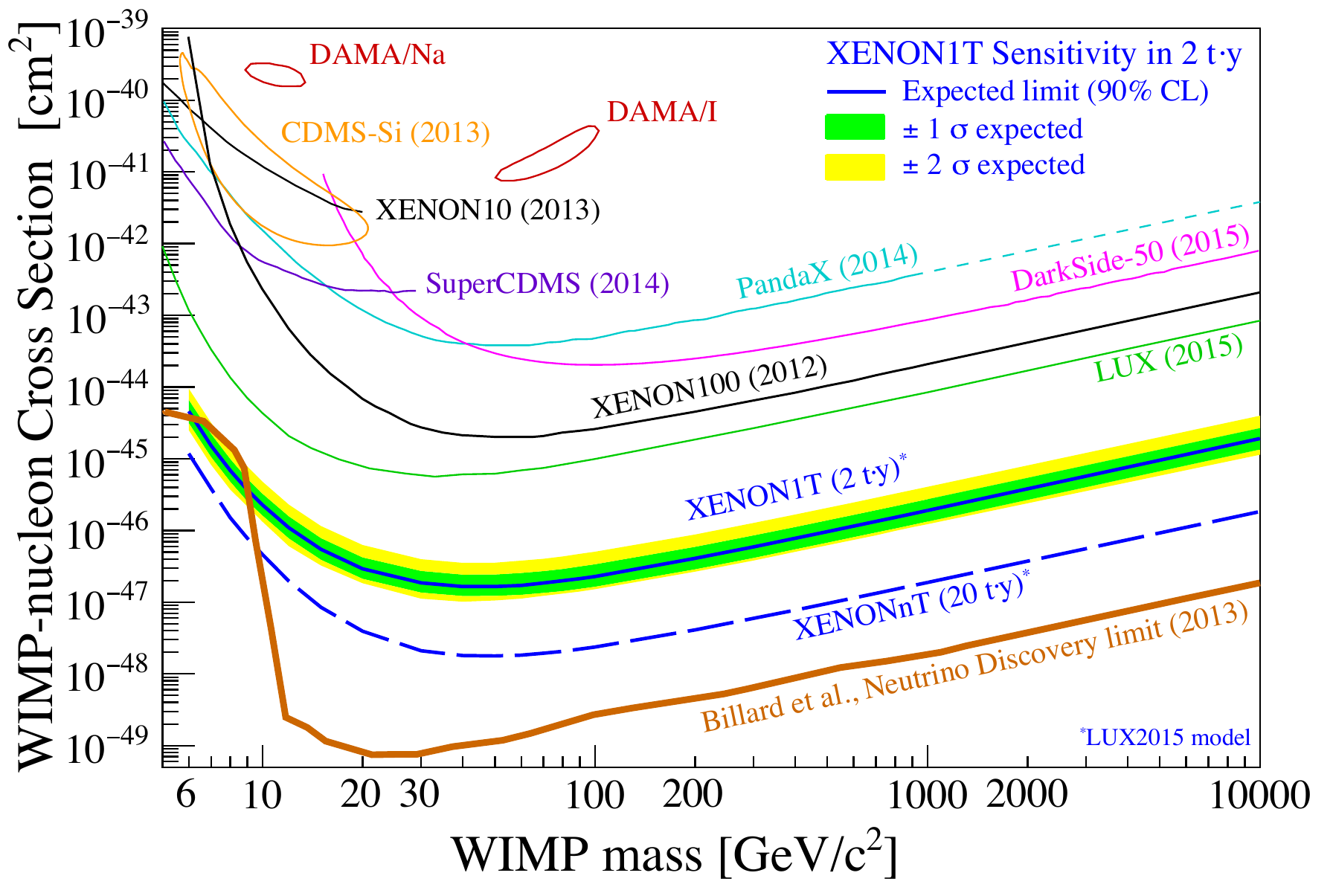}}
\vspace{-0.1cm}
\caption{Current WIMP--nucleon cross-section limits versus the WIMP mass
  and extrapolations (plot taken from XENONnT). The lowest bold line depicts the 
  expected coherent $\nu$--$N$ scattering background.
  \label{fig:DMSummary}}
\end{figure}
Dark matter searches undertake and prepare a healthy experimental programme 
with orders of magnitude improved sensitivity.
Figure~\ref{fig:DMSummary} shows the WIMP--nucleon cross-section limits 
versus the WIMP mass and extrapolations. Shown by the thick red line is the 
irreducible coherent elastic neutrino--nucleon scattering background that is 
expected to be in reach with the next generation experiments.

\section{Gravitational waves}

The LIGO/VIRGO Collaboration, a new popstar in science, reported on February 
11$^{\rm th}$, 2016 an earth-shattering measurement$\,$\cite{LIGO/VIRGO}: 
a huge gravitational wave (GW) signal of a binary black hole merger detected 
simultaneously in the two LIGO sites 
(the VIRGO experiment was not operational at the time of the measurement), 
first noticed by its online burst detection system. This measurement is an 
example of scientific perseverance.$\,$\cite{AlessioRocchi}

\begin{wrapfigure}{R}{0.65\textwidth}
\centering
\vspace{-0.4cm}
\includegraphics[width=0.65\textwidth]{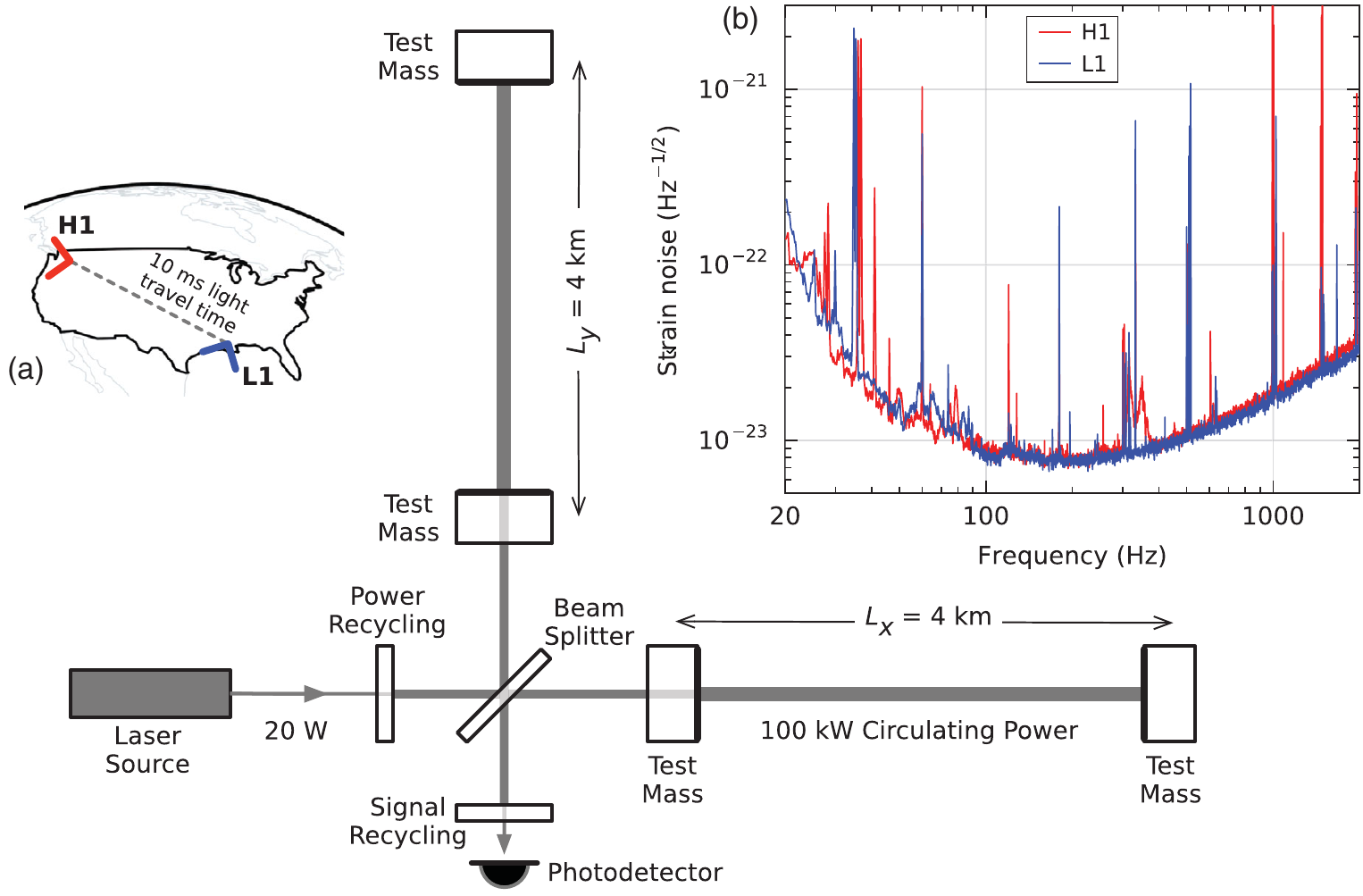}
\vspace{-0.4cm}
\caption[.]{Principle of the LIGO gravitational wave measurement. 
  \label{fig:LIGO-Principle}}
\vspace{-0.0cm}
\end{wrapfigure}
The principle of the measurement is sketched in Fig.~\ref{fig:LIGO-Principle}.
Spin-2 GWs lengthens one arm while shortening the other and vice versa in the LIGO
laser interferometer: $\Delta L(t) = \delta L_x - \delta L_y = h(t) L$.
The optical signal measured is proportional to the strain $h(t)$. There are 
several enhancements to a basic Michelson interferometer in LIGO: test mass 
mirrors multiply the effect of GW on the light phase by a factor of about 300;
a power recycling mirror on the input amplifies the laser light; output signal 
recycling broadens the bandwidth. The test masses are isolated from seismic 
noise and have very low thermal noise. All relevant components of the 
interferometer are isolated against vibrations. The laser light passes through 
vacuum to reduce Raleigh scattering of light off air molecules.
A system of calibration lasers and array of environmental sensors further
helps to reduce systematic uncertainties. Two (better three!) distant 
interferometer are needed to localise a GW and measure its polarisation. 

\begin{figure}[t]
\centerline{\includegraphics[width=\linewidth]{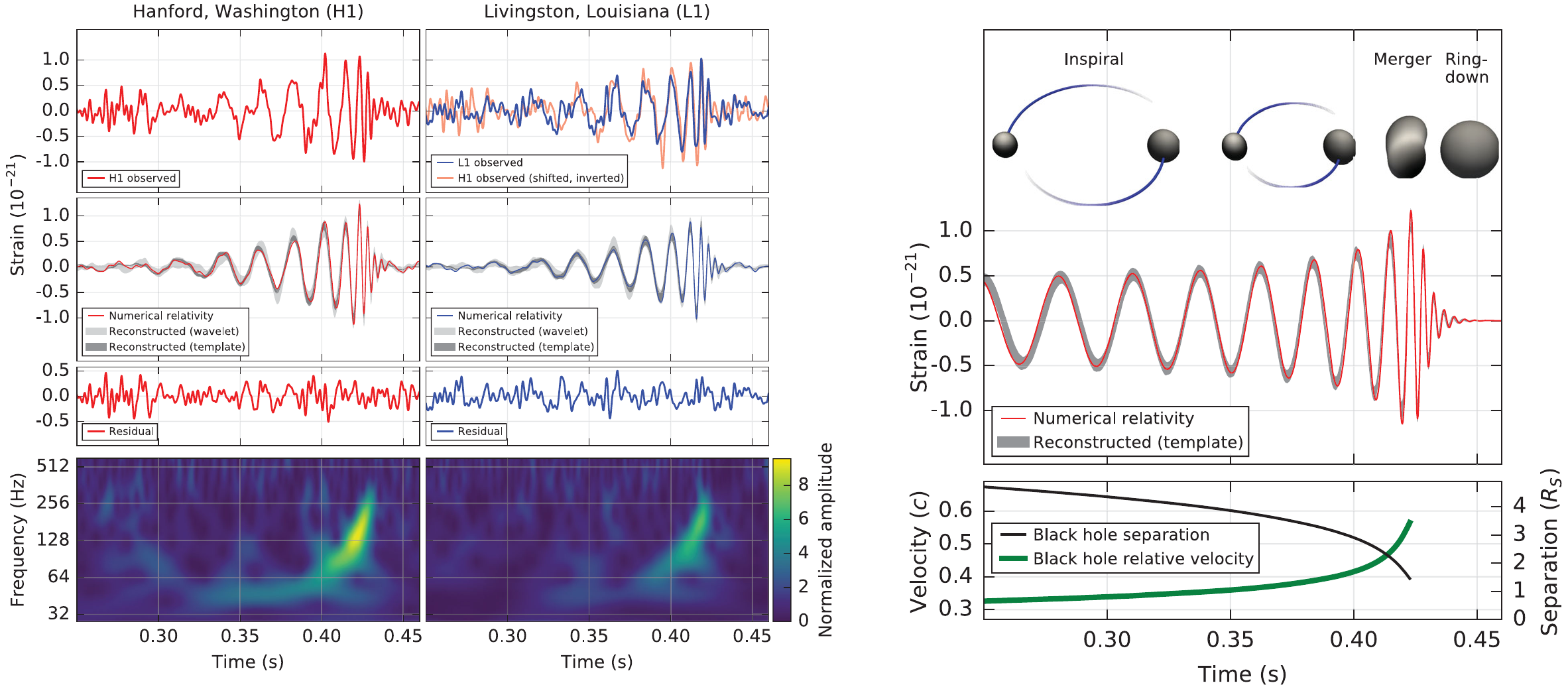}}
\vspace{-0.0cm}
\caption[.]{{\bf Left:} the gravitational-wave event GW150914 observed by the 
  LIGO Hanford (H1, left column panels) and Livingston (L1, right column 
  panels) detectors. Times are shown relative to September 14, 2015 at 
  09:50:45 UTC. The top panels show the measured GW strains. For a better visual 
  comparison the H1 data are shifted in time by 6.9$\;$ms. The panels in the second row 
  show the GW strains projected onto each detector in the 35--350$\;$Hz frequency
  band. The solid line superimpose the fit predictions based on general relativity 
  calculations. The third row shows the residuals after subtracting the filtered 
  numerical relativity waveform from the filtered detector time series, and the 
  bottom panels show a time-frequency representation of the strain data, 
  showing the signal frequency increasing over time. {\bf Right:} estimated 
  gravitational-wave strain amplitude from GW150914 projected onto H1. The bottom
  panel shows the effective black hole separation in units of Schwarzschild radii 
  and the effective relative velocity. Figures and explanations 
  taken from Ref.$\,$\cite{LIGO/VIRGO}.
 \label{fig:LIGO-Measurement}}
\end{figure}

On September 14, 2015 at 09:51 UTC (11:51 CEST), within a total of 16 
days of simultaneous two-detector observational data taken by LIGO Hanford, 
Washington (H1) and LIGO Livingston, Louisiana (L1), the signals shown
in the left panel of Fig.~\ref{fig:LIGO-Measurement} were detected (the 
H1 data are shifted by 6.9$\;$ms to allow for a better comparison). The 
detected GW pattern is an extremely loud event (modified signal-to-noise ratio 
of $\hat\rho_c = 23.6$).
The maximum strain ($10^{-21}$) times the 4$\;$km arm length gives 
a length deformation of $4\cdot10^{-18}\;$m, which is about 0.5\% the size 
of a proton. The measured spectrum can be well reproduced by 
GW calculations after fitting its parameters to the observation. 
The event, dubbed GW150914,
is found to have a significance over background of more than 5.1$\sigma$.
The time series shown in the figure was filtered with a 35--350$\;$Hz 
bandpass filter to suppress large fluctuations outside the detectors' 
most sensitive frequency band, and with band-reject filters to remove 
strong instrumental spectral lines.

The right panel of Fig.~\ref{fig:LIGO-Measurement} shows a sketch of
the posited black hole encounter and coalescence. Thereafter, GW150914 
occurred $1.3\pm0.5$ billion years (410 Mpc) ago. Over a duration of 0.2$\;$s, 
frequency and amplitude of the binary black hole system increased from 
35 to 150$\;$Hz (in about 8 cycles). To reach 75$\;$Hz orbital frequency, 
the objects needed to be very close (about 350$\;$km to each other and 
massive (thus black holes$\;$\footnote{{\em Digression}. There 
are gargantuan black holes in the universe. Many galaxies are expected to 
host supermassive black holes with more than a million times the mass of  
the Sun in its centre, formed during the galaxy creation process. NGC 4889, 
the brightest elliptical galaxy in the Coma cluster (94 Mpc $\sim$ 300 Mly 
from earth), hosts a record black hole of 21 billion times $M_{\odot}$, 
with event horizon diameter of $\sim$130 billion km (Sun: 1.4 million km).}). 
Two black holes of initially 36 and 29 solar masses 
($M_{\odot}$) inspiral with about half the speed of light. The black holes 
merge within tens of ms; the inspiral, merging and ringdown leave a 
characteristic amplitude and frequency GW pattern. The total radiated 
GW energy amounts to $(3.0 \pm 0.5) M_{\odot}$. The direct observation 
of this event follows upon the indirect proof of  GWs from energy loss 
measurements in binary pulsar systems in 1982.\cite{GW-pulsars1,GW-pulsars2}

The observation of GW150914 bundles several discoveries: it is the 
first direct detection of GWs, the first observation of a binary black hole merger,
it shows that relatively heavy stellar-mass black holes ($> 25\;M_{\odot}$) exist 
in nature, it is the observation of the ``no-hair-conjecture'' according to 
which any black hole can be fully characterised by only three classical 
observables (mass, electric charge, angular momentum), it is the 
most relativistic binary event ever seen ($v/c\sim0.5$), and it leads to a
new limit on the graviton mass of $< 1.2\cdot10^{-22}\;$eV.
GW150914 is likely not a unique binary black hole event. The rate is inferred 
to lie between 2 and 400 events per Gpc$^{3}$ and year, which is at the 
higher end of the expectations.  Adding VIRGO will improve the localisation of 
future GW events. New interferometers are upcoming in India and Japan.
Electromagnetic and high-energy neutrino follow-up programmes are 
also in work (no high-energy neutrino coincidence during GW150914 was 
seen by ANTARES and IceCube).

Gravitational waves have joined the club for multi-messenger astronomy 
together with photons, cosmic rays and neutrinos. The paper reporting 
the observation$\,$\cite{LIGO/VIRGO} collected more than 100 citations 
within a month. The LIGO/VIRGO collaboration released a number of
companion papers about detector and analysis details and implications of 
GW150914.$\,$\cite{LIGOPapers}

\section{Flavour Physics}

Flavour physics deals with the study of flavour transitions, mixing and CP violation 
in all its aspects. Precision measurements and the measurement of rare, and 
search for forbidden processes provides sensitivity to new physics beyond 
the current energy frontier of direct production. By these measurements 
it is hoped to acquire insight into the mystery of the observed flavour structure 
(which is related to the BEH sector).

\subsection{Tetraquarks?}

Although hadron spectroscopy is not a topic traditionally discussed at this 
conference, it occurred that the D0 experiment at Tevatron had recently 
reported$\,$\cite{D0Bspi} the observation of a new state in the invariant
mass spectrum of $B_s(\to J/\psi\phi)\pi^\pm$. The new state has a mass and width 
of 5568$\;$MeV and 22$\;$MeV, respectively, and a fiducial yield ratio
$\rho$ (relative to the $B_s$ yield) between about 5 and 12\%. 
It is compatible with a hadronic state with valence quarks of four different 
flavors (tetraquark) made of a diquark-antidiquark pair and quantum 
numbers $J^P=0^+$. 
A prompt cross-check performed by the LHCb 
collaboration$\,$\cite{LHCb-Bspi,JeroenvanTilburg} did not confirm the 
observation in a 20 times larger $B_s$ sample. An upper limit on $\rho$ of 
1--2\% depending on the fiducial region is found.
Other experiments are also looking for this state. The results may depend on 
beam, energy and analysis differences between the experiments. 

\subsection{CKM Matrix}

\begin{wrapfigure}{R}{0.52\textwidth}
\centering
\vspace{-0.4cm}
\includegraphics[width=0.49\textwidth]{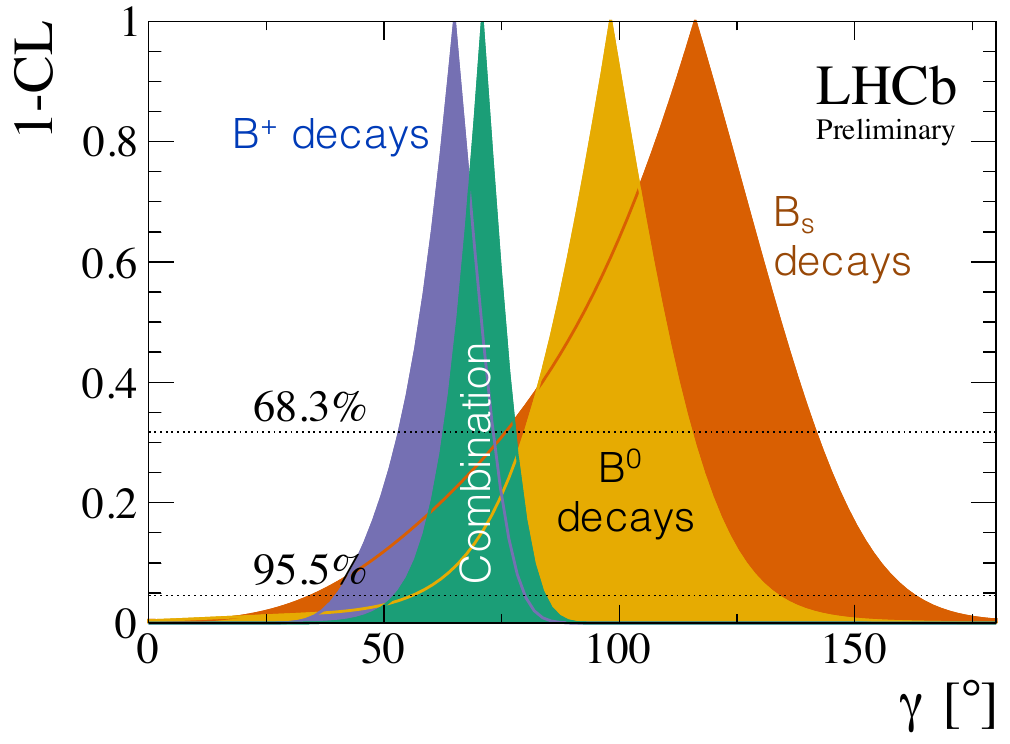}
\vspace{-0.1cm}
\caption[.]{Confidence level versus the CKM angle $\gamma$ as obtained by 
  LHCb$\,$\cite{MalcolmJohn} for analyses involving $B^+$ decays (blue), $B_s$ 
  decays (dark orange), $B^0$ decays (light orange), and their 
  combination$\,$\cite{LHCb-gammaComb} (green).
  \label{fig:LHCb-gamma}}
\vspace{-0.0cm}
\end{wrapfigure}
Among the central topics of flavour physics is the continuing effort to 
overconstrain the CKM matrix and thus test the Standard Model quark-flavour 
sector. LHCb has joined this effort with important 
contributions. A precise measurement of the CKM angle $\gamma$ (through
tree-level processes) together with ${\rm sin}(2\beta)$ (through 
mixing-induced $CP$ violation) or $|V_{ub}|$ (through tree-level processes), 
fixes the apex of the CKM unitarity triangle. All other measurements probe these 
constraints. Among the results reported by LHCb in this area 
are$\,$\cite{JeroenvanTilburg}: a $|V_{ub} / V_{cb}|$ measurement from 
$\Lambda_b  \to p\mu\nu$ with 5\% precision$\,$\cite{LHCb-Vub} 
that is closer to the exclusive $B$-factory numbers for $|V_{ub}|$ 
(which exhibit tension with the larger inclusive numbers), the world's best single 
$\Delta m_d$ measurement$\,$\cite{LHCb-dmd}
 $0.5050 \pm 0.0021\pm 0.0010\;{\rm ps}^{-1}$ 
(the $B$-factories have a combined uncertainty of $0.005\;{\rm ps}^{-1}$), a
${\rm sin}(2\beta)$ measurement$\,$\cite{LHCb-sin2b} 
of $0.731 \pm 0.035 \pm 0.020$ that approaches 
the precision of the $B$-factories, the world's best constraints on $CP$ violation 
in $B^0_{(s)}$ mixing ($a_{sl}^s$, $a_{sl}^d$) in agreement with the 
Standard Model (D0 sees a 3.6$\sigma$ deviation), and a search for $CPT$ 
violation$\,$\cite{LHCb-CPT} 
(difference in mass or width) in the $B^0_{(s)}$ systems together with the 
measurement of sidereal phase dependence of the $CPT$ violating parameter.

LHCb has engaged into a vigorous programme$\,$\cite{MalcolmJohn} 
to determine the CKM angle $\gamma\sim {\rm arg}(-V_{ub}^\star)$. 
It can be measured through interference of $b\to u$ with $b\to c$ tree 
transitions. The hadronic parameters are the amplitude ratio 
$r_B$ and the strong final-state-interaction (FSI) phase $\delta_B$
that need to be determined from data. The extraction of $\gamma$ is then 
theoretically clean, but large statistics are needed due to the CKM suppression 
of some of the involved amplitudes. To fully exploit the available data LHCb 
uses $B^\pm$, $B^0$, $B_s$, and many $D$ decay modes requiring different 
techniques; also $DK^\star$ and $D_s K$ are used. Some modes show large direct
$CP$ asymmetries. It is unfortunately impossible to appropriately discuss  the individual 
measurements in this summary, so we only show the overall results on $\gamma$ 
in Fig.~\ref{fig:LHCb-gamma}. The combined fit, dominated by the measurements
from charged $B^+$ to charm decays, gives$\,$\cite{LHCb-gammaComb}  
$\gamma=70.9^{\,+7.1}_{\,-8.5}\;$deg, which is in agreement with the value from 
the CKM fit (not including the direct $\gamma$
measurements) of$\;$\cite{CKMfitter} $68\pm2\;$deg.


\subsection{{\em CP} violation and mixing in charm decays}

In the neutral charm sector the  mixing probability is extremely low due to CKM 
suppression of order $\lambda^{10}$,  making charm mixing a challenging 
measurement. Mixing-induced or direct $CP$-violation effects are also expected 
to be small so that for both measurements large data yields are needed. 
Owing to a large cross section and hadronic triggers, LHCb has collected a 
huge charm sample during Run-1. A new mixing 
analysis$\,$\cite{AlexPearce} 
presented at this conference used the decay $D^0\to K^-\pi^+\pi^-\pi^+$ to 
determine the strong phase difference needed for the measurement of $\gamma$ 
from $B^+\to D^0(\to K^-\pi^+\pi^-\pi^+)K^+$ decays. It exploits the time-dependent
ratio of wrong-sign ($D^0\to K^+\pi^-\pi^+\pi^-$) to right-sign 
($D^0\to K^-\pi^+\pi^-\pi^+$) events that depends on the charm mixing 
coefficients,  the ratio of Cabibbo-favoured and doubly Cabibbo-suppressed 
amplitudes, and on their interference (hence the sensitivity to the strong 
phase).

\begin{figure}[t]
\centerline{\includegraphics[width=0.6\linewidth]{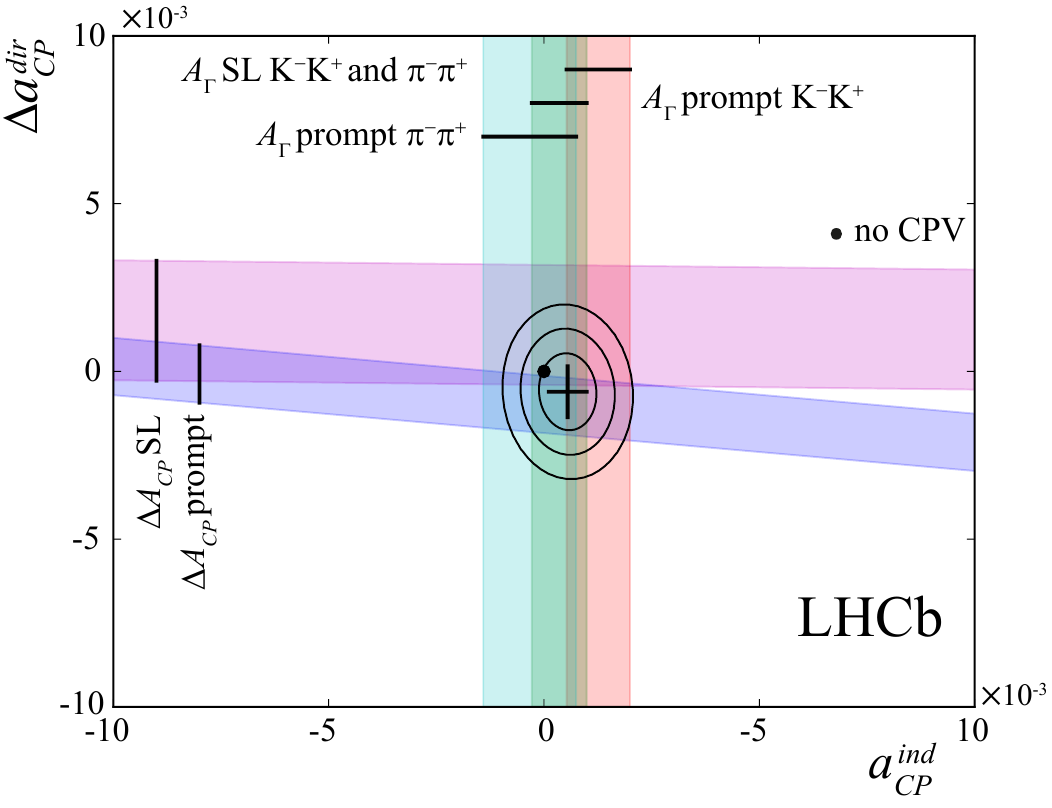}}
\vspace{-0.0cm}
\caption[.]{$CP$ asymmetries measured in $D^0/\overline D^0\to K^+K^-/\pi^+\pi^-$
  decays decomposed into direct and indirect components.$\,$\cite{LHCb-AcpNew}
  The measured values by LHCb are compatible with the no-$CP$-violation hypothesis. 
  \label{fig:LHCb-dAcp}}
\end{figure}
LHCb also presented$\,$\cite{LHCb-CharmMix,AlexPearce} 
a new measurement of the time-integrated $CP$ asymmetry 
$\Delta A_{CP} = A_{CP}(D^0/\overline D^0\to K^+K^-)-A_{CP}(D^0/\overline D^0\to \pi^+\pi^-)$, where the $D^0$ flavour is inferred from the charge of the 
soft pion in the decay $D^{\star+}\to D^0\pi^+$. 
An earlier result by LHCb$\,$\cite{LHCb-Acp} 
using $0.6\;{\rm fb}^{-1}$ of data collected during 
Run-1 exhibited an unexpected $3.5\sigma$ deviation from zero 
($\Delta A_{CP} =  −0.82 \pm 0.21 \pm 0.11$, where the first uncertainty is 
statistical and the second systematic). The new result$\,$\cite{LHCb-AcpNew}  using 
the full $3.0\;{\rm fb}^{-1}$ Run-1 data sample,
$\Delta A_{CP} =  −0.10 \pm 0.08 \pm 0.03$,  does not confirm the earlier
evidence for a deviation$\,$\cite{AlexPearce} (see also Fig.~\ref{fig:LHCb-dAcp}).

\vfill\pagebreak
\subsection{Rare $B$ decays}

\begin{wrapfigure}{R}{0.31\textwidth}
\centering
\vspace{-0.5cm}
\includegraphics[width=0.25\textwidth]{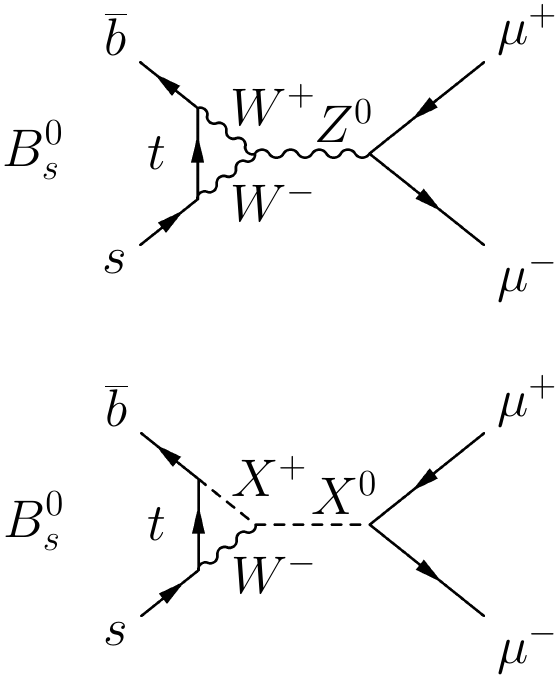}
\vspace{-0.4cm}
\caption[.]{Representative Feynman diagrams for Standard Model (top) and new physics 
  (bottom) contributions to  $B_s\to\mu^+\mu^-$.
  \label{fig:BsmumuFeyn}}
\vspace{-0.4cm}
\end{wrapfigure}
The  rare flavour-changing neutral current decay $B_s\to\mu^+\mu^-$ 
(c.f. diagrams in Fig.~\ref{fig:BsmumuFeyn}) is a prominent channel 
to look for new physics. It has been searched for during almost 30 years 
at many accelerators improving the sensitivity by five orders of magnitude 
before being observed by CMS and 
LHCb$\,$\cite{CMSLHCb-Bsmumu,JohannesAlbrecht,SanjayKumarSwain} 
in November 2014 through a combination of their Run-1 datasets. They found
${\cal B}(B_s\to\mu^+\mu^-)=2.8^{\,+0.7}_{\,-0.6}\cdot10^{-9}$ with a 
significance of 6.2$\sigma$. The corresponding Standard Model 
prediction$\,$\cite{Theory-Bsmumu} $(3.7\pm0.2)\cdot10^{-9}$ is
in agreement with the measurement. 
The CKM suppressed $B_d$ channel$\:$\footnote{The $d$ subscript in $B_d$ 
  is usually omitted.}
was found to be ${\cal B}(B\to\mu^+\mu^-)=3.9^{\,+1.6}_{\,-1.4}\cdot10^{-10}$ 
with a significance of 3.2$\sigma$. This value is larger than the 
prediction$\,$\cite{Theory-Bsmumu} $(1.1\pm0.1)\cdot10^{-10}$.

At this conference,  ATLAS presented their full Run-1 result
for the two channels.$\,$\cite{ATLAS-Bsmumu,SandroPalestini} The analysis 
proceeded similarly to that of CMS and LHCb employing a multivariate Boosted 
Decision Tree (BDT) to suppress hadrons faking muons giving rise to
peaking backgrounds, another BDT to suppress continuum background, and a
two-dimensional fit to continuum-BDT bins and the dimuon mass (unbinned)
to locate the signal. 
The fitted event yields are normalised to $B^+ \to J/\psi K^+$ requiring as 
input the ratio of decay constants $f_s /f_d$ taken from a dedicated ATLAS 
measurement (also requiring theoretical input). As control channels to validate
the cut efficiencies and multivariate analyses serve $B^+ \to J/\psi K^+$ and 
$B_s\to J/\psi\phi$. The expected sensitivity of the analysis for a Standard 
Model branching fraction is 3.1$\sigma$.
\begin{figure}[t]
\centerline{\includegraphics[width=\linewidth]{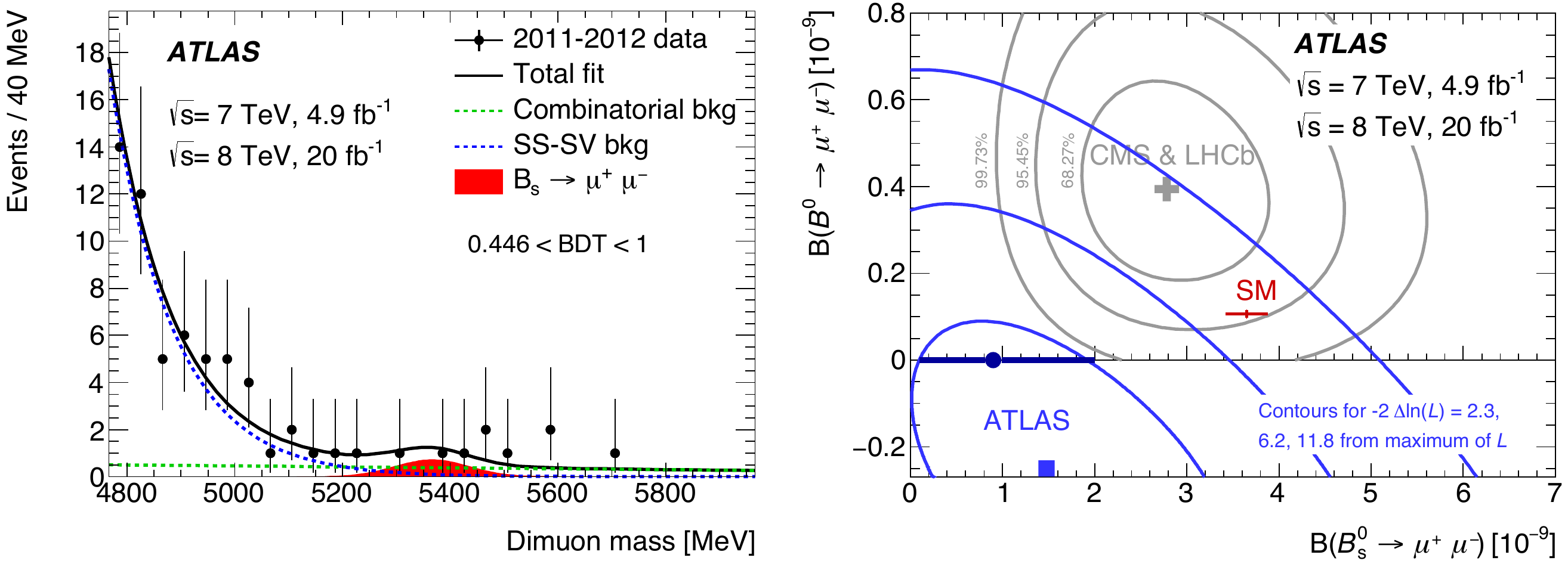}}
\vspace{-0.0cm}
\caption[.]{{\bf Left:} dimuon mass distribution in the most signal-like BDT bin.
  Data and fit results (split into signal and background components) are overlaid.
  {\bf Right:} delta log-likelihood contours in the ${\cal B}(B\to\mu^+\mu^-)$ 
  versus ${\cal B}(B_s\to\mu^+\mu^-)$ plane obtained without imposing the 
  constraint of non-negative branching fractions. Also shown are the combined 
  result by CMS and LHCb, the Standard Model prediction with uncertainties
  and the ATLAS result within the boundary of non-negative branching fractions.
  The figures taken from Ref.$\,$\cite{ATLAS-Bsmumu}.
  \label{fig:ATLAS-Bsmumu}}
\end{figure}
Figure~\ref{fig:ATLAS-Bsmumu} (left panel) shows the dimuon mass 
distribution in the most signal-like BDT bin. No significant signal is seen in 
this or the other two selected BDT bins. Constraining the two branching fractions 
to be non-negative in the fit gives
${\cal B}(B_s\to\mu^+\mu^-)=0.9^{\,+1.1}_{\,-0.8}\cdot10^{-9}$ with an 
upper limit of $3.0\cdot10^{-9}$ at 95\% CL. The  upper limit 
for ${\cal B}(B\to\mu^+\mu^-)$ is $4.2\cdot10^{-10}$. 
The compatibility with the Standard Model amounts to 2.0$\sigma$.
The right panel of Fig.~\ref{fig:ATLAS-Bsmumu} shows the fit result in the 
two-dimensional branching fraction plane together with the combined CMS 
and LHCb result and the Standard Model prediction. Also shown is the 
ATLAS result within the boundary of non-negative branching fractions.

\subsection{Flavour anomalies}

\begin{wrapfigure}{R}{0.29\textwidth}
\centering
\vspace{-0.5cm}
\includegraphics[width=0.25\textwidth]{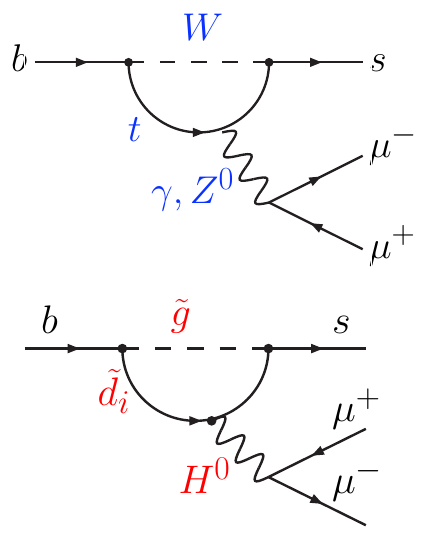}
\vspace{-0.3cm}
\caption[.]{Representative Feynman diagrams for Standard Model (top) and new physics 
  (bottom) contributions to the process $b\to s\mu^+\mu^-$.
  \label{fig:smumu-Feyn}}
\vspace{-0.1cm}
\end{wrapfigure}
\begin{figure}[t]
\centerline{\includegraphics[width=\linewidth]{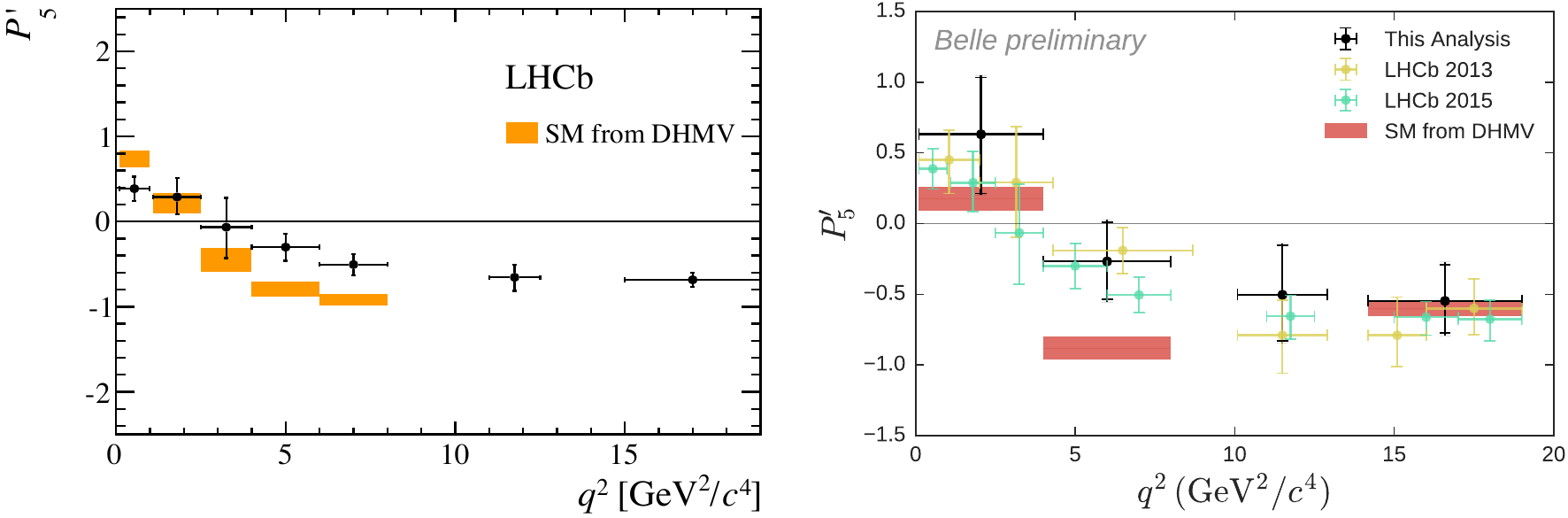}}
\vspace{-0.0cm}
\caption[.]{The angular ratio $P_5^\prime$ versus the invariant mass 
  $q^2$ of the recoiling dimuon system in $B\to K^\star\mu^+\mu^-$
  measured by LHCb$\,$\cite{LHCb-K*mumu} (left panel) and 
  Belle$\,$\cite{Belle-smumu} (right panel). Also shown 
  are selected theoretical predictions.
\label{fig:LHCbBelle-P5prime}}
\end{figure}

Several measurements in the flavour sector exhibit non-significant but interesting
anomalies with respect to theory predictions. A prominent example is given by 
angular coefficients describing the transition $b\to s\mu^+\mu^-$. 
Figure~\ref{fig:smumu-Feyn} shows  Feynman graphs for Standard
Model and putative new physics contributions. 
The LHCb collaboration performed an  angular analysis of the decay
$B\to K^\star\mu^+\mu^-$ using the full Run-1 data sample and 
determining eight independent $CP$-averaged 
observables.$\,$\cite{LHCb-K*mumu,JohannesAlbrecht} A convenient observable 
for comparison with theory is the ratio $P_5^\prime=S_5/\sqrt{F_L(1-F_L)}$
in which the form-factor uncertainty cancels. Figure~\ref{fig:LHCbBelle-P5prime} 
(left panel) shows the distribution of $P_5^\prime$ versus the invariant mass 
$q^2$ of the recoiling dimuon system. The LHCb data points show a 
tension with the chosen theory prediction in the $q^2$ range 
between 4 and 8$\;$GeV$^2$. The Belle collaboration recently performed 
a measurement of that observable which is in agreement with 
the LHCb result but has lower statistical precision$\,$\cite{Belle-smumu} 
(see right panel of Fig.~\ref{fig:LHCbBelle-P5prime}).
Angular and differential branching fraction analyses were also 
performed for $B_s\to\phi\mu^+\mu^-$ (also exhibiting a localised tension 
with the prediction) and a differential branching fraction
analysis for $B^+\to K^+\mu^+\mu^-$. 
A global fit with an effective new physics parameterisation (Wilson coefficients 
$C_9^{\rm NP}$, $C_{10}^{\rm NP}$) can reproduce the observed 
discrepancy pattern.$\,$\cite{LarsHofer}

\begin{figure}[t]
\centerline{\includegraphics[width=\linewidth]{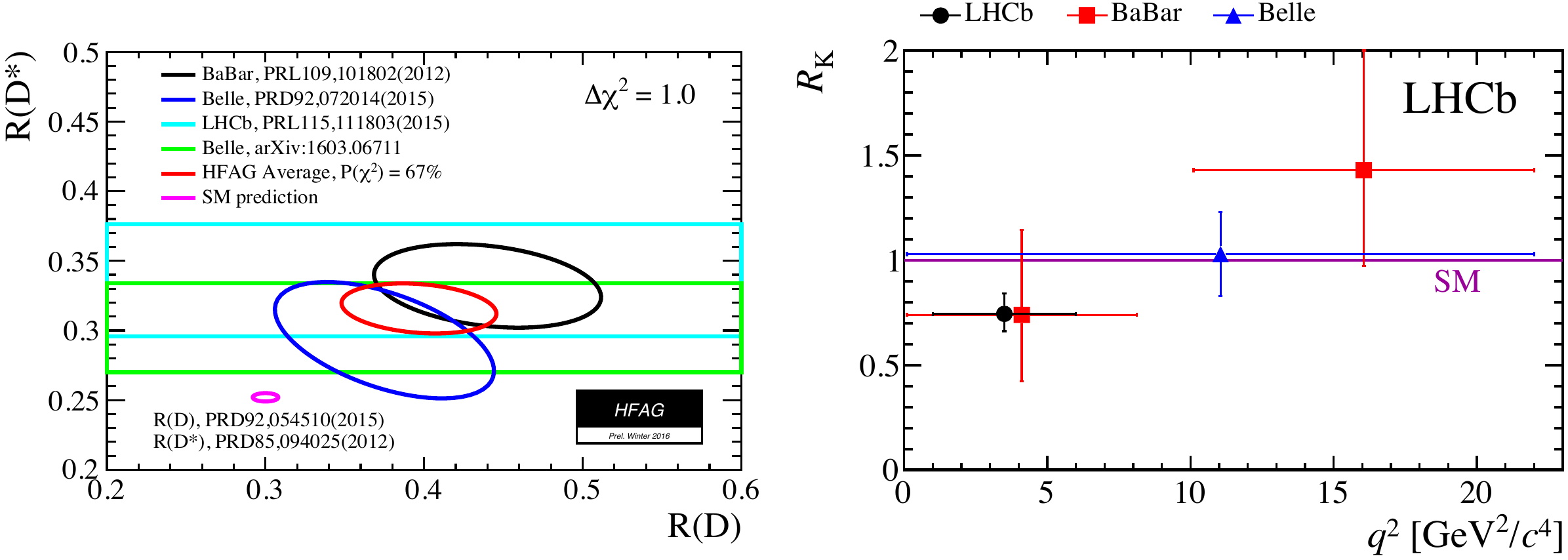}}
\vspace{-0.0cm}
\caption[.]{Ratios of the semileptonic $B$ decays as measured by the $B$-factory
  experiments and LHCb. The left panel shows  the two-dimensional plane 
  $R_{D^{\star}}$ versus $R_{D}$, and the right panel shows $R_K$ versus the 
  invariant mass of the lepton pair. See text for the definitions of the variables.
\label{fig:BfactLHCb-RDK}}
\end{figure}
Less plagued by hard to estimate theoretical uncertainties are lepton 
universality tests. Such tests were performed at the per-mil level at LEP 
and other $e^+ e^-$ colliders not showing any significant discrepancy 
with the expectation of universal lepton coupling. The $B$-factory experiments 
and LHCb have measured ratios of semileptonic $B$ decays that have 
robust Standard Model 
predictions.$\,$\cite{BABAR-RD,Belle-RD,LHCb-RD,JohannesAlbrecht,PabloGoldenzweig} 
Figure~\ref{fig:BfactLHCb-RDK} shows the various measurements 
of the ratios 
$R_{D^{(\star)}}={\cal B}(B^0\to D^{(\star)}\tau\nu)/{\cal B}(B^0\to D^{(\star)}\ell\nu)$
(left panel) and 
$R_{K}={\cal B}(B^+\to K^+\mu^+\mu^-)/{\cal B}(B^+\to K^+e^+e^-)$ (right).
It includes a new  measurement by the Belle 
experiment$\,$\cite{Belle-RDstar} using semileptonic
tagging of the recoil $B$ (as opposed to fully hadronic reconstruction). Belle finds
$R_{D^\star} = 0.302 \pm 0.030 \pm 0.011$ with the first uncertainty being
statistical and the second systematic. The Standard Model expectation 
is $0.252 \pm 0.003$. Belle also studies additional kinematic distributions that have 
new physics sensitivity. The Heavy Flavour Averaging Group (HFAG)
has computed a new combination of  $R_{D^{\star}}$ that includes the latest Belle 
result, giving$\,$\cite{HFAG-RD} $R_{D^\star} = 0.316 \pm 0.016 \pm 0.010$
which is $3.3\sigma$ away from the Standard Model value. The two-dimensional 
combination with $R_D$ increases the significance to $4.0\sigma$. 

\subsection{Charged lepton flavour violation}

A very active field of new physics searches looks for decays that do not conserve
the charged lepton flavour. The predictions of such processes
within the Standard Model and including massive neutrinos are immeasurably 
small so that any signal would be a clean sign of new physics. Searches for
charged lepton flavour violation have a long history. The canonical channels
are $\mu\to e\gamma,\,3e$, $\mu N\to e N$ conversion and 
$\tau\to\mu\gamma, \,3\mu$ reaching down to branching fractions of 
order $10^{-13}$ ($10^{-8}$) for the former (latter) 
channels.$\,$\footnote{The MEG collaboration just released$\,$\cite{MEG-newest}  
  their final limit ${\cal B}(\mu\to e\gamma) < 4.2\cdot10^{-13}$ 
  at 90\% CL, based on the full 2009--2013 dataset (totalling $7.5\cdot10^{14}$ 
  stopped muons on target). An upgrade programme MEG II is underway.} 
Forthcoming  $\mu$-to-$e$ conversion experiments planned in Japan and the US 
have spectacular perspectives with several orders of magnitude improved sensitivity
compared to the current state of the art. 

The NA48/2 experiment at CERN has performed a new 
preliminary analysis$\,$\cite{KarimMassri} of their 2003--2004 data 
sample to search for lepton number violation in the decay 
$K^+ \to \pi^- \mu^+\mu^+$. The main background 
in this channel stems from $K^+ \to \pi^- \pi^+\pi^+$ followed by two 
$\pi^+\to\mu^+\nu$ decays. No excess of events was observed 
giving the strong limit ${\cal B}(K^+ \to \pi^- \mu^+\mu^+)<8.6\cdot10^{-11}$
at 90\% CL. NA48/2 also studied the dimuon invariant mass spectrum of the
opposite-charge $K^+ \to \pi^+ \mu^+\mu^-$ data sample for resonances 
which were not seen. With the new NA62/2 experiment at CERN starting data 
taking a sensitivity of $10^{-12}$ for charged-lepton flavour violation in this
channel is expected. The search for the decay $\pi^0 \to e\mu$ is 
expected to reach a sensitivity of $10^{-11}$.

\subsection{Rare kaon decays}

\begin{wrapfigure}{R}{0.45\textwidth}
\centering
\vspace{-0.0cm}
\includegraphics[width=0.43\textwidth]{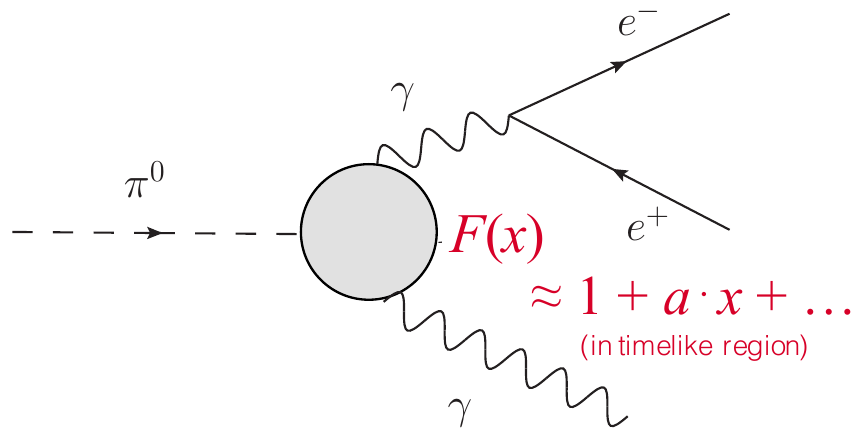}
\vspace{-0.1cm}
\caption[.]{Feynman graph of the Dalitz decay $\pi^0\to e^+e^-\gamma$ 
  used to determine the slope of the timelike transition form factor.
  \label{fig:NA62-DalitzDecayFeyn}}
\vspace{-0.2cm}
\end{wrapfigure}
The NA62 collaboration presented an important  preliminary
measurement$\,$\cite{GiuseppeRuggiero} using their 
2007 dataset of the timelike transition-form-factor (TFF) slope $a$ in 
$F(x)\approx 1+a\cdot x+\dots$ (c.f. Fig.~\ref{fig:NA62-DalitzDecayFeyn}) with 
$\pi^0\to e^+e^-\gamma$ Dalitz decays (1.2\% branching fraction), 
using about 5 billion triggered $\pi^0$ from $K^\pm\to\pi^\pm\pi^0$
decays and a total of about 20 billion $K^\pm$ in the decay  region. The TFF
is an input to model the muon $g-2$ light-by-light scattering 
contribution.$\;$\footnote{Other experimental information relevant for that contribution  
stems from spacelike measurements of the process
$e^+e^-\to e^+e^-\gamma^\star\gamma^\star\to e^+e^-\pi^0$
by CELLO, CLEO and BABAR.}
A challenge for the $F(X)$ extraction is the proper treatment of the QED
radiative corrections that are included in the Monte Carlo (MC) simulation used. 
A fit using MC-based templates gives $a= (3.70\pm0.53\pm0.36)\cdot10^{-2}$,
which exceeds in precision previous measurements by factors.

The NA62/2 collaboration presented the latest  commissioning 
status$\,$\cite{GiuseppeRuggiero} on the way to a first measurement of 
the ultra-rare decay $K^\pm\to\pi^\pm\nu\nub$. That decay was  
observed at BNL by the E949 experiment with a measured branching 
fraction of $(17 \pm 11) \cdot 10^{-11}$ with $(8.4 \pm 1.0) \cdot 10^{-11}$ 
predicted. The goal by NA62/2 is a branching fraction measurement with 
10\% precision (assuming Standard Model rate). The experimental requirements 
are 5 trillion $K^\pm$ decays (giving about 50 signal events) per year, which 
could already be reached in 2016, and a similar order for the background 
suppression (dominated by $K^\pm \to\pi^\pm\pi^0$) to select less than 10
background events per year in the signal regions. 
The most sensitive discriminating variable is the missing 
mass $m^2_{\rm miss}=(p_{K^\pm} - p_{\pi^\pm})^2$, which is positive and 
monotonously falling for signal while it can be negative or peaked for 
backgrounds. The commissioning results showed that the detector 
performance is close to the design requirements already. 

\section{Electroweak precision physics}

High precision measurements of electroweak observables and the global fit 
to these were a masterpiece of the LEP era. It led to constraints on the 
top-quark and Higgs-boson masses before these particles were discovered 
at, respectively, the Tevatron and the LHC. The direct mass measurements were 
then found in agreement with the indirect constraints.  The discovery of the 
Higgs boson overconstrains the electroweak fit and dramatically improves 
its predictability. The fit has thus turned into a powerful test of the Standard Model. 
The current predictions of the observables most benefiting from the known 
Higgs boson mass, split into the various uncertainty terms, are$\,$\cite{Gfitter}: 
$
  m_W=80.3584  \pm 0.0046_{m_t} \pm 0.0030_{\delta_{\rm theo} m_t} 
                         \pm 0.0026_{m_Z} \pm 0.0018_{\Delta\alpha_{\rm had}} 
                         \pm  0.0020_{\as} \pm 0.0001_{m_H} 
                         \pm 0.0040_{\delta_{\rm theo} m_W}\;{\rm GeV},
$
and
$
  \sinleff = 0.231488 \pm 0.000024_{m_t} \pm 0.000016_{\delta_{\rm theo} m_t} 
                               \pm 0.000015_{m_Z} \pm 0.000035_{\Delta\alpha_{\rm had}}
                               \pm 0.000010_{\as} \pm 0.000001_{m_H} 
                               \pm 0.000047_{\delta_{\rm theo} \sinfeff}
$.
Their total uncertainties of $8 \;{\rm MeV}$ and $7\cdot10^{-5}$, respectively, 
undercut the (world average) experimental errors of$\;$\cite{MWWA,sin2thW} 
15$\;$MeV and $16\cdot10^{-5}$, respectively. 

The LHC experiments, as do CDF and D0 since long and continuing, are investing 
efforts into precision measurements of the electroweak observables $m_W$, 
$m_{\rm top}$, and ${\rm sin}^2\theta^{\rm eff}_W$. All are extremely challenging. 
In this respect, it is worth pointing out that the LHC Run-1 is not over yet. It 
represents a high-quality,  very well understood data sample for precision 
measurements.

\subsection{Top-quark mass}

There has been significant progress on the top-quark mass measurements
at the LHC achieving similar precision as those performed by the Tevatron 
experiments. The  currently most accurate LHC number is the 
CMS combination$\,$\cite{CMS-mtopcombined} giving
$m_{\rm top}=172.44\pm0.13\pm0.47\;$GeV, where the first uncertainty 
is statistical and the second systematic. The most recent Tevatron combination 
gives$\,$\cite{Tevatron-mtop} $m_{\rm top}=174.34\pm0.37\pm0.52\;$GeV 
with a tension of $2.4\sigma$ or more with the CMS result.  

While these kinematic mass measurements provide the best current precision on 
$m_{\rm top}$ and must be continued, it is also apparent that they approach 
a difficult systematic uncertainty regime from, mostly, the $b$-quark
fragmentation. A way to improve$\,$\cite{BenjaminStieger} could be to choose
more robust observables with respect to the leading systematic effects
at the possible price of loosing statistical power. The dilepton kinematic 
endpoint is an experimentally clean 
observable, which has however large theoretical uncertainties. More robust could 
be the selection of charmonium states$\,$\cite{CMS-mtopJ/psi} 
or charmed mesons originating from a $b$-hadron
produced in one of the $b$-jets. A clean but rare signature. ATLAS and CMS have also 
invested work into the indirect determination of the top mass from inclusive and 
differential cross-section measurements. These are promising approaches
benefiting from theoretically well defined observables, which are however
not yet competitive with the kinematic methods. They 
also stronger depend on the assumption that no new physics contributes to 
the measured cross sections. The currently best top pole mass determination from 
CMS using a precise Run-1 $e\mu$-based cross-section measurement 
is$\,$\cite{CMS-topXS} $173.8^{\,+1.7}_{\,-1.8}\;$GeV in agreement with the direct 
(kinematic) measurements. 

\subsection{Weak mixing angle}

The CDF, D0,$\,$\cite{Tevatron-sinth} and 
LHC experiments$\,$\cite{ATLAS-sinth,CMS-sinth,LHCb-sinth} 
have extracted the weak mixing angle from $Z/\gamma^\star$ polarisation 
measurements.$\,$\cite{ArieBodek,WilliamJamesBarter} 
The total uncertainty on ${\rm sin}^2\theta^{\rm eff}_W$ at the Tevatron 
are dominated by statistical effects, that of LHCb has similar statistical and 
systematic contributions, while for ATLAS and CMS parton density function 
(PDF) uncertainties are dominant.  A data-driven ``PDF replica rejection'' 
method applied by CDF allows to reduce the sensitivity to PDF and to 
update the measurement when improved PDF sets are available.
Overall, these are complex measurements (in particular with respect to the 
physics modelling) that are important to pursue also in view of a better 
understanding of $Z/\gamma^\star$  production at hadron colliders. 
The precision obtained is however not yet competitive with that of LEP/SLD. 

 \subsection{$W$ boson mass}

\begin{wrapfigure}{R}{0.55\textwidth}
\centering
\vspace{-0.5cm}
\includegraphics[width=0.53\textwidth]{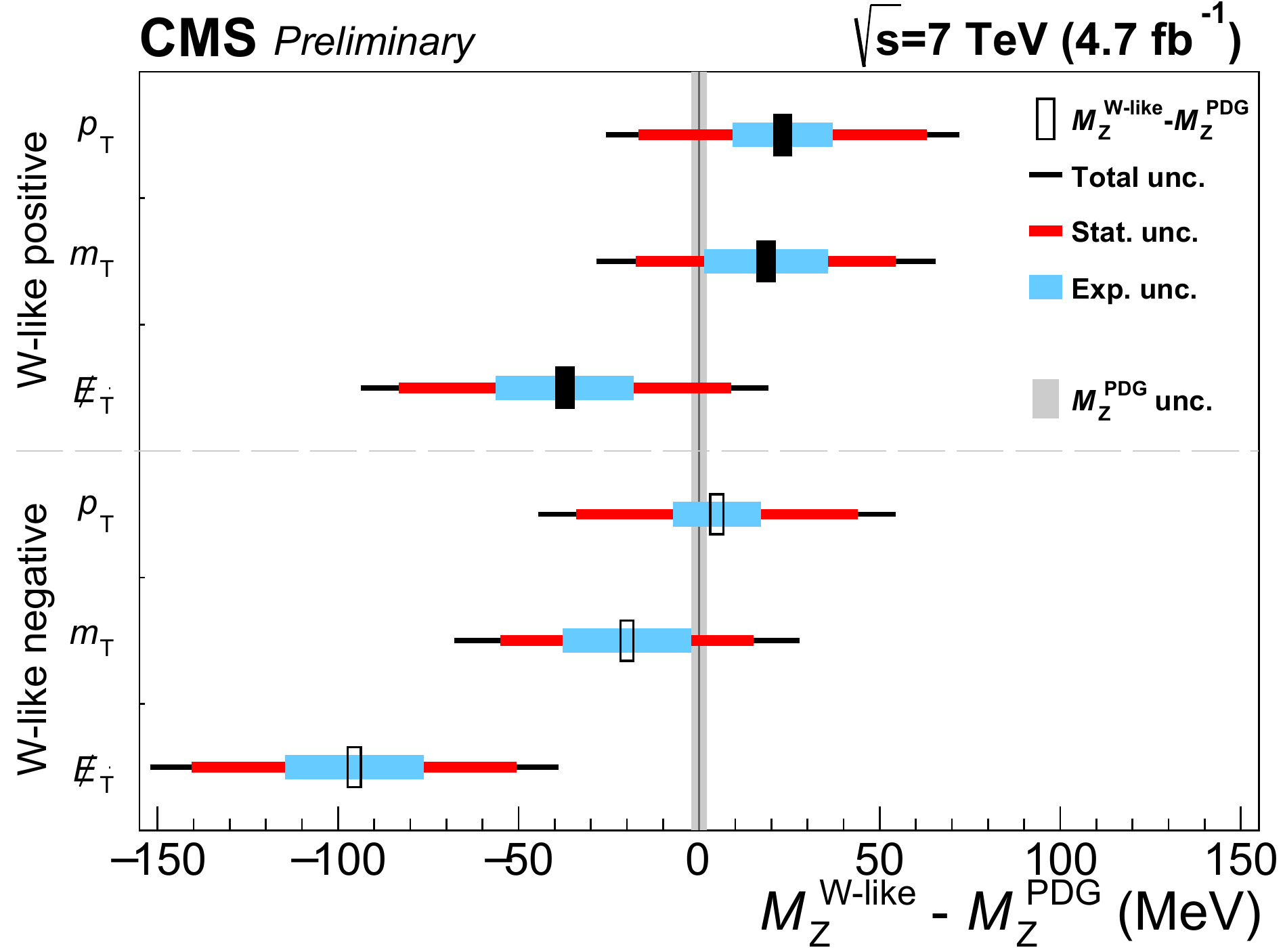}
\vspace{-0.1cm}
\caption[.]{Difference between the fitted $W$-like $Z$ mass and the LEP
 measurement for each $m_W$ probe and $W$ charge. 
  \label{fig:CMS-Wmass}}
\vspace{-0.1cm}
\end{wrapfigure}
ATLAS, CMS and LHCb have presented progress towards a first measurement
of the $W$ mass at the LHC using the leptonic $W$ boson decay, which relies 
on an excellent understanding of the final state.$\,$\cite{MariarosariaD'Alfonso} 
The observables used to probe $m_W$ 
are the transverse momentum of the lepton ($p_{T,\ell}$), the transverse momentum 
of the neutrino ($p_{T,\nu}$), measured from the transverse recoil of the event, 
and the transverse mass of the lepton-neutrino system ($m_T$).
The measurement requires a high-precision momentum and energy scale calibration
(including the hadronic recoil) obtained from $Z$, $J/\psi$ and $\Upsilon$ data,
and excellent control of the signal efficiency and background modelling.
The biggest challenge is posed by the physics modelling. The production is 
governed by PDF and initial state interactions (perturbative and non-perturbative),
that can be constrained by $W^+$, $W^-$, $Z$, and $W+c$ data, and the use of 
NNLO QCD calculations including soft gluon resummation. The experimental $m_W$ 
probes are very sensitive to the $W$ polarisation (and hence to PDF, including 
its strange density). Electroweak corrections are sufficiently well known.

The experiments are in a thriving process of addressing the above issues. Many 
precision measurements (differential $Z$, $W + X$ cross sections, polarisation 
analysis, calibration performance, etc.) are produced on the way with benefits  
for the entire physics programme. Theoretical developments are also mandatory. 
Altogether this is a long-term effort.

CMS presented for the first time a $m_Z$ measurement using a $W$-like 
$Z \to \mu^+\mu^-$ analysis where one muon is replaced by a neutrino that contributes
to the missing transverse momentum in the 
event.$\,$\cite{CMS-Wmass,MariarosariaD'Alfonso} 
It represents a proof-of-principle, although differences with the full $m_W$ analysis 
remain in the event selection, the background treatment and most of the theory 
uncertainties, (\dots). CMS used the 7$\;$TeV dataset to take benefit from the 
lower number of pileup interactions. The momentum scale and resolution 
calibration for that  measurement relies on 
$J/\psi$ and $\Upsilon$ data. Track-based missing transverse momentum is used 
and the $W$ transverse recoil is calibrated using $Z+{\rm jets}$ events. The results
for the different probes and the positive and negative $W$-like cases are shown 
in Fig.~\ref{fig:CMS-Wmass}. Agreement with the LEP measurement is found. 
The uncertainties, depending on the probe used, are: 
statistical: 35--46$\;$MeV, total  systematic: 28--34$\;$MeV, 
QED radiation: $\sim$23$\;$MeV (dominant), lepton calibration: 12--15$\;$MeV.

\section{The LHC at 13 TeV --- Standard Model physics}

A huge milestone was achieved in 2015 with a record proton--proton collision 
energy of 13$\;$TeV and high-energy lead--lead collision. After a rocky start, 
the LHC delivered an integrated proton--proton luminosity of $4.2\;{\rm fb}^{-1}$
with a peak instantaneous luminosity of $5.2\cdot10^{33}\;{\rm cm}^{-2}{\rm s}^{-1}$.
The majority of the data were produced with 25$\;$ns bunch crossing distance
(as opposed to 50$\;$ns at the beginning of the run).  This amount of data 
already improves the reach for many new physics searches. The year 2015 
has also been rewarding for the experiments with many results available for 
the summer conferences, a huge amount of results released for the CERN
end-of-year seminars, and many more at this conference. LHC 
running in 2016 has already started and is expected$\,$\cite{JörgWenninger} 
to reach up to $25\;{\rm fb}^{-1}$ integrated luminosity over the year 
with peak luminosity of about $10^{34}\;{\rm cm}^{-2}{\rm s}^{-1}$.

The integrated luminosity collected by the experiments in 2015 for physics analysis 
amounts to 3.3--3.6$\;{\rm fb}^{-1}$ for ATLAS (depending on the data quality 
requirements applied), 2.2--3.3$\;{\rm fb}^{-1}$ for CMS (0.8$\;{\rm fb}^{-1}$ was
taken in a solenoid-off configuration), and 0.32$\;{\rm fb}^{-1}$ for LHCb after 
luminosity levelling to suppress pileup interactions. The luminosity monitors of 
the experiments were calibrated with dedicated beam-separation scans to preliminary
5.0\% (ATLAS), 2.7\% (CMS), 3.8\% (LHCb) relative precision. The average 
number of pileup interactions in ATLAS and CMS were
$\langle \mu\rangle_{50\;\rm ns} \simeq 20$, 
$\langle \mu\rangle_{25\;\rm ns} \simeq 13$  (for comparison 
$\langle \mu\rangle_{8\;\rm TeV} \simeq 21$), 
and in LHCb $\langle \mu\rangle_{\rm levelled} \simeq 1.7$.

\subsection{Inclusive W and Z production}

Inclusive $W$ and $Z$ boson events represent a rich physics laboratory
with strong PDF dependence (the $W^+ / W^-$ ratio is sensitive to low-$x$ up and 
down valence quarks, the $W^\pm / Z$ ratio constrains the strange density), and 
as probes for QCD and electroweak physics. Their leptonic decays  also serve as
standard candles to calibrate the electron and muon 
performance of the detectors. 

\begin{figure}[t]
\centerline{\includegraphics[width=\linewidth]{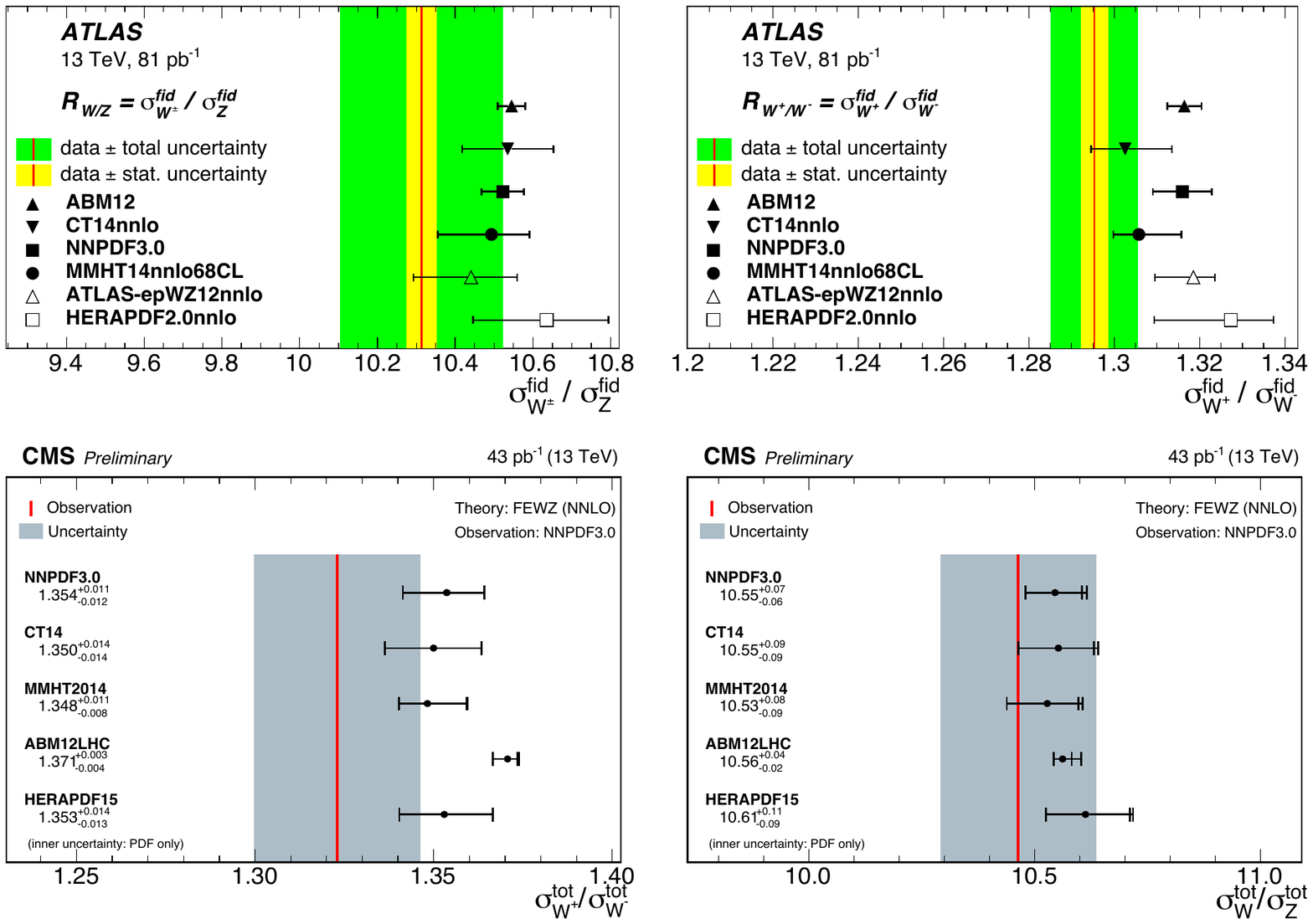}}
\vspace{-0.0cm}
\caption[.]{Ratios of measured fiducial (top, ATLAS) and total (bottom, CMS) 
  cross sections of $W^+$ to $W^-$ (left) and $W^\pm$ to $Z$ (right) 
  production compared to predictions using various PDF sets.
\label{fig:ATLASCMS-W-Z}}
\end{figure}
ATLAS, CMS and LHCb have studied single 
gauge boson production at 7, 8 and 13$\;$TeV, where LHCb covers a complementary 
phase space in $x, Q^2$ owing to its forward acceptance ($2.0 < |\eta| < 4.5$).
Initial 13$\;$TeV inclusive $Z$ ($W^\pm$) cross section measurements were performed 
by all three experiments (ATLAS and CMS), who find overall agreement with the 
Standard Model predictions.$\,$\cite{WilliamJamesBarter,ATLAS-W/Z,CMS-W/Z,LHCb-Z} 
Figure~\ref{fig:ATLASCMS-W-Z} shows ratios of cross sections from 
ATLAS (top panels) and CMS (bottom panels) compared to various PDF sets.
Systematic uncertainties cancel to some extent in these ratios so that already 
a precision of better than 3\% is achieved. Similar experiment-versus-theory 
patterns are observed for both experiments. 

Among the Run-1 results presented were measurements of $p_T(Z)$ at 8$\;$TeV 
from ATLAS$\,$\cite{ATLAS-pTZ8} (also CMS$\,$\cite{CMS-pTZ8} and 
LHCb$\,$\cite{LHCb-pTZ8}) showing that soft gluon resummation is 
needed at low $p_T$ to describe the data. 
NNLO calculations lie systematically below the data at high 
$p_T$. Charge asymmetry results are found to be well 
predicted by theory. High-rapidity $W$ and $Z$ cross sections measured by LHCb 
are well predicted by NNLO theory. A full angular analysis of $Z\to\mu^+\mu^-$
production and decay at 8$\;$TeV that is sensitive to the $Z$ polarisation and 
decay structure was performed by CMS.$\,$\cite{CMS-Zangular}

\subsection{Diboson production}

Diboson production is an important sector of LHC physics, intimately 
related to electroweak symmetry breaking. ATLAS and CMS studied diboson 
production at 7, 8, 13$\;$TeV. Detailed inclusive, fiducial and differential 
cross-section analyses were performed at 8$\;$TeV, and first 13$\;$TeV results
were released.$\,$\cite{TieshengDai} 
Theoretical predictions at NNLO accuracy are needed to match the data.

\begin{figure}[t]
\centerline{\includegraphics[width=\linewidth]{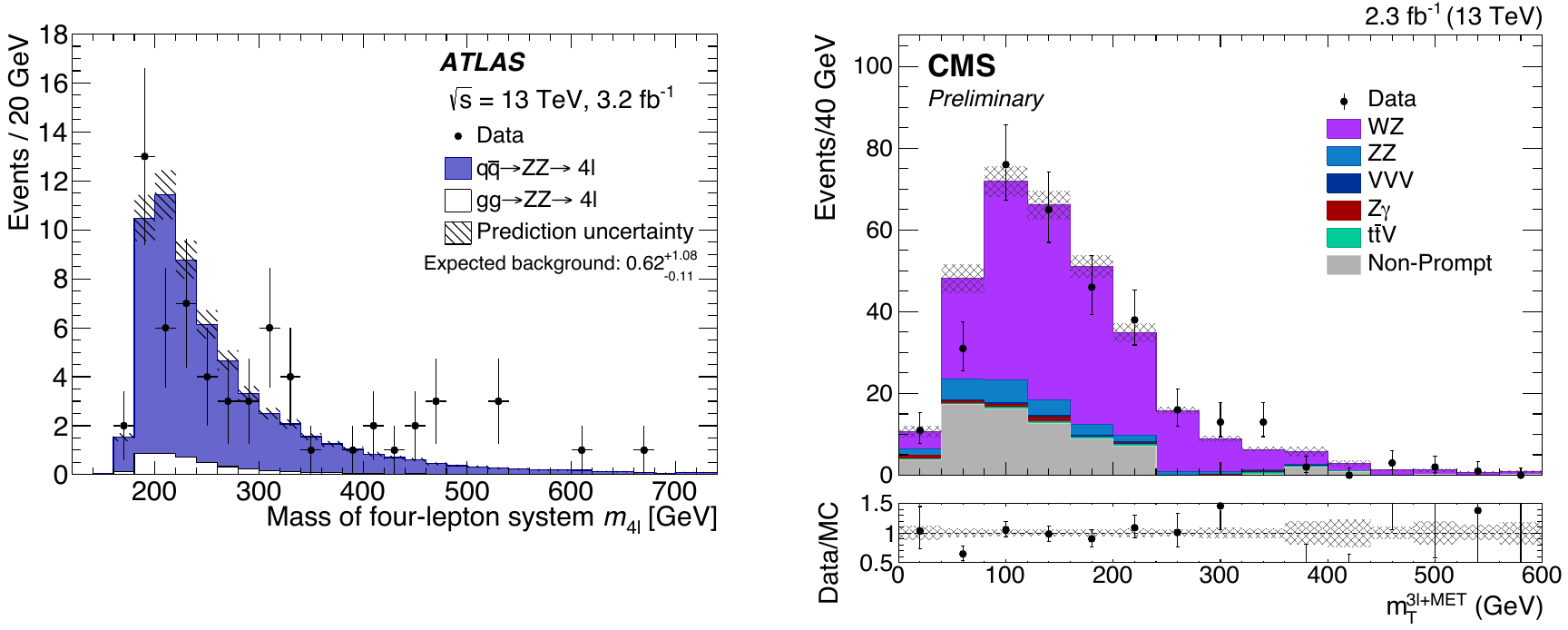}}
\vspace{-0.0cm}
\caption[.]{Detector level distributions of the four-lepton invariant mass (left, 
  ATLAS$\,$\cite{ATLAS-ZZ}) and the three-lepton plus missing transverse
  momentum transverse mass (right, CMS$\,$\cite{CMS-WZ}) after 
  corresponding diboson selections. 
  \label{fig:ATLASCMS-diboson}}
\end{figure}

The $ZZ$ cross section at 13$\;$TeV was measured by ATLAS$\,$\cite{ATLAS-ZZ}
and CMS,$\,$\cite{CMS-ZZ} $WZ$ by CMS$\,$\cite{CMS-WZ}: all agree with 
the Standard Model predictions (see Fig.~\ref{fig:ATLASCMS-diboson} for 
selected detector-level distributions). The $WW$ cross section at 8$\;$TeV, 
measured by both experiments,$\,$\cite{ATLAS-WW8,CMS-WW8} agrees 
with the NNLO prediction improved by soft $p_T$ resummation. A detailed 
recent analysis of $WZ$ production at 8$\;$TeV by ATLAS$\,$\cite{ATLAS-WZ8} 
shows deviations from the NLO prediction, which is not unexpected. A recent 
NNLO calculation moves the theory towards the data.$\,$\cite{WZ-NNLO}
Measurements of $Z\gamma$ cross sections at 8$\;$TeV by 
ATLAS$\,$\cite{ATLAS-Zy} and CMS$\,$\cite{CMS-Zy} are matched by 
NNLO predictions. First evidence for vector-boson scattering (VBS)
was reported in 2014 by ATLAS$\,$\cite{ATLAS-WWqq} and 
by CMS$\,$\cite{CMS-WWqq} in the $W^\pm W^\pm qq$ 
channel. New  8$\;$TeV VBS searches were released in the $(W/Z)\gamma qq$ 
(CMS$\,$\cite{CMS-Vyqq8}) and $WZ qq$ (ATLAS$\,$\cite{ATLAS-WZ8}) modes,
not yet leading to an observation of this process. The triboson process  
$Z\gamma\gamma$ was observed by CMS,$\,$\cite{CMS-Vyy} evidence 
for $W\gamma\gamma$  was reported by ATLAS$\,$\cite{ATLAS-Wyy} 
and CMS.$\,$\cite{CMS-Vyy} The various diboson analyses 
provide a large set of anomalous coupling limits.

\subsection{Top-quark physics}

\begin{wrapfigure}{R}{0.25\textwidth}
\vspace{-0.5cm}
\centering
\includegraphics[width=0.22\textwidth]{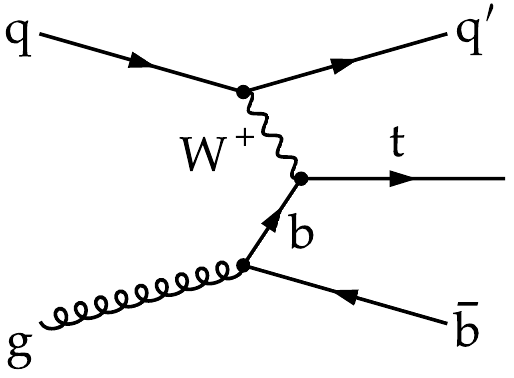}
\vspace{-0.1cm}
\caption[.]{Feynman diagram for electroweak single top quark production. 
  \label{fig:SingleTop-t}}
\vspace{+0.5cm}
\end{wrapfigure}
\begin{figure}[t]
\centerline{\includegraphics[width=\linewidth]{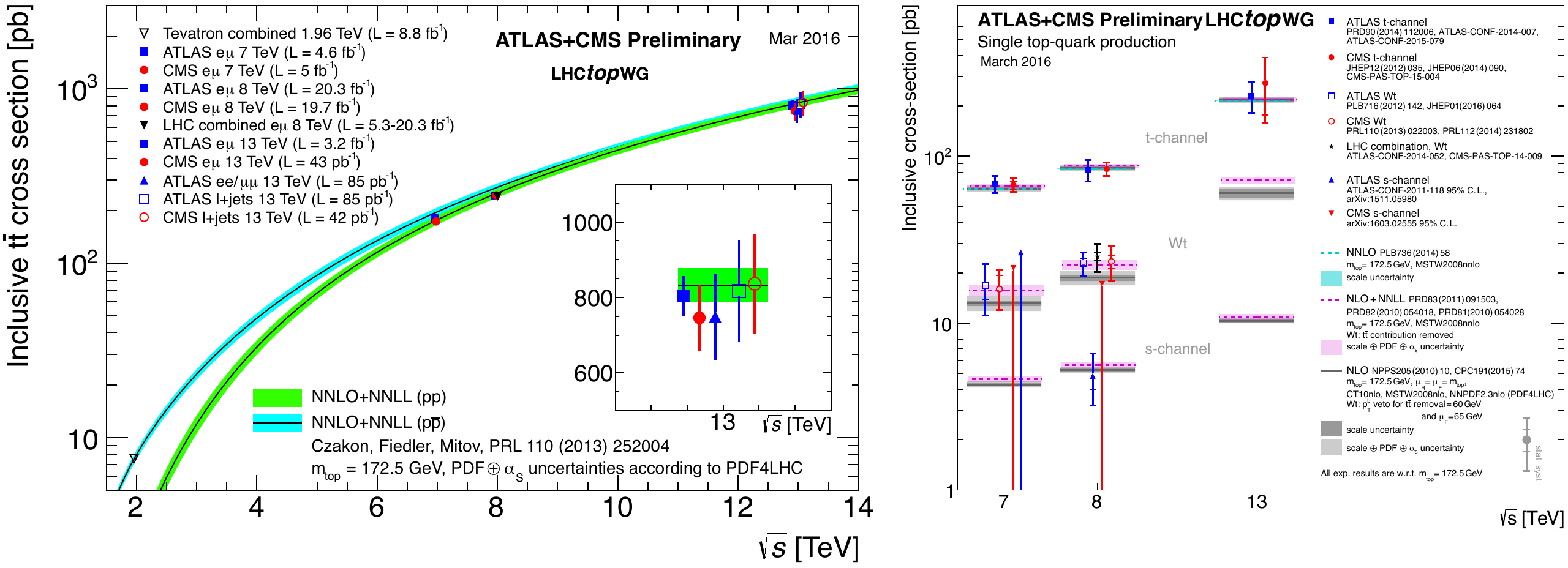}}
\vspace{-0.1cm}
\caption[.]{{\bf Left:} 
  inclusive $pp(p\overline p)\to t\overline t$ production cross section as measured
  by ATLAS, CMS and the Tevatron experiments versus centre-of-mass  energy. Also shown
  are the Standard Model predictions at NNLO+NNLL perturbative order.
  {\bf Right:} single-top production measured by ATLAS and CMS in $pp$ collisions 
  compared to theoretical predictions. 
  \label{fig:ATLASCMS-topXS}}
\end{figure}

The cross section of top-antitop quark pair production at 13$\;$TeV is predicted in the 
Standard Model to increase by a factor of 3.3 over that at 8$\;$TeV. ATLAS and CMS 
have already studied top production in many ways$\,$\cite{PedroFerreiradaSilva} 
at 13$\;$TeV benefiting from a fast analysis turn around in 2015. 
The robust dilepton $e\mu$ final state provides the most precise inclusive results 
at all proton--proton centre-of-mass energies. The inclusive $t\overline t$ production 
cross sections as measured  by ATLAS, CMS and the Tevatron experiments versus 
centre-of-mass energy (see Refs.$\,$\cite{ATLAS-ttXS,CMS-ttXS} for the
13$\;$TeV results), and compared to theory predictions are shown in the left panel of
Fig.~\ref{fig:ATLASCMS-topXS} Differential cross-section measurements 
at 13$\;$TeV show reasonable modelling, though some deviations at large 
jet multiplicity are seen.$\,$\cite{ATLAS-ttdiffXS,CMS-ttdiffXS}

ATLAS and CMS have also measured  t-channel  single top-quark
production$\,$\cite{ATLAS-tchan,CMS-tchan} (see Fig.~\ref{fig:SingleTop-t}) 
that is predicted to increase in rate by a factor of 2.5 at 13$\;$TeV over 8$\;$TeV. 
The cross-section measurements are consistent with this prediction within 
still sizable experimental uncertainties.$\,$\cite{PedroFerreiradaSilva} 
A summary of the measurements and comparison to theory is given in 
Fig.~\ref{fig:ATLASCMS-topXS} (right panel).

\begin{wrapfigure}{R}{0.51\textwidth}
\centering
\vspace{-0.4cm}
\includegraphics[width=0.50\textwidth]{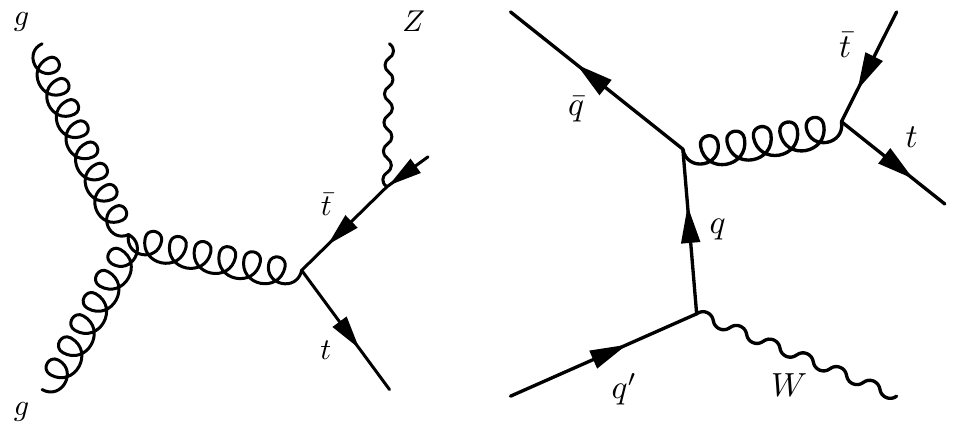}
\vspace{-0.1cm}
\caption[.]{Feynman diagram for top pair production in association with a $Z$ (left)
  or a $W$ boson (right).
   \label{fig:ttV}}
\vspace{-0.4cm}
\end{wrapfigure}
Of particular interest is the measurement of top-pair production in association 
with bosons ($ttZ$ and $ttW$, see Fig.~\ref{fig:ttV} for representative leading 
order Feynman graphs). These channels are important in their own right (in 
particular $ttZ$, which directly probes the top coupling to the $Z$ boson), 
but they also represent 
irreducible backgrounds in $ttH$ and many new physics searches. Because of 
different production processes their respective 13$\;$TeV to 8$\;$TeV cross-section 
ratios are 3.6 ($ttZ$) and 2.4 ($ttW$). ATLAS and CMS showed first 13$\;$TeV results
that combine several multilepton final 
states.$\,$\cite{ATLAS-ttV,CMS-ttV,EmmanuelMonnier} 
The most challenging part of the analysis is the estimate of the reducible 
background due to prompt-lepton misidentification which must be measured 
in the data. At 8$\;$TeV, both processes were observed and found consistent
with the Standard Model predictions (the measured $ttW$ cross section was 
about 1$\sigma$ high in both ATLAS and CMS). The preliminary results 
for 13$\;$TeV are:
$\sigma(ttZ)=0.92 \pm 0.30 \pm 0.11\; \rm pb$, 
$\sigma(ttW)=1.38 \pm 0.70 \pm 0.33\; \rm pb$ (ATLAS),
and  $\sigma(ttZ)=1.07^{\,+0.35\,+0.17}_{\,-0.31\,-0.14}\; \rm pb$ (CMS). They 
agree with the Standard Model predictions computed at NLO, with the 
$ttW$ measurement being again on the high side of the prediction. 

\begin{wrapfigure}{R}{0.47\textwidth}
\centering
\vspace{-0.4cm}
\includegraphics[width=0.45\textwidth]{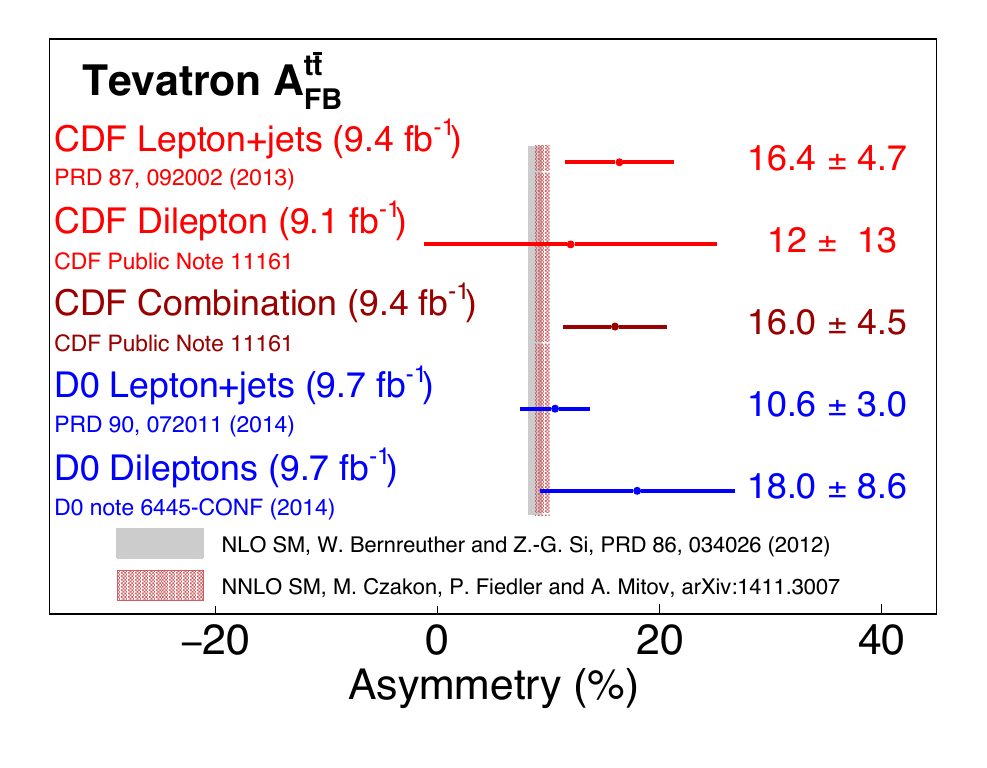}
\vspace{-0.1cm}
\caption[.]{Summary of $A_{\rm FB}(t\overline t)$ measurements by CDF and D0 compared
  to the Standard Model prediction. 
   \label{fig:Tevatron-topAFB}}
\vspace{-0.0cm}
\end{wrapfigure}
The current amount of 13$\;$TeV data is not yet sufficient to probe top decay 
properties beyond those of the LHC Run-1. Instead, new measurements
at 8$\;$TeV and from the Tevatron experiments were 
presented.$\,$\cite{EmmanuelMonnier,ChristianSchwanenberger} 
The Tevatron top forward-backward asymmetry 
measurements,$\,$\cite{CDF-topAfb,D0-topAfb} $A_{\rm FB}(t\overline t)$, 
and its NNLO Standard Model prediction have converged towards each other resolving 
the previous tension in this observable (see Fig.~\ref{fig:Tevatron-topAFB}). 
The measured top charge asymmetries at the
LHC are found in agreement with the Standard Model predictions. The D0 experiment
has released a new measurement of $P$ and $CP$-odd observables, where the $CP$-odd 
one was found compatible with zero as expected.$\,$\cite{D0-CP} 
Top-antitop spin correlations have been established at the LHC and were 
used by ATLAS to put bounds on ``stealth'' supersymmetric
top partners, so-called top squarks or stop.$\,$\cite{ATLAS-topSpin} 
A 4.2$\sigma$ evidence for spin correlations was presented by 
D0.$\,$\cite{D0-topSpin} Highly suppressed FCNC processes such as 
$t \to gq, Zq, Hq$ were probed by ATLAS and CMS with no signal seen. 

\vfill\pagebreak
\subsection{Higgs boson physics}

\begin{wrapfigure}{R}{0.6\textwidth}
\centering
\vspace{-0.4cm}
\includegraphics[width=0.58\textwidth]{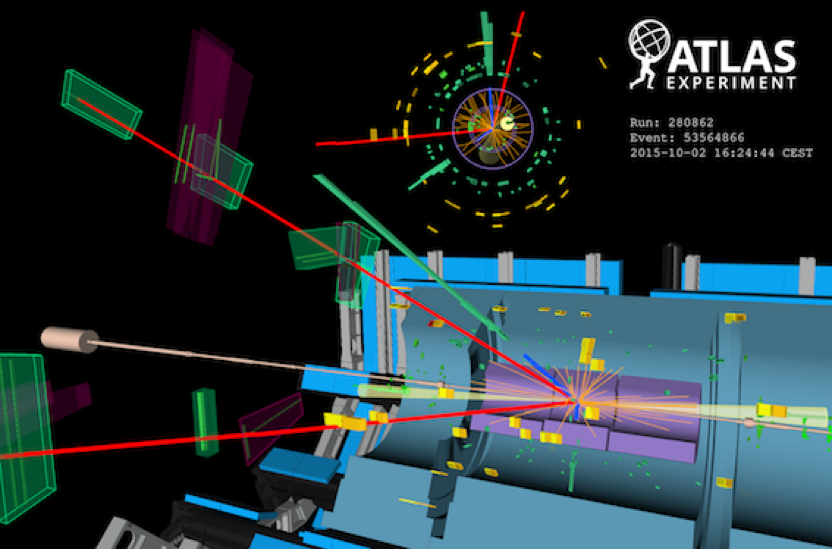}
\vspace{-0.0cm}
\caption[.]{Display of a $H \to ee\mu\mu$ candidate from 13$\;$TeV proton--proton 
  collisions measured by ATLAS. 
  The invariant mass of the four lepton system is 129 GeV, the dielectron (dimuon) 
  invariant mass is 91 (29) GeV, the pseudorapidity difference between the two jets is 
  6.4, the di-jet invariant mass is 2$\;$TeV. This event is consistent with VBF production 
  of a Higgs boson decaying to four leptons.
  \label{fig:HiggsVBFDisplay}}
\vspace{-0.1cm}
\end{wrapfigure}
\begin{figure}[t]
\centerline{\includegraphics[width=0.8\linewidth]{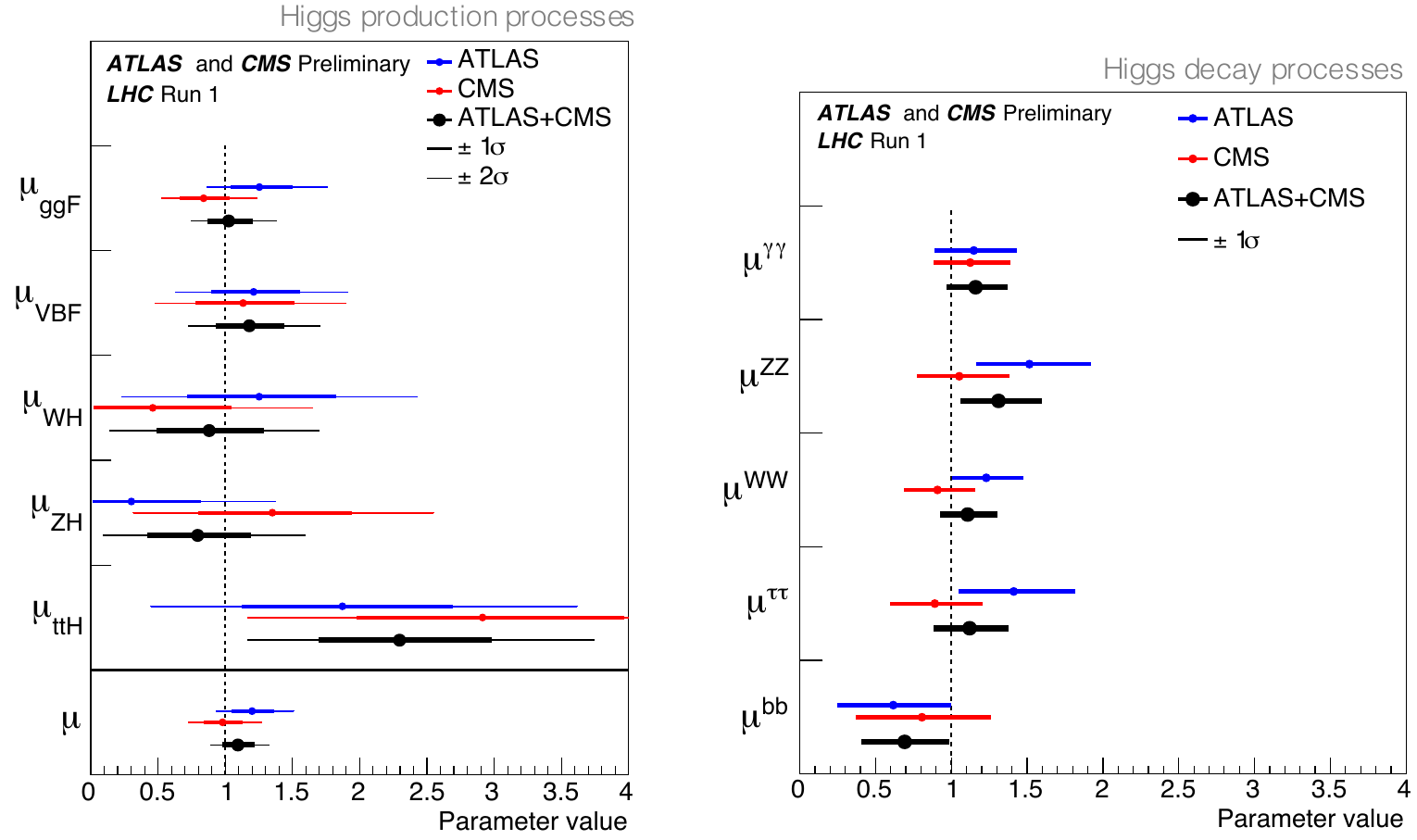}}
\vspace{-0.0cm}
\caption[.]{Higgs boson production signal strengths (left panel) and decay 
  signal strengths (right panel) from the preliminary Run-1 combination of 
  ATLAS and CMS Higgs coupling measurements.$\,$\cite{ATLASCMS-HiggsComb} 
  Also shown are the results 
  for each experiment. The error bars indicate the $1\sigma$ (thick lines) and 
  $2\sigma$ (thin lines on the left panel) intervals. The measurements of 
  the global signal strength $\mu$ are also shown. 
  \label{fig:ATLASCMS-HiggsCouplings}}
\end{figure}

In 2015 ATLAS and CMS accomplished a preliminary combination of their 
Run-1 Higgs boson measurements.$\,$\cite{ATLASCMS-HiggsComb,LidiaDell'Asta} 
Among improved constraints on all couplings 
it established the observation with more than 5$\sigma$ significance of the decay 
$H \to \tau\tau$ and the Higgs boson production through vector-boson 
fusion (VBF). The resulting ratios of measured to predicted signal strengths are 
shown for the production and decay channels in Fig.~\ref{fig:ATLASCMS-HiggsCouplings},
where for the production (decay) channels the corresponding decay (production) 
modes are assumed to be Standard Model like. No significant deviation from one
is observed, albeit a somewhat higher than expected $ttH$ cross section is seen. 
The expected increase in Higgs boson cross section at 13$\;$TeV compared to 8$\;$TeV
is between 2 and 2.4 for $VH$, $ggH$ and VBF, but 3.9 for $ttH$. A luminosity 
of 3.3$\;{\rm fb}^{-1}$ at 13$\;$TeV already attains roughly 80\% of the Run-1 
sensitivity for the latter mode. 

Both ATLAS and CMS have finalised their Run-1 searches for lepton flavour violation 
in Higgs boson decays.$\,$\cite{ATLAS-HiggsLFV,CMS-HiggsLFV} 
While $H \to \mu e $ is severely bound from other flavour physics
measurements, $H \to \tau\mu,  \tau e$ are only weakly constrained. 
CMS released early 2015 a $H \to \tau\mu$ result with a slight 
(2.4$\sigma$) excess. ATLAS has completed its full analysis (including a search for 
$H \to \tau e$) for this conference. The results for ${\cal B}(H\to\tau\mu)$ are 
$0.53 \pm 0.51\%$ ($<1.43\%$ at 95\% CL) for ATLAS, 
$0.84^{\,+0.39}_{\,-0.37}\%$ ($<1.51\%$) for CMS, and 
${\cal B}(H\to\tau e)=-0.3\pm0.6\%$ ($<1.04\%$) for ATLAS.

\begin{figure}[t]
\centerline{\includegraphics[width=\linewidth]{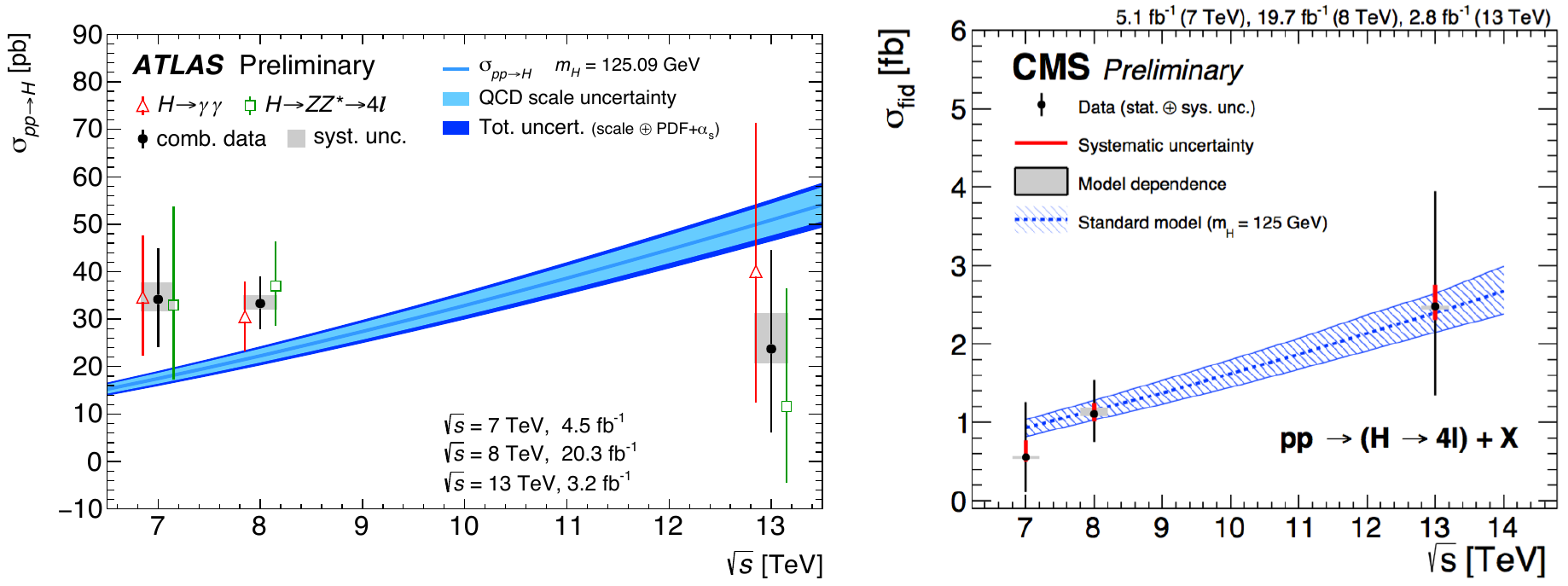}}
\vspace{-0.0cm}
\caption[.]{{\bf Left: } combined inclusive $H\to4\ell,\;\gamma\gamma$ cross sections
  versus proton--proton CM energy as measured by ATLAS and compared to the NNLO
  theoretical prediction. {\bf Right}: CMS fiducial $H\to4\ell$ cross section versus CM energy. 
\label{fig:ATLASCMS-HiggsXS}}
\end{figure}

Although the sensitivity is yet marginal for inclusive Higgs boson production,  ATLAS
and CMS have looked in their 13$\;$TeV datasets for $H\to4\ell$ and 
$H\to\gamma\gamma$ events.$\,$\cite{ATLAS-Higgs,CMS-Higgs,SethZenz} 
The observed signal yields are consistent with the theoretical predictions. 
Figure~\ref{fig:ATLASCMS-HiggsXS} shows the measured inclusive Higgs
boson production cross sections versus the proton--proton centre-of-mass 
energy for ATLAS (left) and CMS (right).

\begin{wrapfigure}{R}{0.27\textwidth}
\centering
\vspace{-0.4cm}
\includegraphics[width=0.24\textwidth]{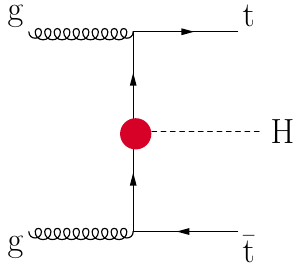}
\vspace{-0.1cm}
\caption[.]{Feynman diagram for Higgs boson production in association with 
  two top quarks probing the top--Higgs coupling strength.
   \label{fig:ttHFeyn}}
\vspace{-0.4cm}
\end{wrapfigure}
The CMS collaboration released in record time first 13$\;$TeV results for $ttH$ 
production searches,$\,$\cite{CMS-ttH,JohannesHauk} 
which is the only currently accessible 
channel that directly measures the top--Higgs coupling (c.f. Feynman graph
in Fig.~\ref{fig:ttHFeyn}). All major Higgs boson decay channels, $\gamma\gamma$, 
multileptons, and $bb$, were analysed. In particular the latter two channels represent
highly complex analyses. The multilepton mode targets Higgs boson decays to 
$\tau\tau$, $WW\to2\ell2\nu$, and $ZZ\to 2\ell2\nu,\,4\ell$ together with at 
least one top-quark decaying leptonically. It requires at 
least two leptons with the same charge, which greatly reduces Standard Model 
backgrounds. The dominant remaining backgrounds are misidentified prompt leptons
and $ttV$ production. The $H\to bb$ mode is analysed in the one and two lepton 
channels. Here, the biggest challenge represents $t\overline t$ production associated 
with heavy flavour quarks ($c$ or $b$) originating mostly from  gluon splitting, which 
is poorly known and needs to be constrained from data simultaneously with the 
signal. Figure$\,$\cite{CMS-ttH} shows representative plots for the 
three $ttH$ channels. The results for the relative signal strengths are: 
$\mu_{ttH(\to\,\gamma\gamma)} = 3.8^{\,+4.5}_{\,-3.6}$, 
$\mu_{ttH(\to\,{\rm leptons})} = 0.6^{\,+1.4}_{\,-1.1}$, and
$\mu_{ttH(\to,bb)} = -2.0\pm1.8$. No significant excess was observed.
\begin{figure}[t]
\centerline{\includegraphics[width=\linewidth]{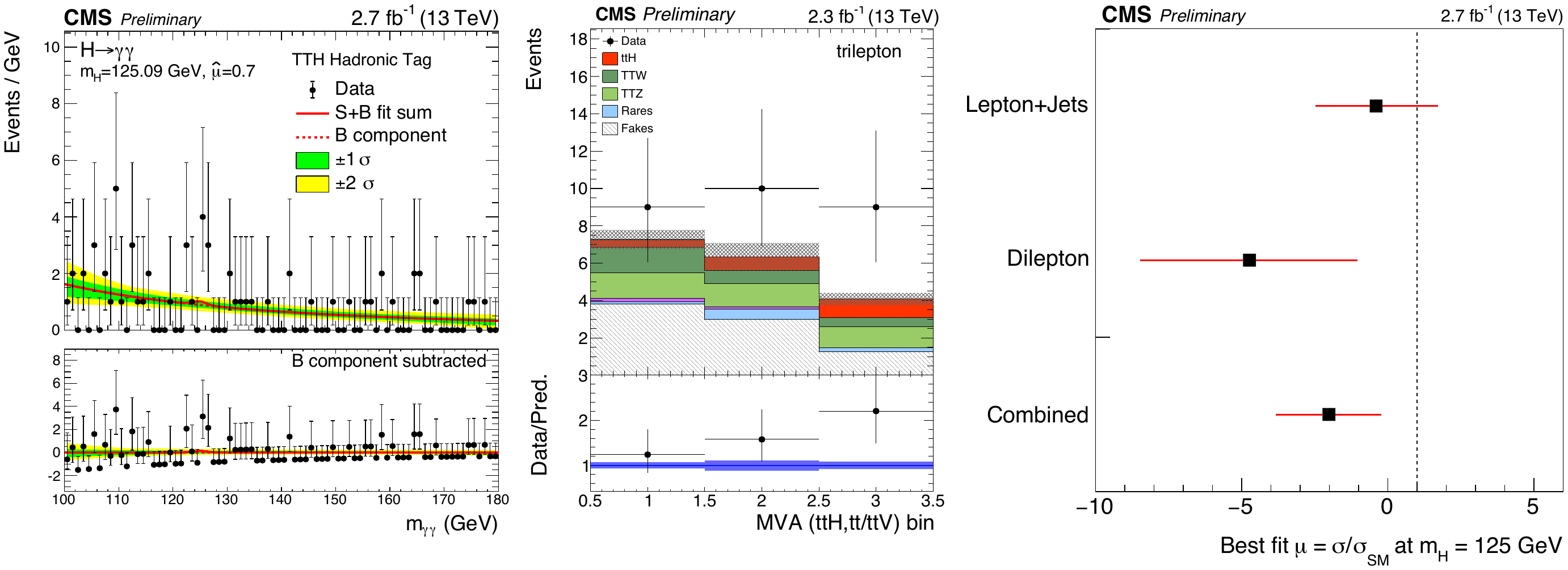}}
\vspace{-0.1cm}
\caption[.]{CMS analyses in the search for $ttH$ production at 13$\;$TeV. The left
  panel shows the diphoton invariant mass in the hadronic channel with at least 
  5 jets and on $b$-tag, the middle panel the BDT output in the trilepton channel 
  of the multilepton search, and the right panel shows the relative signal 
  strengths obtained in the single and dilepton analyses targeting $ttH(\to bb)$,
  and their combination. 
\label{fig:CMS-ttH}}
\end{figure}

\section{The LHC at 13 TeV --- Searches for new physics}

Many of the high mass and higher cross section searches for new physics already
benefit from the 2015 13$\;$TeV data sample to extend their sensitivity. It 
represents thus a fresh start after the negative beyond the Standard Model 
searches from Run-1. The legacy of Run-1 also contained a small number of anomalies
that needed to be verified in the Run-2 data. Only 13$\;$TeV searches are
discussed in the following. 

\subsection{Additional Higgs bosons}

The observed 125$\;$GeV Higgs boson completes the four degrees of freedom of
the Standard Model BEH doublet. Nature may have chosen a more complex scalar
sector of, e.g., two BEH doublets, which extends the scalar sector by four more
Higgs bosons, of which two are neutral (one $CP$-even and one $CP$-odd) and 
the other two are charged. Both ATLAS and CMS have searched$\,$\cite{AllisonMcCarn}
for such additional Higgs bosons 
in Run-1 and Run-2. For $H\to \tau\nu$ ($H / A \to\tau\tau$), the sensitivity of the 
new data exceeds that of Run-1 for masses larger than 250$\;$GeV (700$\;$GeV). 
The search for $A \to Z(\to\ell\ell,\nu\nu) h_{125}(\to bb)$ features improved sensitivity 
beyond about 800$\;$GeV. Searches for $H  \to ZZ(\to \ell\ell qq,\,\nu\nu qq,\,4\ell)$ 
and $WW (\to \ell\nu qq)$ target the $>1\;$TeV mass range where the bosons are 
boosted and their hadronic decays are  reconstructed with jet substructure techniques.
The search for a resonance decaying to $hh_{125} (\to bb\gamma\gamma)$ had a 
small excess in Run-1 at about 300$\;$GeV, which could not yet be excluded 
at 13$\;$TeV. Also performed were searches for resonant and non-resonant 
$hh_{125} (\to bb\tau\tau)$ production. None of these many searches exhibited an 
anomaly so far in the 13$\;$TeV data.

\subsection{New phenomena with high-transverse-momentum jets and leptons}

\begin{figure}[t]
\centerline{\includegraphics[width=\linewidth]{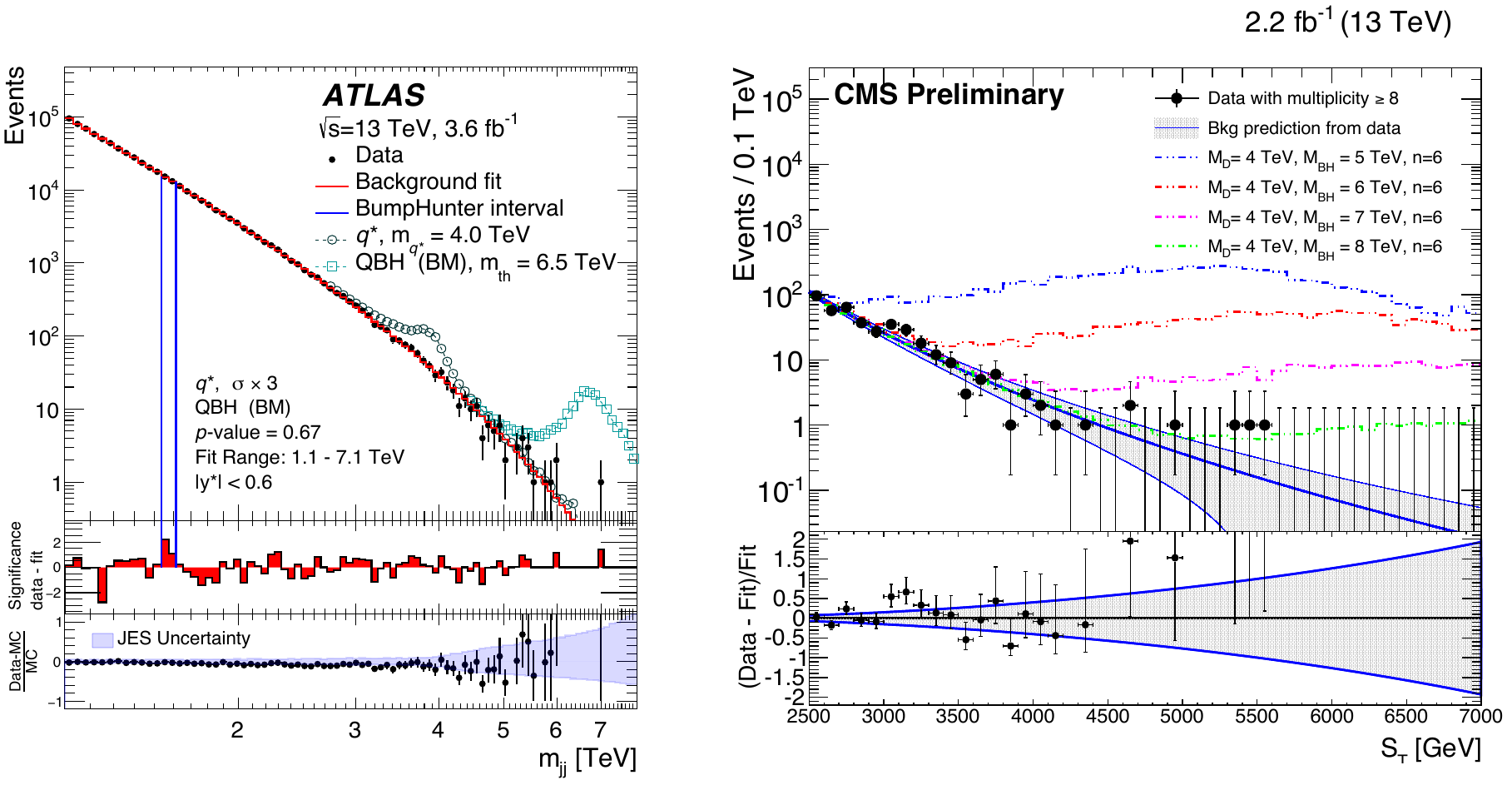}}
\vspace{-0.1cm}
\caption[.]{Dijet invariant mass distribution measured by ATLAS (left) and $S_T$ 
  spectrum in multijet events measured by CMS (right). The data are compared to 
  fits using smoothly falling functions. Also shown are distributions for benchmark
  signal models. 
\label{fig:ATLASCMS-dijet}}
\end{figure}

Among the first searches performed at any significant increase of collision energy
are those for heavy strongly interacting new phenomena.$\,$\cite{ClemensLange} 
Figure~\ref{fig:ATLASCMS-dijet} shows on the left panel the ATLAS dijet 
invariant mass spectrum$\,$\cite{ATLAS-dijetRes,CMS-dijetRes} and on the 
right panel the CMS multijet $S_T$ (defined as the scalar sum of the jet transverse 
momenta) distribution.$\,$\cite{ATLAS-multijet,CMS-multijet} The 
measured spectra are compared to phenomenological fits using smoothly falling 
functions as expected from the QCD continuum. No significant deviation from 
these fits is seen in the data. 
The experiments have also looked at dijet angular distributions versus the 
dijet mass which add further sensitivity to phenomena described by effective contact 
interactions. An ATLAS analysis$\,$\cite{ATLAS-lepjets} looked for new physics
in the $\sum p_T$ spectrum of events with at least one high-$p_T$ lepton and 
jets. ATLAS and CMS have also looked for resonances decaying to heavy-flavour 
quarks,$\,$\cite{ATLAS-hfRes,CMS-hfRes,PieterEveraerts} 
$X\to b\overline b,t\overline t$. None of these searches exhibited an anomaly. 
Second generation scalar lepto-quark$\;$\footnote{Lepto-quarks are hypothetical 
particles carrying both lepton and baryon numbers.} pair production was searched 
for by CMS$\,$\cite{CMS-LQ}  in the 
($\mu q$--$\mu q$) final state excluding such particles below a mass of 1.2$\;$TeV
in case of 100\% branching fraction to $\mu q$. 

\begin{wrapfigure}{R}{0.58\textwidth}
\centering
\vspace{-0.0cm}
\includegraphics[width=0.56\textwidth]{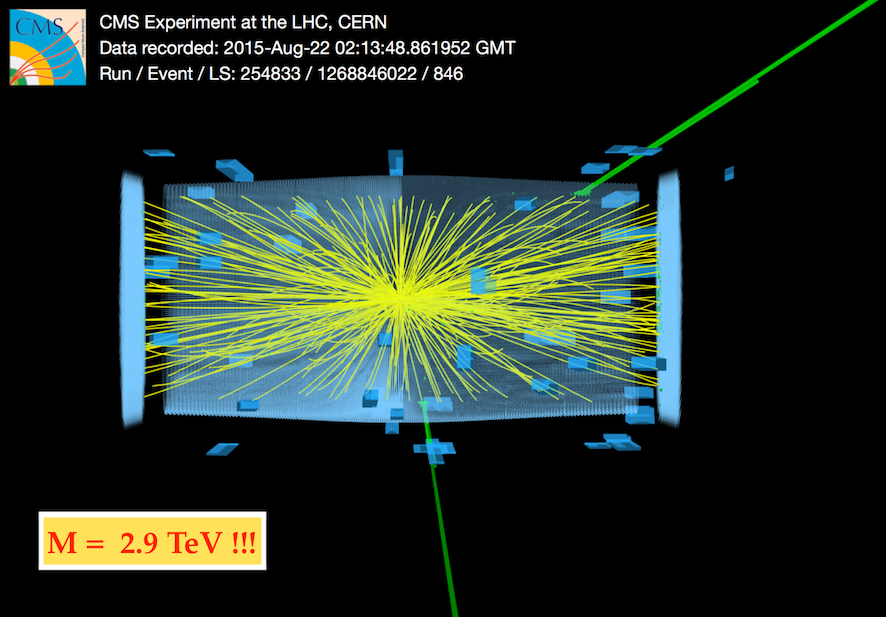}
\vspace{-0.1cm}
\caption[.]{Display of the highest-mass dilepton pair measured by CMS at 
  2.9$\;$TeV mass. Each electron candidate has 1.3$\;$TeV $E_T$, 
  and the two candidates are back-to-back in azimuthal angle. For comparison, 
  the highest-mass Run-1 events found by CMS were at 1.8$\;$TeV ($ee$), 
  1.9$\;$TeV ($\mu\mu$).
  \label{fig:CMS-DileptonDisplay}}
\vspace{-0.1cm}
\end{wrapfigure}
\begin{figure}[t]
\centerline{\includegraphics[width=\linewidth]{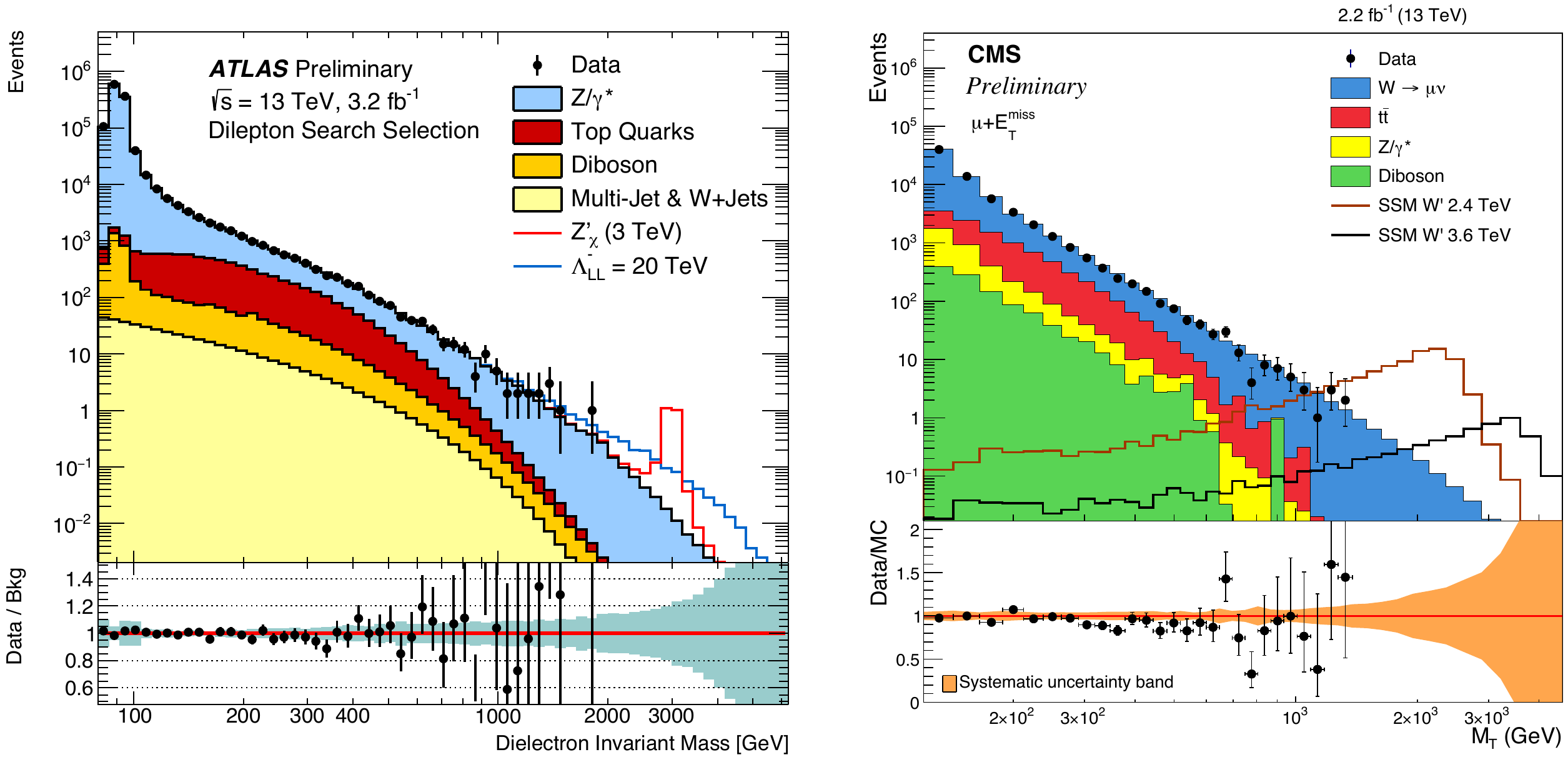}}
\vspace{-0.1cm}
\caption[.]{Dielectron mass distribution measured by ATLAS (left) and 
  the muon--neutrino transverse mass distribution from CMS (right). 
  The data are compared to predictions from (mostly) 
  MC simulation. Also shown are distributions for benchmark
  signal models. 
\label{fig:ATLASCMS-dilepton}}
\vspace{0.7cm}
\centerline{\includegraphics[width=\linewidth]{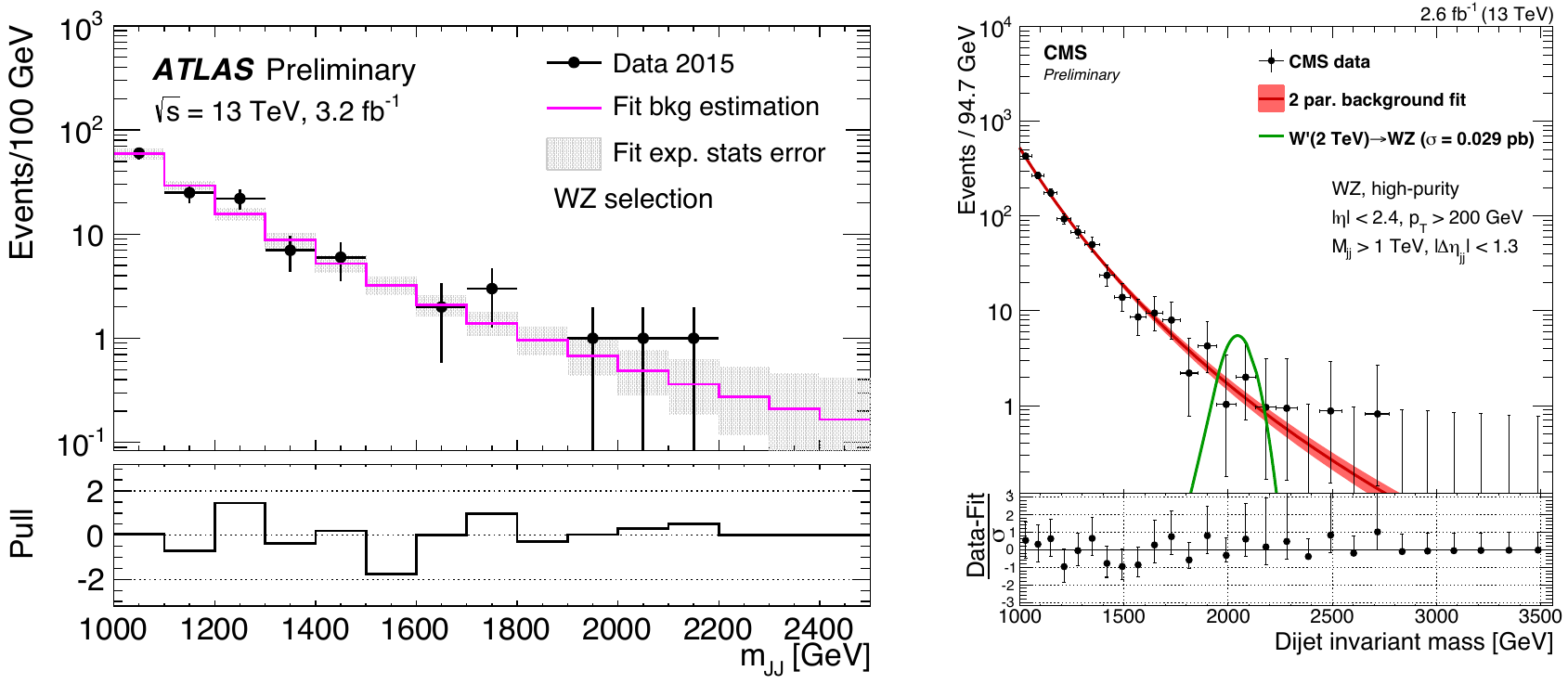}}
\vspace{-0.1cm}
\caption[.]{Dijet invariant mass distributions in the search for 
  a heavy resonance decaying to a $W$ and a $Z$ boson, each of which 
  decays hadronically into a merged jet. The left plot shows ATLAS, the right CMS. 
\label{fig:ATLASCMS-dibosonResonance}}
\vspace{-1.2cm}
\end{figure}

Important canonical searches involve charged and charged--neutral dilepton 
pairs.$\,$\cite{JamesCatmore} Excellent high-mass Drell-Yan background 
modelling is crucial here, which requires to pair detailed differential cross-section 
measurements with these searches. High-$p_T$ muon 
reconstruction challenges the detector alignment in particular for the complex
ATLAS muon spectrometer structure.$\,$\cite{ATLAS-muonperf}
Figure~\ref{fig:ATLASCMS-dilepton}
shows the ATLAS dielectron mass distribution$\,$\cite{ATLAS-dilepRes}
(left panel) and the CMS$\,$\cite{CMS-dilepRes} 
transverse mass between the muon and the missing transverse momentum
(measuring the neutrino from the transverse balancing of the event, right
panel). Figure~\ref{fig:CMS-DileptonDisplay} shows a display of the 
highest-mass dilepton events measured by CMS in the 2015 data. 
No anomaly was found. Limits for the traditional sequential Standard Model 
$Z^\prime$ ($W^\prime$) benchmark are set at 3.4$\;$TeV (4.4$\;$TeV) (for comparison: 
2.9$\;$TeV (3.3$\;$TeV) at 8$\;$TeV). ATLAS and CMS$\;$
also looked into high-mass $e\mu$ (lepton flavour violation) 
production.$\,$\cite{ATLAS-emuRes,CMS-emuRes} 
The main background here are dilepton top-antitop events. Again, no anomaly was seen.

\vfill\pagebreak

{\footnotesize\mbox{}}
\subsection{Diboson resonances ($VV$, $Vh$, $hh$)}

Also  13$\;$TeV searches for a diboson resonance were promptly released by 
ATLAS and CMS.$\,$\cite{MaxBellomo} The high $p_T$ of the bosons boosts 
the hadronic decay products so that jets are merged and 
analysed using substructure techniques to suppress strong-interaction 
continuum background. An excess of events was seen at 8$\;$TeV in the fully hadronic 
$X\to VV$ ($V=W,Z$) resonance searches$\,$\cite{ATLAS-WZRes8,CMS-WZRes8}
around 2$\;$TeV (globally 2.5$\sigma$ for ATLAS in the $WZ$ mode), 
which was however not observed in the other weak gauge boson decay 
channels of similar sensitivity. Figure~\ref{fig:ATLASCMS-dibosonResonance} shows 
the 13$\;$TeV dijet mass distributions after substructure analysis for 
ATLAS$\,$\cite{ATLAS-VVRes13} (left) and CMS$\,$\cite{CMS-VVRes13} (right).
There is no hint of an excess around 2$\;$TeV, but the current statistical 
precision is not large enough$\,$\cite{MaruzioPierini} 
to fully exclude the anomaly seen at 8$\;$TeV. Other diboson resonance searches
also do not exhibit discrepancies from the Standard Model expectation. 

\vfill\pagebreak
\subsection{Supersymmetry}

\begin{figure}[t]
\centerline{\includegraphics[width=\linewidth]{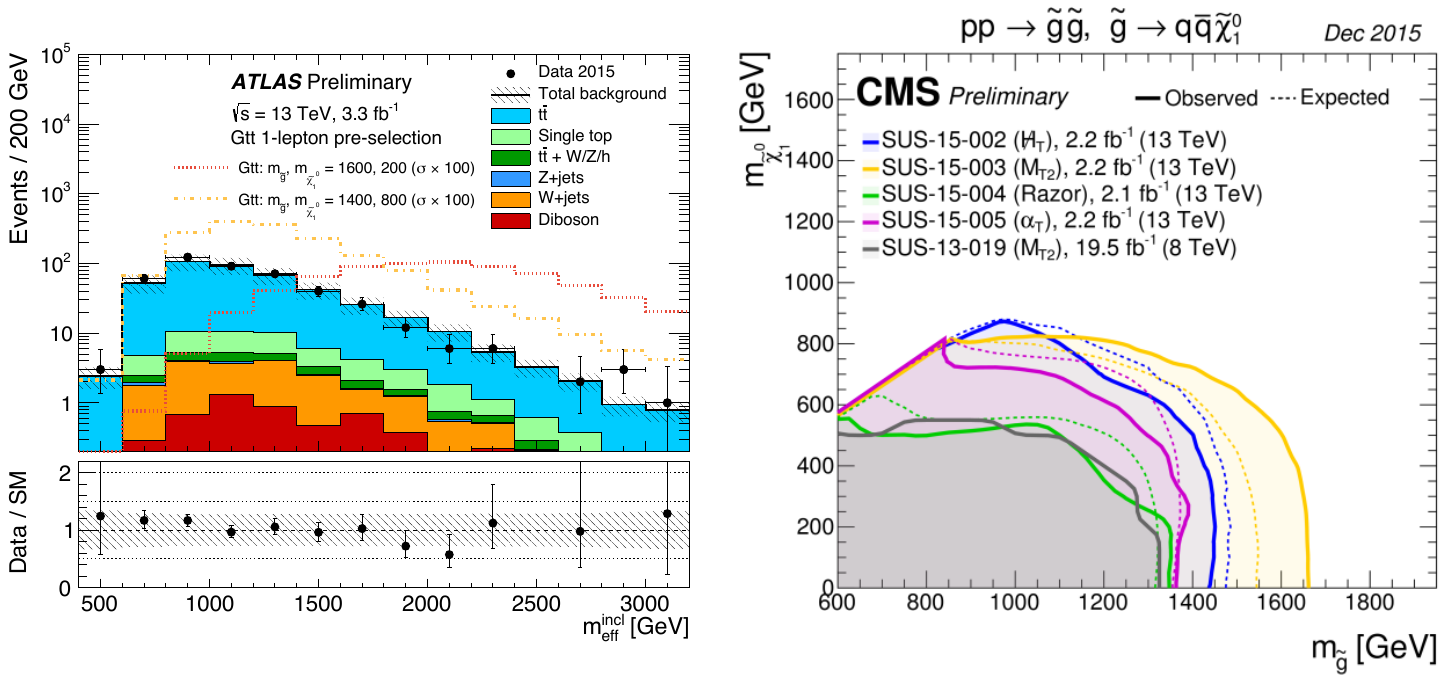}}
\vspace{-0.0cm}
\caption[.]{{\bf Left:} effective mass distribution (scalar sum of selected jet
  transverse momenta and 
  the event's missing transverse momentum) measured in the search targeting gluino 
  pair production and decay through intermediate top squarks. Data and Standard
  Model background estimations are shown together with representative signal 
  benchmark distributions. {\bf Right:} lower limits in the gluino--lightest neutralino
  mass plane for the four-jet + MET 
  simplified gluino production models used to interpret the 
  jets plus missing transverse momentum searches. 
\label{fig:ATLASCMS-SUSYstrong}}
\end{figure}
\begin{figure}[b]
\centerline{\includegraphics[width=0.9\linewidth]{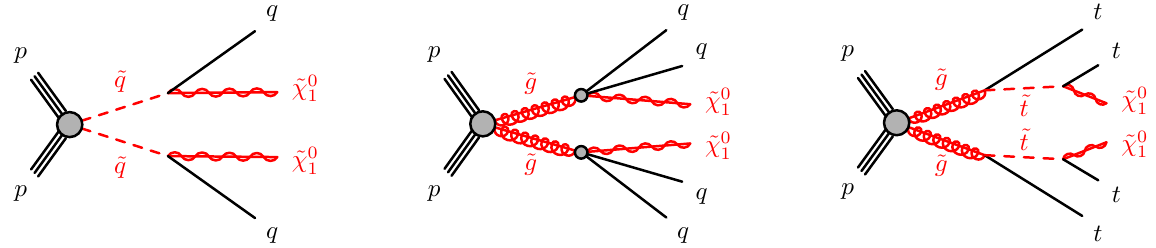}}
\vspace{-0.0cm}
\caption[.]{Simplified models for supersymmetric 
  squark and gluino pair production at the LHC. The right-hand process
  can occur if the top squark is lighter than the first and second generation
  squarks as is possible in models with large squark mixing. 
  \label{fig:StrongSUSYFeyn}}
\end{figure}
\begin{wrapfigure}{R}{0.45\textwidth}
\centering
\vspace{-0.5cm}
\includegraphics[width=0.43\textwidth]{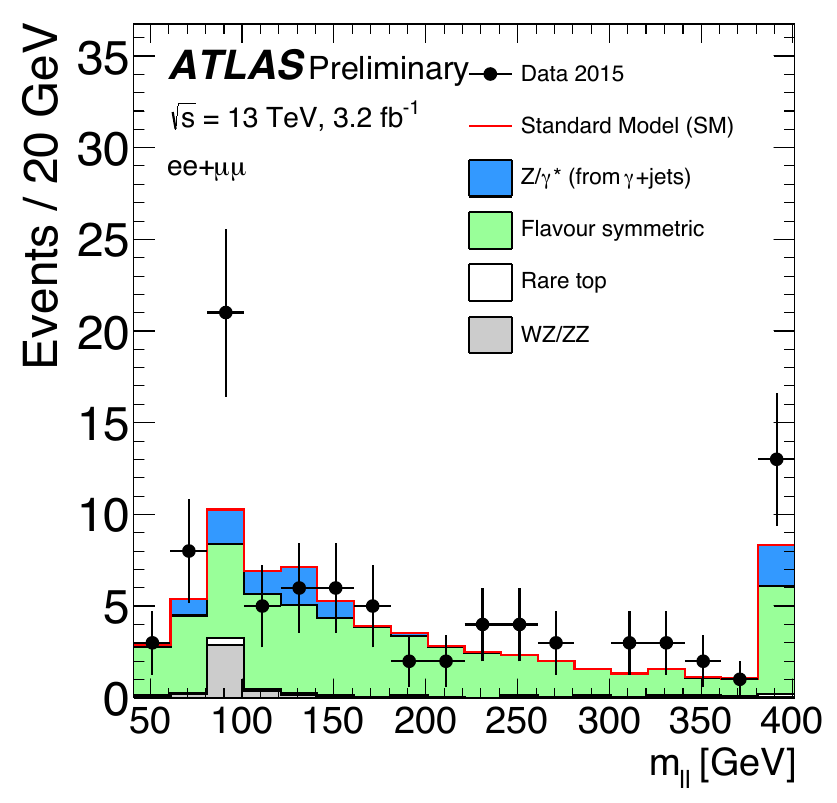}
\vspace{-0.2cm}
\caption[.]{Dilepton mass distribution observed by ATLAS in a 13$\;$TeV 
  a $Z$ plus jets plus missing transverse momentum search in the dilepton 
  final state compared to the Standard Model expectation. 
  \label{fig:ATLAS-SUSYZMET}}
\vspace{-0.1cm}
\end{wrapfigure}
Both ATLAS and CMS have updated their most sensitive searches for
high-cross-section strong supersymmetric squark and gluino production using the 
13$\;$TeV data sample$\,$\cite{ChrisYoung} (c.f. Fig.~\ref{fig:StrongSUSYFeyn} 
for representative simplified models). Although the jets plus missing transverse 
momentum searches benefit from improved background modelling, owing to 
more accurate MC generators and improved tuning, discrepancies in the extreme 
phase space regions of these searches remain and are corrected using scale factors 
determined in data control regions. A total of seven early hadronic 13$\;$TeV analyses 
were performed by ATLAS and CMS in time of the conference selecting up to ten jets
and up to three $b$-tagged jets.$\,$\cite{ATLAS-SUSY-jetsmet,CMS-SUSY-jetsmet} 
None of these searches observed a significant excess of events in the 
signal regions. Representative distributions and limits are shown in 
Fig.~\ref{fig:ATLASCMS-SUSYstrong}. In the simplified models used to interpret
the searches gluino masses up to 1.7$\;$TeV could be excluded in case of 
light or moderate-mass neutralinos, exceeding the Run-1 limits by up to 300$\;$GeV. 
 
Inclusive supersymmetry searches also involved events with leptons, where single
lepton, dilepton (on and off the $Z$ mass resonance) and same-charge dilepton signatures 
with additional jets and missing transverse momentum were 
studied.$\,$\cite{HenningKirschenmann} Such signatures can occur for example 
when gluinos (assumed to be Majorana fermions) decay via intermediate charginos 
or higher neutralino states or  via top squarks. These searches benefit from 
reduced background levels than in the fully hadronic cases, but often also have 
lower signal efficiencies due to small leptonic branching fractions. A total of eight 
searches were 
presented$\,$\cite{ATLAS-SUSY-lepjetsmet,ATLAS-SUSY-ZMET,CMS-SUSY-lepjetsmet,CMS-SUSY-ZMET} with 
only one (non-significant) anomaly seen. The ATLAS search in $Z\to\ell\ell$
plus jets plus missing transverse momentum final states$\,$\cite{ATLAS-SUSY-ZMET}
exhibits a modest excess 
of $2.2\sigma$ in a signal region that had already  shown a $3.0\sigma$ excess in the
corresponding 8$\;$TeV ATLAS analysis.$\,$\cite{ATLAS-SUSY-ZMET8} 
Figure~\ref{fig:ATLAS-SUSYZMET} shows
the observed dilepton mass distribution in data compared to the Standard Model 
expectation. This excess was not confirmed by 
CMS at 8$\;$TeV,$\,$\cite{CMS-SUSY-ZMET8} 
neither at 13$\;$TeV.$\,$\cite{CMS-SUSY-ZMET} 
A small excess seen by CMS at 8$\;$TeV  
(2.6$\sigma$) in the off-$Z$ dilepton mass region$\,$\cite{CMS-SUSY-ZMET8} 
was not confirmed in the 13$\;$TeV data.

Direct production of pairs of third generation bottom and top squarks was also 
already studied by both ATLAS and 
CMS.$\,$\cite{ATLAS-SUSY-3rd,CMS-SUSY-3rd,PieterEveraerts} 
The sensitivity of these searches does only moderately exceed that of Run-1 
due to the relatively low cross sections
of third generation scalar squark production (top-squark pair production has an about
six times lower cross section as the corresponding top-quark pair production at equal 
mass). A total of ten 13$\;$TeV analyses targeted this production and also 
that of vector-like quark production. Vector-like quarks are hypothetical fermions 
that transform as triplets under colour and that have left- and right-handed 
components with same colour and electroweak quantum numbers. For these
searches, signatures for pair or single production and decays to $bW$, $tZ$ and 
$tH$ were studied.$\,$\cite{ATLAS-VLQ,CMS-VLQ} Also considered were 
exotic $X_{5/3} \to tW$ particles. No anomaly was seen in these searches.

ATLAS searched for top squark pair production with $R$-parity violating decays
governed by non-zero $\lambda^{\prime\prime}_{323}$ 
couplings to a pair of $bs$ quarks$\,$\cite{ATLAS-RPV}
that leads to a four-jet final state with no missing transverse momentum. 
Employing a hadronic trigger and a data-driven background 
determination, top squark masses between 250$\;$GeV and 345$\;$GeV 
were excluded at 95\% CL.

\begin{figure}[t]
\centerline{\includegraphics[width=\linewidth]{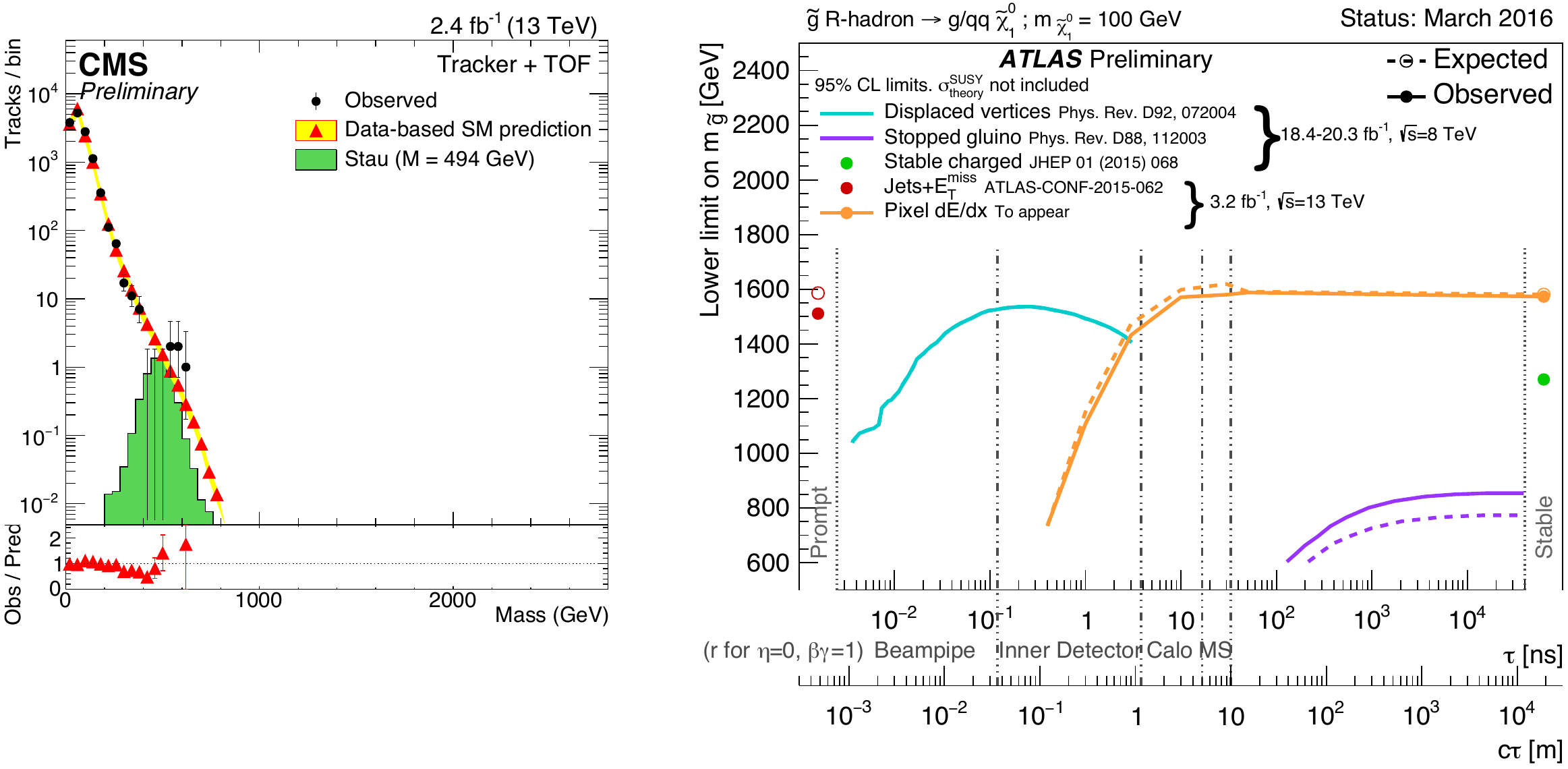}}
\vspace{-0.0cm}
\caption[.]{{\bf Left:} distribution of the reconstructed particle mass in 
  CMS compared to the background expectation determined from data. 
  Also shown is the expected distribution of a long-lived tau slepton signal benchmark. 
  {\bf Right:} limits on the gluino mass versus lifetime obtained by ATLAS. 
  The dots on the left (right) of the plot indicate the limits obtained on promptly 
  decaying (stable) gluinos. Varying searches cover the full lifetime spectrum.
  \label{fig:ATLASCMS-SUSYLLP}}
\end{figure}
The production of long-lived massive particles as it can occur due to large virtuality, 
low coupling and/or mass degeneracy in a cascade decay, e.g., via the 
scale-suppressed colour triplet 
scalar from unnaturalness presented at this conference,$\,$\cite{TonyGherghetta} 
is an important part of the LHC search 
programme.$\,$\cite{RevitalKopeliansky} ATLAS and CMS presented searches for heavy 
long-lived supersymmetric particles at 13$\;$TeV using measurements of the 
specific ionisation loss in the tracking detectors and time-of-flight in the calorimeters
and muon systems.$\,$\cite{ATLAS-LLP,CMS-LLP} 
Figure~\ref{fig:ATLASCMS-SUSYLLP} shows on the left the 
distribution of the reconstructed particle mass in CMS compared to the background
expectation determined from data together with the distribution of a signal 
benchmark. The right plot shows limits on the gluino mass versus its 
lifetime obtained by ATLAS from several analyses covering the full lifetime 
spectrum. 

\subsection{Dark matter production}

\begin{figure}[p]
\centerline{\includegraphics[width=\linewidth]{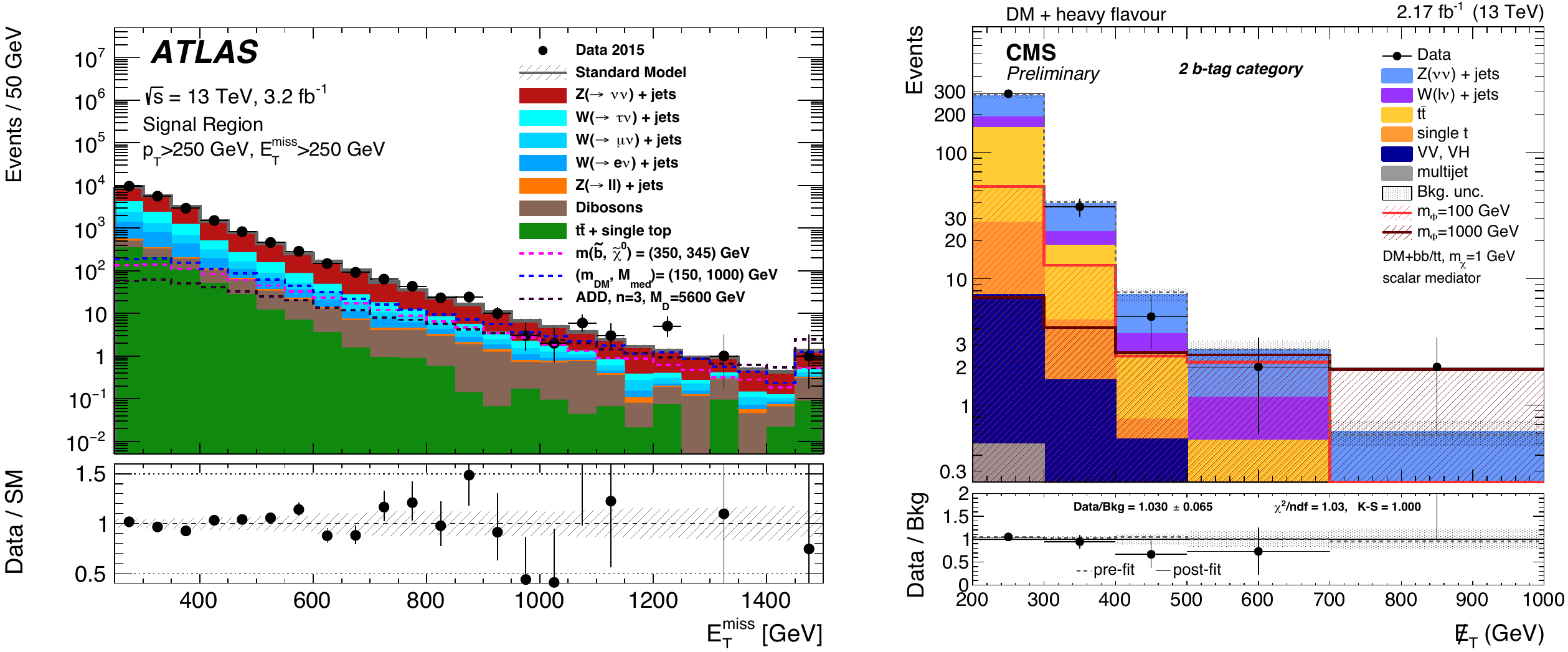}}
\vspace{-0.0cm}
\caption[.]{{\bf Left:} distribution of missing transverse momentum measured
  by ATLAS at 13$\;$TeV in a ``mono-jet'' search.The dominant backgrounds
  stem from leptonic $Z$ and $W$ plus jets events. Also shown are distributions 
  for new physics  benchmark models. 
  {\bf Right:} distribution of missing transverse momentum measured
  by CMS at 13$\;$TeV in a search for heavy-flavour quarks produced in 
  association with large missing transverse momentum.
  \label{fig:ATLASCMS-DM}}
\vspace{1.5cm}
\centerline{\includegraphics[width=0.7\linewidth]{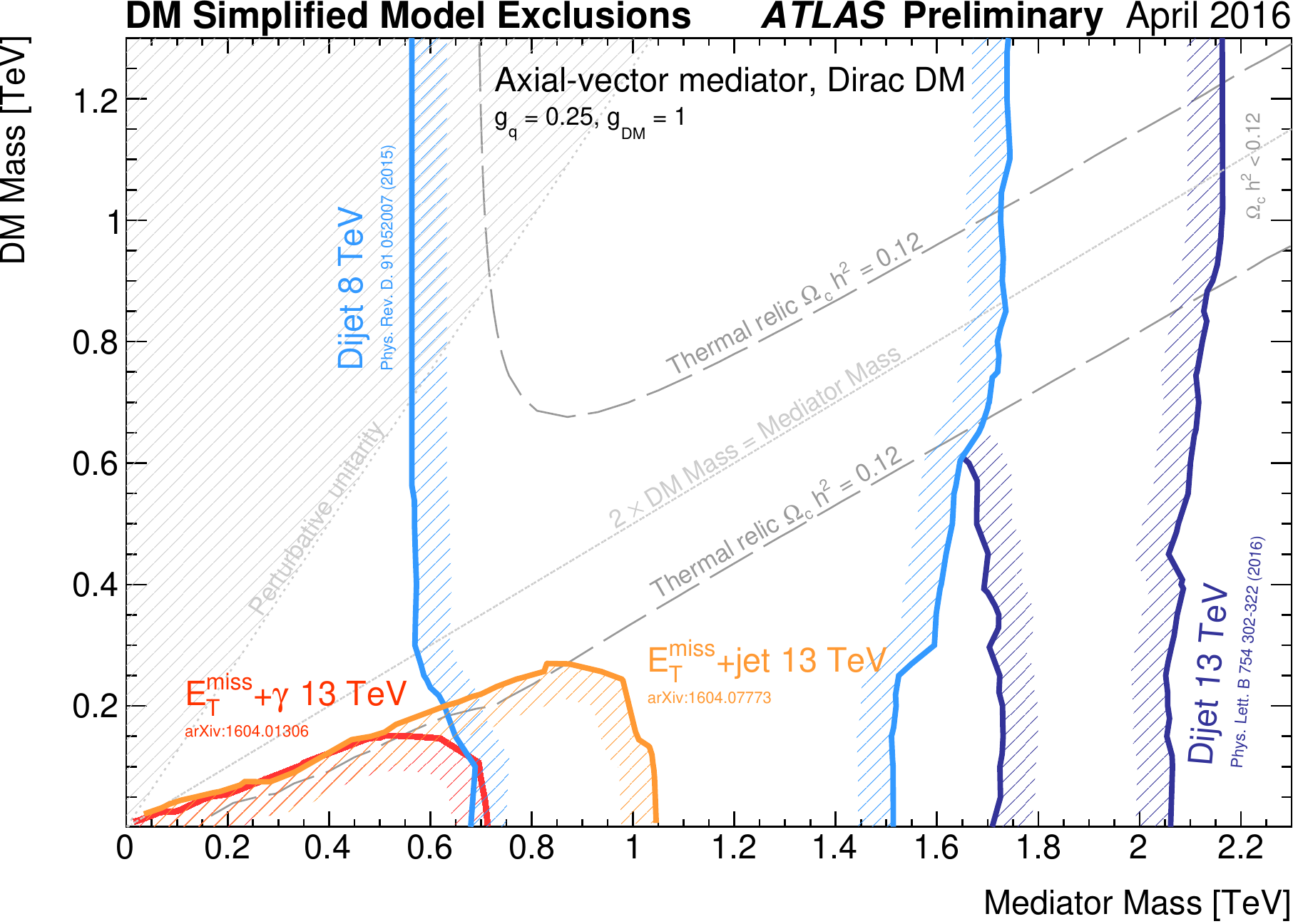}}
\vspace{0.1cm}
\caption[.]{Regions in a dark matter (DM) mass--mediator mass plane excluded at 
  95\% CL by a selection of ATLAS DM searches, for one possible 
  interaction between the Standard Model and DM, the lepto-phobic 
  axial-vector mediator described in Ref.$\,$\cite{LHC-DMforum}
  Dashed curves labelled ``thermal relic'' indicate combinations of DM 
  and mediator mass that are consistent with the cosmological 
  DM density  and a standard thermal history. Between the 
  two curves, annihilation processes described by the simplified model 
  deplete the relic density. A dotted curve indicates the kinematic 
  threshold where the mediator can decay on-shell into DM. 
  Points in the plane where the model is in tension with  perturbative 
  unitary considerations are indicated by the shaded triangle at the 
  upper left. The exclusion regions, relic density contours, and unitarity 
  curve are not applicable to other choices of coupling values or model. 
  See Ref.$\,$\cite{ATLAS-DM-plot} for more information.
  \label{fig:ATLAS_DarkMatter_Summary}}
\end{figure}
If dark matter particles (assumed to be weakly interacting and massive, WIMPs) 
interact with quarks and/or gluons they can be directly pair produced in the proton
collisions at the LHC.$\,$\cite{MatteoCremonesi,LHC-DMforum} 
Since the WIMPs remain undetected, to trigger the events a large boost via initial 
state jet or photon radiation (or other recoiling particles) is needed leading to 
large missing transverse momentum (MET) from the recoiling WIMP pair. 
The final state signature depends on the unknown details of the proton--WIMP 
coupling requiring a large range of ``$X$ + MET'' searches. The most 
prominent and among the most sensitive of these is the so-called ``mono-jet'' 
search, which extends to a couple of high-$p_T$ jets recoiling against the MET. 
Large irreducible Standard Model backgrounds in this channel stem from 
$Z(\to\nu\nu)+{\rm jets}$  and $W(\to\ell\nu)+{\rm jets}$ events, 
where in the latter case the charged lepton is either undetected or a hadronically 
decaying tau lepton. These backgrounds are determined in data control 
regions requiring accurate input from theory to transfer the measured normalisation 
scale factors to the signal regions. Several 13$\;$TeV results were already 
released by ATLAS and CMS: jets + MET,$\,$\cite{ATLAS-monojet,CMS-monojet}
photon + MET,$\,$\cite{ATLAS-monophoton}
$Z/W$ + MET,$\,$\cite{ATLAS-V+MET,CMS-V+MET} and 
$bb/tt$ + MET.$\,$\cite{CMS-HF+MET} None has so far shown an 
anomaly.

Figure~\ref{fig:ATLASCMS-DM} shows missing transverse momentum 
distributions measured by ATLAS and CMS in jets + MET and $bb/tt$ + MET
searches, respectively. Figure~\ref{fig:ATLAS_DarkMatter_Summary} shows for 
a specific benchmark model (see figure caption) ATLAS exclusion regions in the 
DM versus the model's mediator mass plane as obtained from the jets + MET
and photon + MET analyses as well as from the dijet resonance search. These 
searches have complementary sensitivity.

\subsection{Diphoton resonance}

Searches for a new resonance in the diphoton mass spectrum were performed 
by ATLAS$\,$\cite{ATLAS-diphotonRes8-exo,ATLAS-diphotonRes8-higgs}
and CMS$\,$\cite{CMS-diphotonRes8} in Run-1 looking for a low to medium 
mass scalar resonance, or a medium to high mass spin-two resonance 
motivated by strong gravity models. Diphoton spectra were also analysed 
in view of high-mass tail anomalies due to new nonresonant phenomena.
Searches involving at least three photons were used during Run-1 to look for 
new physics in Higgs or putative $Z^\prime$ decays.$\,$\cite{ATLAS-multiphoton} 

Preliminary analyses of the 13$\;$TeV diphoton data presented at the 
2015 end-of-year seminars showed enhancements at around 750$\;$GeV invariant
diphoton mass in both ATLAS and CMS. This conference featured updated analyses, 
still preliminary, of the 2015 and for ATLAS also the $8\;$TeV Run-1 
datasets.$\,$\cite{ATLAS-diphotonRes,CMS-diphotonRes,MarcoDelmastro,PasqualeMusella} \\[0.0cm]

{\bf ATLAS}$\,$\cite{ATLAS-diphotonRes} 
performed dedicated searches for a spin-zero and a spin-two diphoton resonance. 
The main difference between these searches are the photon acceptance requirements: 
for the spin-zero case these are $E_T(\gamma_1) > 0.4\cdot m_{\gamma\gamma}$, 
$E_T(\gamma_2) > 0.3\cdot m_{\gamma\gamma}$, where the indices 1, 2 indicate the 
leading and subleading photon. In the spin-two case, the fixed 
requirement $E_T(\gamma_{1/2}) > 55\;$GeV is applied. The differences are 
motivated by the photon decay angle behaviour in the centre-of-mass
of the resonance, resulting in more 
low-$E_T$ forward photons in the spin-two case. Photons are tightly identified 
and isolated giving a typical photon purity of about 94\%. The background 
modelling is empirical in the spin-zero analysis, and theoretical in the spin-two 
case for the dominant irreducible diphoton contribution (the small misreconstructed 
photon background is determined from data and extrapolated to high mass). This 
choice is motivated by the high mass reach of the spin-two search. 

\begin{figure}[t]
\centerline{
  \includegraphics[width=0.5\linewidth]{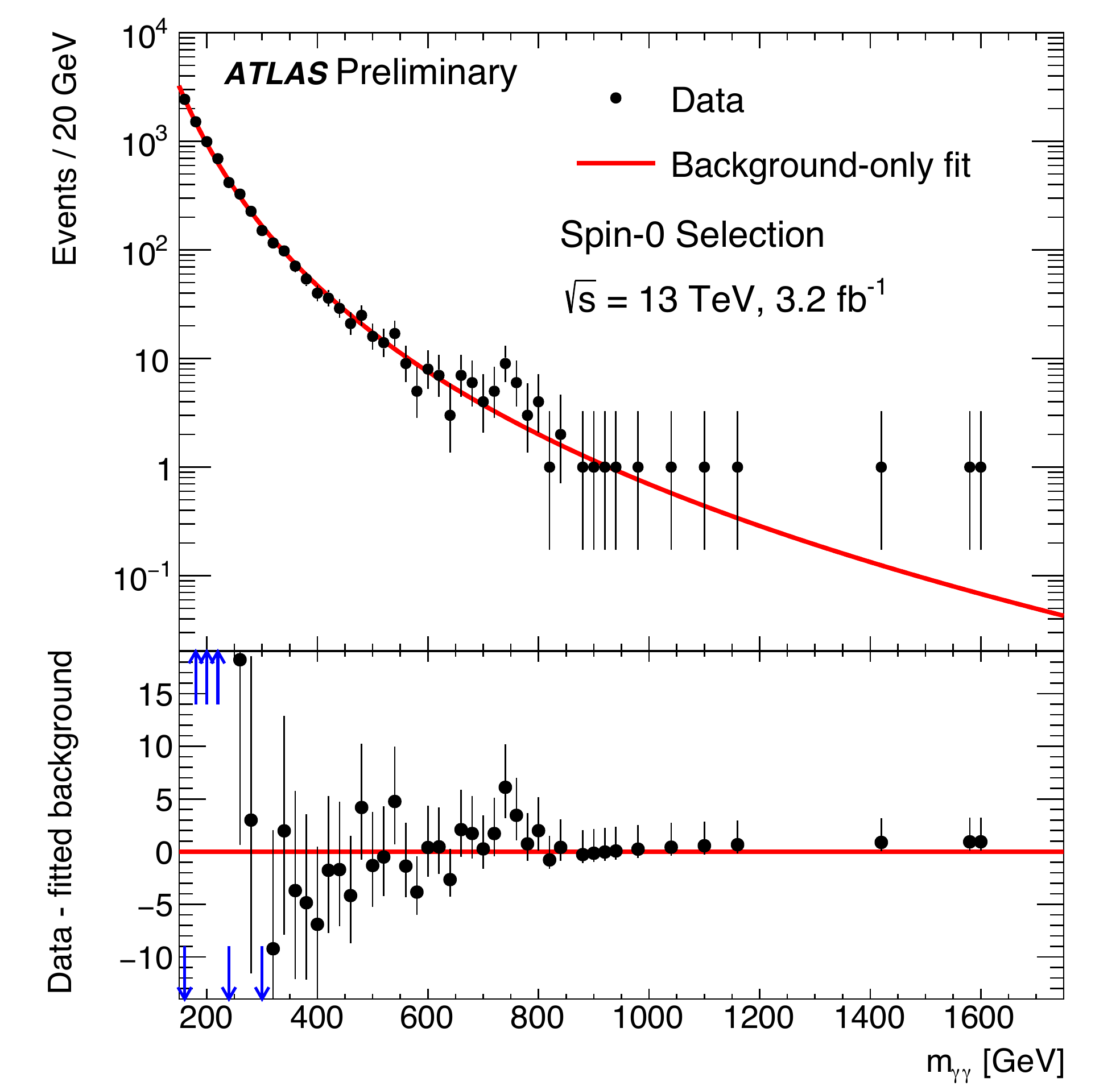}
  \includegraphics[width=0.5\linewidth]{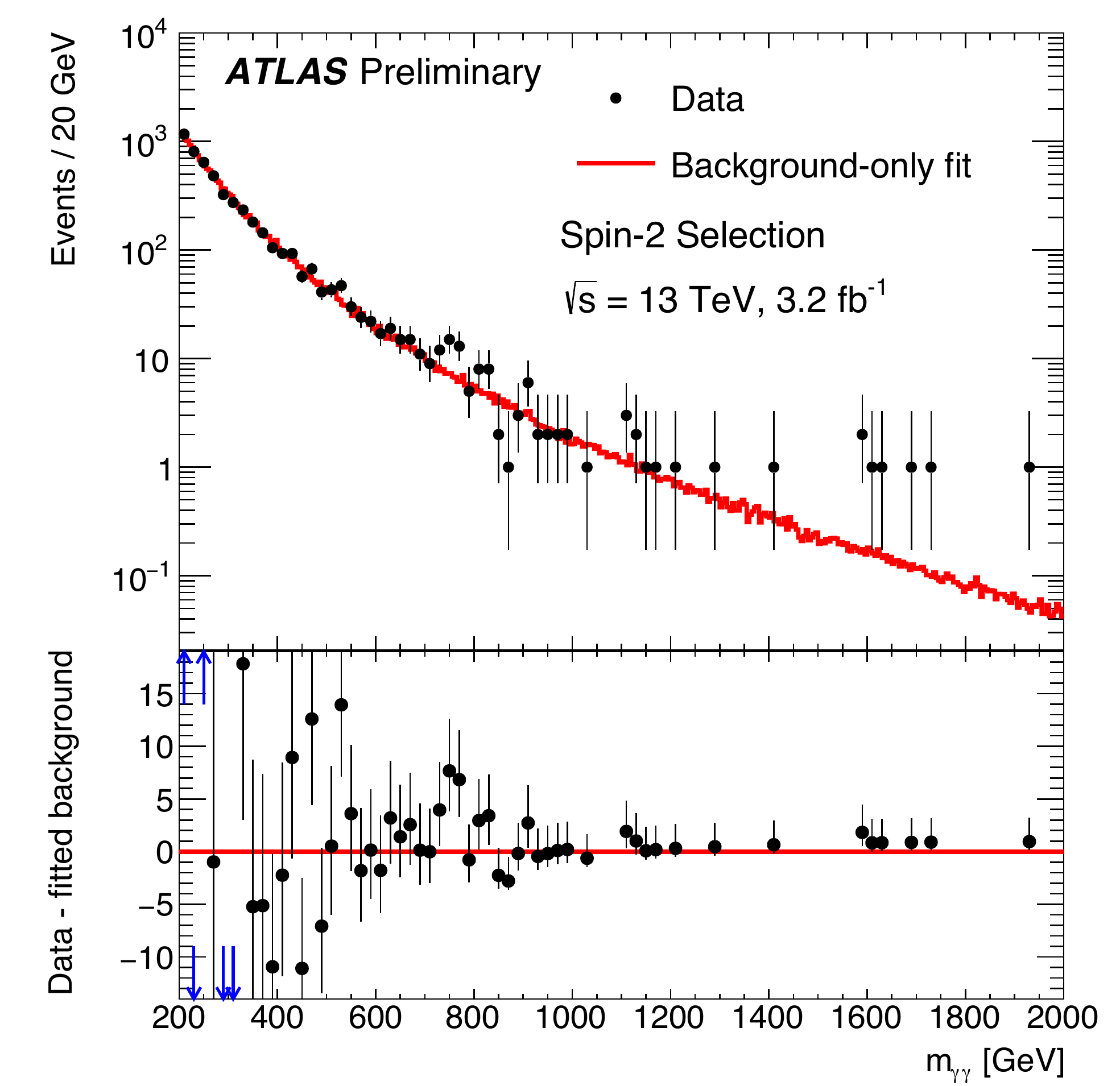}
}
\vspace{0.5cm}
\centerline{
  \includegraphics[width=0.5\linewidth]{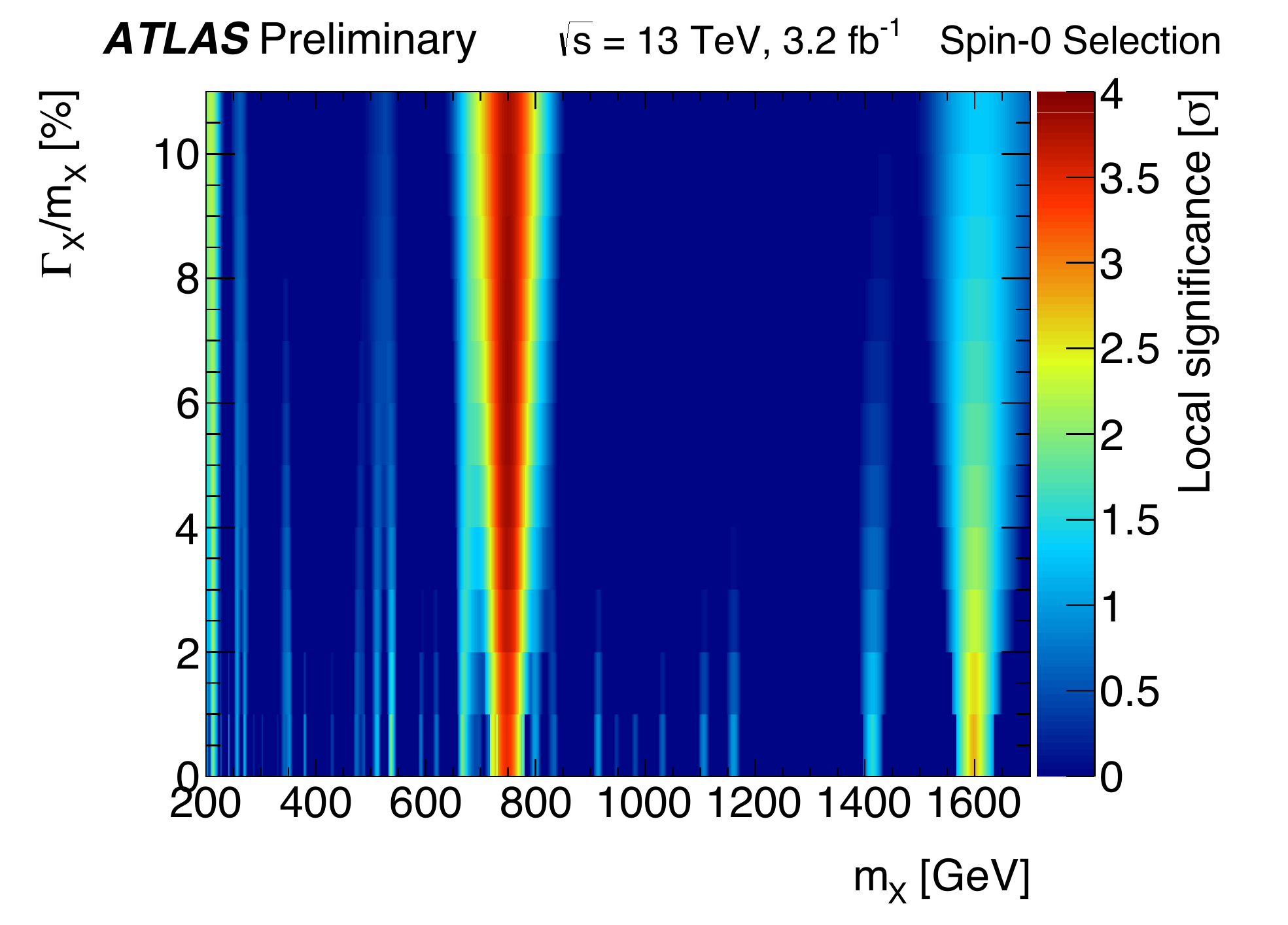}
  \includegraphics[width=0.5\linewidth]{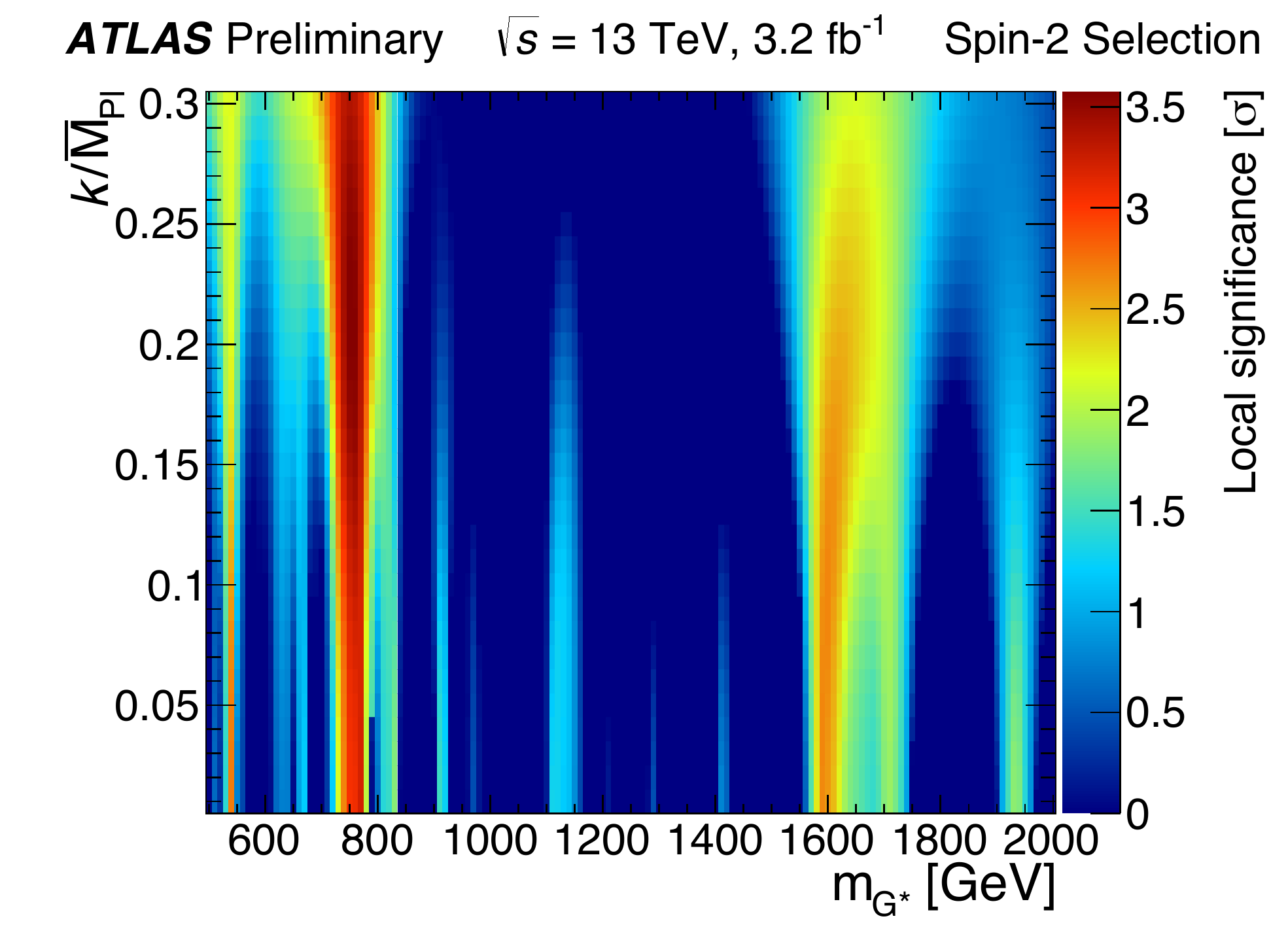}
}
\vspace{-0.0cm}
\caption[.]{Top panels: diphoton invariant mass spectra observed with the 
  spin-zero (left) and spin-two (right) selections compared to background 
  estimations. The total numbers of selected
  events entering the plots are 2878 (5066) for the spin-zero (spin-two) cases. 
  The bottom panels show the   local significances obtained when scanning a 
  signal plus background model with varying signal mass and width. Left for 
  the spin-zero and right for the spin-two selections.
  \label{fig:ATLAS-diphotonRes}}
\end{figure}

The top panels of Fig.~\ref{fig:ATLAS-diphotonRes} 
show the diphoton invariant mass spectra observed with the spin-zero (left) and 
spin-two (right) selections together with the background estimations. 
The bottom panels show the local significances obtained 
when scanning a signal plus background model with varying signal mass and width. 
The lowest compatibility of the data with the background-only hypothesis is found 
for the spin-zero case at around 750$\;$GeV and a signal width of about 45$\;$GeV 
(6\% relative to the mass). The p-value of that point is found to correspond to
a local significance of $Z=3.9\sigma$. Taking into account the statistical trials 
factor$\:$\footnote{We emphasise that the trials factor parametrising the 
statistical ``look-elsewhere effect'' is a reality that must be taken 
into account when interpreting these results. The main results put forward by the 
experiments are therefore the global significance numbers.}
inherent in the signal mass and width scan reduces the significance 
to globally $2.0\sigma$. The corresponding values for the spin-two case are: largest
local significance at around 750$\;$GeV and relative width of 7\%, local / global
significances of $3.6\sigma$ / $1.8\sigma$. ATLAS compared the event properties 
in the excess interval (700--800$\;$GeV) with those in the upper and lower sidebands
and did not find statistically significant differences. ATLAS also updated
its 8$\;$TeV diphoton resonance searches to the latest 
photon calibration and 13$\;$TeV analysis methods, finding a modest 1.9$\sigma$ 
excess at 750$\;$GeV mass and assuming 6\% signal width in the spin-zero analysis,
and no excess in the spin-two analysis. Assuming the putative resonance to be produced
by gluon fusion the production cross section is expected to increase by a factor 
of 4.7 between 8$\;$TeV and 13$\;$TeV. The compatibility between the 
observations in the two datasets is then estimated to be at the 1.2$\sigma$ 
level for the spin-zero analysis. 
Would the resonance be produced by light quark--antiquark annihilation,
the cross-section scale factor would reduce$\;$\footnote{Annihilation of heavy 
quarks would lead to a larger expected 13$\;$TeV to 8$\;$TeV cross-section 
ratio of 5.1 for $c\overline c$ and 5.4 for $b\overline b$.}  to 2.7 leading to a 
compatibility at the 2.1$\sigma$ level between the two datasets. The corresponding 
numbers for the spin-two analysis and production via gluon fusion / light 
quark--antiquark annihilation are 2.7$\sigma$ / 3.3$\sigma$.

ATLAS also searched for a resonant signal in the $Z\gamma$ final 
state$\,$\cite{ATLAS-ZgamRes,AllisonMcCarn} using leptonic and hadronic $Z$ 
boson decays and empirical background fits. No significant excess was seen 
in either spectrum. \\[0.0cm]

\begin{figure}[p]
\centerline{\includegraphics[width=0.8\linewidth]{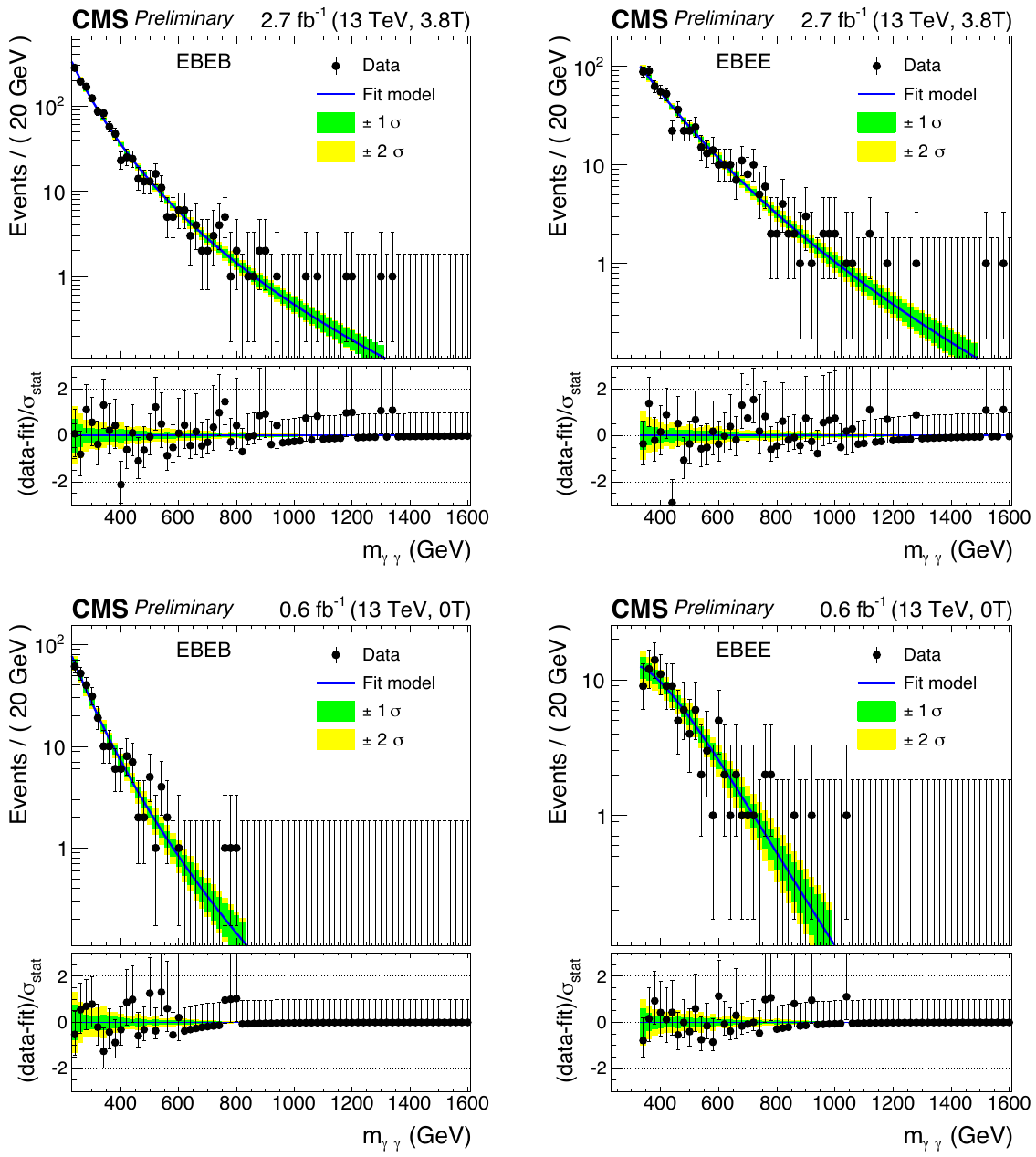}}
\vspace{-0.0cm}
\caption[.]{Diphoton invariant mass distributions measured by CMS in the 
  solenoid-on (top panels) and solenoid-off datasets (bottom panels). 
  The left (right) panels show the barrel-barrel (barrel-endcap) categories. 
  Also shown are the background predictions obtained from the fits to data.   
  \label{fig:CMS-diphotonRes}}
\vspace{1cm}
\centerline{\includegraphics[width=0.9\linewidth]{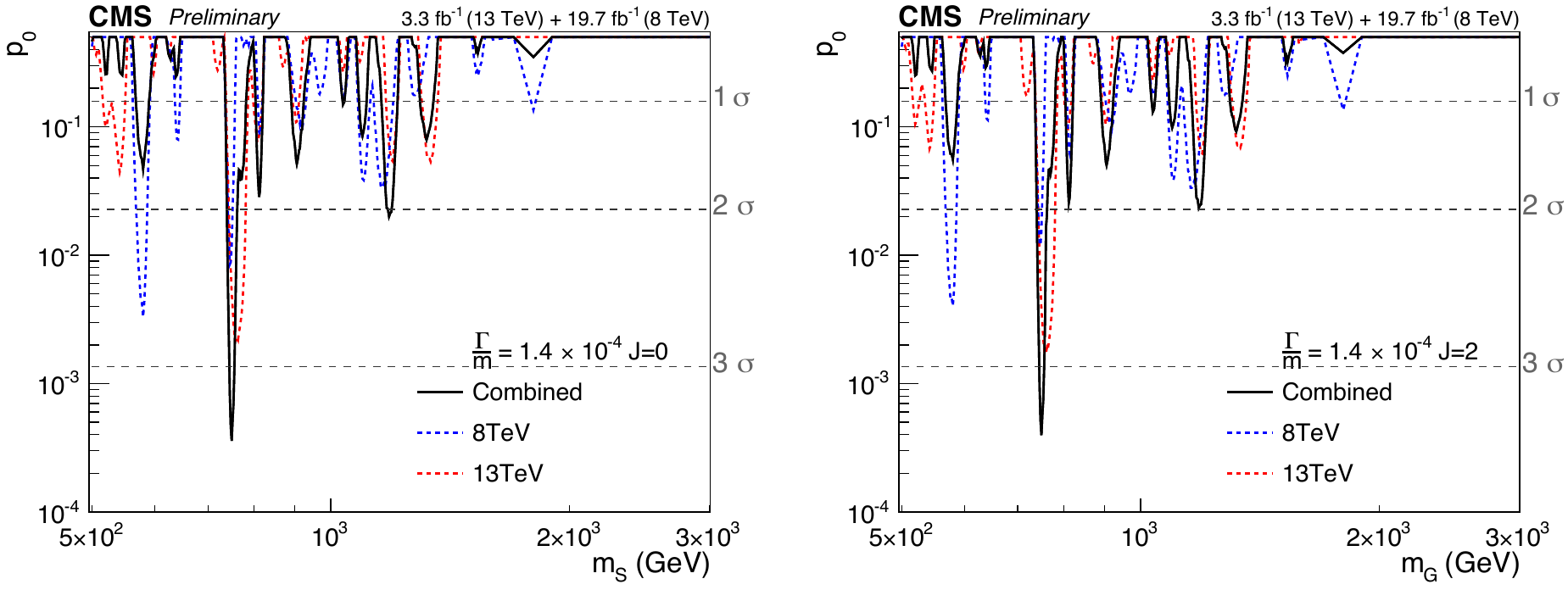}}
\vspace{-0.0cm}
\caption[.]{Local p-value versus mass for a narrow-width signal model as found 
  by CMS for the spin-zero (left) and spin-zero (right) hypotheses. The red (blue) 
  dotted lines give the individual 13$\;$TeV (8$\;$TeV) results and the black solid
  lines their statistical combinations. 
  \label{fig:CMS-diphotonResSig}}
\end{figure}
{\bf CMS}$\,$\cite{CMS-diphotonRes}  
searched in an agnostic way for a spin-zero or spin-two resonance. The 13$\;$TeV 
analysis was updated for this conference 
with a refined electromagnetic calorimeter calibration 
leading to an about 30\% improved mass resolution above 
$m_{\gamma\gamma} \sim500\;$GeV. CMS also included 0.6$\;$fb$^{-1}$ of 
solenoid-off data. Photons are selected with $E_T(\gamma_{1/2}) > 75\;$GeV
and requiring at least one photon to lie in the barrel (absolute pseudorapidity 
lower than 1.44). The analysis is split into  barrel-barrel and barrel-endcap 
categories that are fit simultaneously. Dedicated energy and efficiency calibrations 
were developed for the solenoid-off data giving a slightly lower photon identification  
efficiency and better energy resolution compared to the solenoid-on data. Also 
the primary vertex finding probability is reduced, which affects the diphoton mass
resolution. The backgrounds in all categories are fit using empirical functions. 

Figure~\ref{fig:CMS-diphotonRes} shows the diphoton invariant mass distributions 
in the four data categories (barrel/endcap, solenoid-on/off). The resulting p-values
versus mass for the narrow-width hypothesis (preferred by the data) are shown 
in Fig.~\ref{fig:CMS-diphotonResSig} for the spin-zero (left panel) and spin-two 
(right panel) cases. In these plots, the red dotted line shows the 13$\;$TeV dataset,
the blue dotted line the 8$\;$TeV dataset, and the black solid line their combination
computed according to the signal model assumed. 
The lowest p-values are found at around 750$\;$GeV mass (760$\;$GeV for the 
13$\;$TeV data alone). The corresponding local / global 
significances are 3.4$\sigma$ / 1.6$\sigma$, reducing to 2.9$\sigma$ / $<1$$\sigma$
for the 13$\;$TeV data alone.  \\[0.0cm]

The upcoming restart of the LHC is expected to clarify the current uncertainty 
in the interpretation of these findings.

\vfill\pagebreak
\section{Conclusions}

\begin{wrapfigure}{R}{0.11\textwidth}
\centering
\vspace{-0.4cm}
\includegraphics[width=0.11\textwidth]{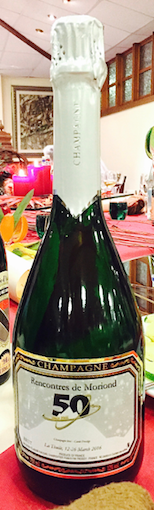}
\vspace{-0.9cm}
\end{wrapfigure}
The 51$^{\rm st}$ edition of the Moriond Electroweak and Unified Theories conference  
has been memorable. It exhibited once again the challenges today's experimental 
physics takes on and overcomes to perform groundbreaking measurements.

The discovery of the Higgs boson required the construction of a huge accelerator 
and ultra-sophisticated particle detectors to produce Higgs-boson events and find 
them in several channels buried under 10$^{12}$ times larger backgrounds. 
The direct observation 
of gravitational waves required to measure over 4$\;$km length  a relative 
deformation two hundred times smaller than the size of a proton. Similar things 
can be said about neutrino physics, dark matter searches, etc. 

Accomplishing these measurements requires great ideas, visionary leadership, 
long-term support by governments and society, innovative and highest quality 
hardware and software, computing resources, operational and maintenance 
support, precise and unbiased analysis, and above all: dedication. 

Given what we have seen this week, I have no worry. We live in an extraordinary 
period for fundamental experimental research in physics. 
Congratulations to the 50$^{\rm th}$ anniversary of the conference. There will be 
ample material for an exciting next half a century! 

{\em \small
I thank the organisers for preparing a fascinating conference week
and for giving me the opportunity to present the experimental summary. 
}

\vfill\pagebreak
{\small \tableofcontents}
\vfill\pagebreak
\section*{References}

\input Bibliography

%% file: Definitions.tex
\def\be{\begin{equation}}
\def\ee{\end{equation}}
\def\bea{\begin{eqnarray}}
\def\eea{\end{eqnarray}}

\newcommand{\seffsf}[1]{\sin\!^2\theta^{#1}_{{\rm eff}}}

\newcommand{\sinfeff}{\ensuremath{\seffsf{f}}\xspace}
\newcommand{\sinleff}{\ensuremath{\seffsf{\ell}}\xspace}

\newcommand{\as}{\ensuremath{\alpha_{\scriptscriptstyle S}}\xspace}

\newcommand{\nub}{{\overline\nu}\xspace}
\newcommand{\nue}{{\nu_e}\xspace}
\newcommand{\nueb}{{\overline\nu_e}\xspace}
\newcommand{\numu}{{\nu_\mu}\xspace}
\newcommand{\numub}{{\overline\nu_\mu}\xspace}
\newcommand{\nutau}{{\nu_\tau}\xspace}

\newcommand{\zn}{0\nu\beta\beta\xspace}
\newcommand{\tn}{2\nu\beta\beta\xspace}

%% file: Bibliography.tex
\bibliographystyle{unsrt}

\end{document}